%% file: CN.tex
\documentclass[british,fleqn,fleqn,usenatbib]{mnras}
\usepackage[latin9]{inputenc}
\PassOptionsToPackage{fleqn}{amsmath}
\PassOptionsToPackage{british}{babel}
\usepackage{array}
\usepackage{amsmath}
\usepackage{amssymb}
\usepackage{graphicx}

\makeatletter

\title[Thick-disc binary stars]{Binary stars in the Galactic thick disc}
\providecommand{\tabularnewline}{\\}
\usepackage[british]{babel}             % British English hyphenation
\usepackage{newtxtext}                  % Good fonts
\usepackage[slantedGreek]{newtxmath}    %   "    "   (slanted Greek)

\usepackage[T1]{fontenc}                % Good font encoding
\usepackage{graphicx}                   % Including figures
\usepackage[fleqn]{amsmath}

\DeclareMathSymbol{,}{\mathpunct}{operators}{"2C}

\usepackage{mleftright}
\renewcommand{\left}{\mleft}
\renewcommand{\right}{\mright}

\let\jnl@style=\rmfamily 
\def\ref@jnl#1{{\jnl@style#1}}% 
\providecommand\aj{\ref@jnl{AJ}}%        % Astronomical Journal 
\providecommand\araa{\ref@jnl{ARA\&A}}%  % Annual Review of Astron and Astrophys 
\providecommand\apj{\ref@jnl{ApJ}}%    % Astrophysical Journal ++
\providecommand\apjl{\ref@jnl{ApJL}}     % Astrophysical Journal, Letters 
\providecommand\apjs{\ref@jnl{ApJS}}%    % Astrophysical Journal, Supplement 
\providecommand\ao{\ref@jnl{ApOpt}}%   % Applied Optics ++
\providecommand\apss{\ref@jnl{Ap\&SS}}%  % Astrophysics and Space Science 
\providecommand\aap{\ref@jnl{A\&A}}%     % Astronomy and Astrophysics 
\providecommand\aapr{\ref@jnl{A\&A~Rv}}%  % Astronomy and Astrophysics Reviews 
\providecommand\aaps{\ref@jnl{A\&AS}}%    % Astronomy and Astrophysics, Supplement 
\providecommand\azh{\ref@jnl{AZh}}%       % Astronomicheskii Zhurnal 
\providecommand\baas{\ref@jnl{BAAS}}%     % Bulletin of the AAS 
\providecommand\icarus{\ref@jnl{Icarus}}% % Icarus
\providecommand\jrasc{\ref@jnl{JRASC}}%   % Journal of the RAS of Canada 
\providecommand\memras{\ref@jnl{MmRAS}}%  % Memoirs of the RAS 
\providecommand\mnras{\ref@jnl{MNRAS}}%   % Monthly Notices of the RAS 
\providecommand\pra{\ref@jnl{PhRvA}}% % Physical Review A: General Physics ++
\providecommand\prb{\ref@jnl{PhRvB}}% % Physical Review B: Solid State ++
\providecommand\prc{\ref@jnl{PhRvC}}% % Physical Review C ++
\providecommand\prd{\ref@jnl{PhRvD}}% % Physical Review D ++
\providecommand\pre{\ref@jnl{PhRvE}}% % Physical Review E ++
\providecommand\prl{\ref@jnl{PhRvL}}% % Physical Review Letters 
\providecommand\pasp{\ref@jnl{PASP}}%     % Publications of the ASP 
\providecommand\pasj{\ref@jnl{PASJ}}%     % Publications of the ASJ 
\providecommand\qjras{\ref@jnl{QJRAS}}%   % Quarterly Journal of the RAS 
\providecommand\skytel{\ref@jnl{S\&T}}%   % Sky and Telescope 
\providecommand\solphys{\ref@jnl{SoPh}}% % Solar Physics 
\providecommand\sovast{\ref@jnl{Soviet~Ast.}}% % Soviet Astronomy 
\providecommand\ssr{\ref@jnl{SSRv}}% % Space Science Reviews 
\providecommand\zap{\ref@jnl{ZA}}%       % Zeitschrift fuer Astrophysik 
\providecommand\nat{\ref@jnl{Nature}}%  % Nature 
\providecommand\iaucirc{\ref@jnl{IAUC}}% % IAU Cirulars 
\providecommand\aplett{\ref@jnl{Astrophys.~Lett.}}%  % Astrophysics Letters 
\providecommand\apspr{\ref@jnl{Astrophys.~Space~Phys.~Res.}}% % Astrophysics Space Physics Research 
\providecommand\bain{\ref@jnl{BAN}}% % Bulletin Astronomical Institute of the Netherlands 
\providecommand\fcp{\ref@jnl{FCPh}}%   % Fundamental Cosmic Physics 
\providecommand\gca{\ref@jnl{GeoCoA}}% % Geochimica Cosmochimica Acta 
\providecommand\grl{\ref@jnl{Geophys.~Res.~Lett.}}%  % Geophysics Research Letters 
\providecommand\jcp{\ref@jnl{JChPh}}%     % Journal of Chemical Physics 
\providecommand\jgr{\ref@jnl{J.~Geophys.~Res.}}%     % Journal of Geophysics Research 
\providecommand\jqsrt{\ref@jnl{JQSRT}}%   % Journal of Quantitiative Spectroscopy and Radiative Trasfer 
\providecommand\memsai{\ref@jnl{MmSAI}}% % Mem. Societa Astronomica Italiana 
\providecommand\nphysa{\ref@jnl{NuPhA}}%     % Nuclear Physics A 
\providecommand\physrep{\ref@jnl{PhR}}%       % Physics Reports 
\providecommand\physscr{\ref@jnl{PhyS}}%        % Physica Scripta 
\providecommand\planss{\ref@jnl{Planet.~Space~Sci.}}%  % Planetary Space Science 
\providecommand\procspie{\ref@jnl{Proc.~SPIE}}%      % Proceedings of the SPIE 

\providecommand\actaa{\ref@jnl{AcA}}%  % Acta Astronomica
\providecommand\caa{\ref@jnl{ChA\&A}}%  % Chinese Astronomy and Astrophysics
\providecommand\cjaa{\ref@jnl{ChJA\&A}}%  % Chinese Journal of Astronomy and Astrophysics
\providecommand\jcap{\ref@jnl{JCAP}}%  % Journal of Cosmology and Astroparticle Physics
\providecommand\na{\ref@jnl{NewA}}%  % New Astronomy
\providecommand\nar{\ref@jnl{NewAR}}%  % New Astronomy Review
\providecommand\pasa{\ref@jnl{PASA}}%  % Publications of the Astron. Soc. of Australia
\providecommand\rmxaa{\ref@jnl{RMxAA}}%  % Revista Mexicana de Astronomia y Astrofisica

\providecommand\maps{\ref@jnl{M\&PS}}% Meteoritics and Planetary Science
\providecommand\aas{\ref@jnl{AAS Meeting Abstracts}}% American Astronomical Society Meeting Abstracts
\providecommand\dps{\ref@jnl{AAS/DPS Meeting Abstracts}}% American Astronomical Society/Division for Planetary Sciences Meeting Abstracts

\providecommand\cac{\ref@jnl{Computational Astrophysics and Cosmology}}

\usepackage[export]{adjustbox}
\usepackage{xcolor}

\newcommand{\change}[1]{#1}
\newcommand{\changemore}[1]{#1}

\usepackage[T1]{fontenc}
\usepackage{aecompl}
\makeatother

\usepackage{babel}
\begin{document}
% in the main text
\author[R.G.Izzard~et~al.]
{Robert~G.~Izzard,$^{1}$
Holly~Preece,$^{1,2}$
Paula~Jofre,$^{1,3}$
Ghina~M.~Halabi,$^{1}$
\newauthor 
Thomas~Masseron,$^{1,4,5}$
and
Christopher~A.~Tout$^{1}$
\\ 
$^{1}$Institute of Astronomy, Madingley Road, Cambridge, CB3 0HA, United Kingdom.\\
$^{2}$Armagh Observatory, College Hill, Armagh, BT61 9DG, United Kingdom.\\
$^{3}$N\'ucleo de Astronom\'ia, Facultad de Ingenier\'ia, Universidad Diego Portales,  Av. Ej\'ercito 441, Santiago, Chile.\\ 
$^{4}$Instituto de Astrofísica de Canarias, E-38205 La Laguna, Tenerife, Spain.\\
$^{5}$Departamento de Astrofísica, Universidad de La Laguna, E-38206 La Laguna, Tenerife, Spain.
}
\input{csnames}
\newcommand\data[1]{{\expandafter\csname data#1\endcsname}}
\input{CN_extra_data_macros}
% this is submission version 3

\date{Received ...; Accepted...}

\maketitle
\global\long\def\pc{\mathrm{\,per\,cent}}
\global\long\def\msun{\mathrm{M}_{\odot}}
\global\long\def\rsun{\mathrm{R}_{\odot}}
\global\long\def\lsun{\mathrm{L}_{\odot}}
\global\long\def\sun{\odot}
\global\long\def\CN{\left[\mathrm{C}/\mathrm{N}\right]}
\global\long\def\FeH{\left[\mathrm{Fe}/\mathrm{H}\right]}
\global\long\def\aFe{\left[\alpha/\mathrm{Fe}\right]}
\global\long\def\APO{\textnormal{\ensuremath{\mathrm{APOKASC}}}}
\global\long\def\APObold{\textbf{\ensuremath{\mathrm{\mathbf{APOKASC}}}}}
\global\long\def\APOGEE{\textnormal{\ensuremath{\mathrm{APOGEE}}}}
\global\long\def\CFe{\left[\mathrm{C}/\mathrm{Fe}\right]}
\global\long\def\NFe{\left[\mathrm{N}/\mathrm{Fe}\right]}
\global\long\def\Mlimit{1.3\mathrm{\,M_{\odot}}}
\global\long\def\agemin{5}
\global\long\def\agemax{10}
\global\long\def\Zsolar{0.014}
\global\long\def\Zthickdisc{0.008}
\global\long\def\defaultZ{\Zthickdisc}
\global\long\def\STARS{\normalfont\textsc{{stars}}}
\global\long\def\HTCondor{\normalfont\textsc{{htcondor}}}
\global\long\def\binaryc{\normalfont\textsc{binary\_c}}
\global\long\def\Binaryc{\normalfont\textsc{binary\_c}}
\global\long\def\APOpc{14\pc}
\global\long\def\Gyr{\mathrm{Gyr}}
\global\long\def\CNdef{\log_{10}\left(N_{\mathrm{C}}/N_{\mathrm{C},\odot}\right)-\log_{10}\left(N_{\mathrm{N}}/N_{\mathrm{N},\odot}\right)}
\global\long\def\FeHdef{\log_{10}\left(N_{\mathrm{Fe}}/N_{\mathrm{Fe}\odot}\right)-\log_{10}\left(N_{\mathrm{H}}/N_{\mathrm{H}\odot}\right)}
\global\long\def\BSE{\normalfont\textsc{bse}}
\global\long\def\SSE{\normalfont\textsc{sse}}
\global\long\def\logg{\log_{10}\left(g/\mathrm{cm}^{2}\mathrm{s}^{-1}\right)}
\global\long\def\loggsquare{\log_{10}\left[g/\mathrm{cm}^{2}\mathrm{s}^{-1}\right]}
 \global\long\def\hyphen{\text{--}}
\begin{abstract}
The combination of asteroseismologically-measured masses with abundances
from detailed analyses of stellar atmospheres challenges our fundamental
knowledge of stars and our ability to model them. Ancient red-giant
stars in the Galactic thick disc are proving to be most troublesome
in this regard. They are older than $5\,\mathrm{Gyr}$, a lifetime
corresponding to an initial stellar mass of about $1.2\mathrm{\,M_{\odot}}$.
So why do the masses of a sizeable fraction of thick-disc stars exceed
$\Mlimit$, with some as massive as $2.3\mathrm{\,M_{\odot}}$? We
answer this question by considering duplicity in the thick-disc stellar
population using a binary population-nucleosynthesis model. We examine
how mass transfer and merging affect the stellar mass distribution
and surface abundances of carbon and nitrogen. We show that a few
per cent of thick-disc stars \change{can interact in binary star
systems and become} more massive than $\Mlimit$. Of these stars,
most are single because they are merged binaries. Some stars more
massive than $\Mlimit$ form in binaries by wind mass transfer. We
compare our results to a sample of the $\APO$ data set and find reasonable
agreement except in the number of these thick-disc stars more massive
than $\Mlimit$. This problem is resolved by the use of a logarithmically-flat
orbital-period distribution and a large binary fraction.
\end{abstract}
Stars: binaries: general \textendash{} Stars: abundances \textendash{}
Galaxy: disc \textendash{} Galaxy: abundances \textendash{} Galaxy:
stellar content

\section{Introduction}

\label{sec:Introduction}

The stellar evolution of low-mass, those lighter than about $2\mathrm{\,M_{\odot}}$,
single stars is reasonably well understood. Stars form in radiating,
collapsing clouds of mostly hydrogen and helium. Compression heats
the centre of the clouds until nuclear burning begins in their cores
and stars are born. Stars burn hydrogen on the main sequence then
ascend the red-giant branch when hydrogen-burning moves into a shell
surrounding a helium core \citep*{1952ApJ...116..463S}. Subsequently,
helium ignites and a stage of core helium-burning, also called the
red-clump, follows. Helium subsequently burns in a shell as the star
ascends the asymptotic giant branch. Surface mass loss terminates
single-star evolution at this stage leaving a carbon-oxygen core which
cools into a white dwarf. The main parameters that drive the evolution
of single, low-mass stars are the total stellar mass, $M$, and the
metallicity, $Z$, which is the mass fraction of all elements heavier
than hydrogen or helium. The ratio $\FeH=\FeHdef$ is often used as
a proxy for the metallicity, where $N_{i}$ is the surface number
density of species $i$.

The Milky Way contains several populations of low-mass stars that
have different metallicities, kinematics and star-formation histories
\citep{2016ARA&A..54..529B}. The thin disc is a relatively high-metallicity
population in a large disc surrounding the centre of the Galaxy. Our
Sun is a thin-disc star with $Z=Z_{\odot}\approx\Zsolar$ or $\FeH=0$
by definition \citep{Asplund2009,2010ASSP...16..379L}. At the centre
of the Galaxy is the bulge, which contains many old stars at low,
solar and higher metallicity. The Galactic halo surrounds the Milky
Way out to long distances and contains very old, low-metallicity stars.
The remaining stellar population is the Galactic thick disc \citep{Gilmore1983}.
It is thick because its stars have greater spatial velocities. The
population is kinetically warmer and more extended out of the Galactic
plane, than the thin disc. Generally, thick-disc stars are metal-poor
compared to the thin disc but there is substantial overlap in the
metallicity distributions of the discs \citep{2011MNRAS.412.1203N}.
Thick-disc stars are enhanced in alpha elements such as magnesium,
as measured through the ratio $\aFe$. In the solar neighbourhood
there are fewer stars in the thick than in the thin disc \citep{2017MNRAS.464.2610F}.
Even though the thick disc is a minor stellar component locally, its
properties are vital clues to understanding the structure and formation
of the Milky Way and galaxies in general \citep{2012MNRAS.426..690B,2016ARA&A..54..529B,2017ApJ...834...27M}.
The origin of the thick disc remains the subject of debate. Galactic
formation models predict few young stars in the thick disc because
it formed rapidly soon after the Milky Way was born. The ages of thick-disc
stars are thus vital to disentangle the process of galaxy formation
and its application to the Milky Way.

The thick disc is at least $5\,\Gyr$ old\textbf{ }and hence its stars
are of low mass \citep{2016ApJ...831..139M}. The main-sequence lifetime
of a $Z=0.008$, $M=1.15\mathrm{\,M_{\odot}}$, single star is about
$5.2\,\Gyr$, while a $Z=0.008$, $M=0.96\mathrm{\,M_{\odot}}$, single
star lives for $10.0\,\Gyr$ (based on \citealp{1995MNRAS.274..964P}'s
models). After the main sequence, stars spend about one tenth of the
main-sequence lifetime ascending the red giant branch. The initial
masses of $5-10\,\mathrm{Gyr}$ stars lie in a narrow range of mass
from about $0.95$ to $1.2\mathrm{\,M_{\odot}}$ with a small spread
owing to a variation in metallicity. In this regard, the thick-disc
is similar to a globular cluster in which stars are born roughly at
the same time with one metallicity. The properties of thick-disc red
giants thus map stellar evolution corresponding to a narrow range
of initial parameters.

Recent advances in observational surveys have improved our understanding
of the thick-disc stellar population. The \emph{Kepler} mission has
allowed the measurement of stellar masses by asteroseismology \citep[e.g.][]{Miglio2012}.
Combining stellar masses with follow-up spectroscopy using the Apache
Point Observatory Galactic Evolution Experiment (APOGEE), the APOGEE\textendash Kepler
Asteroseismic Science Consortium ($\APO$) data of \citet{2014ApJS..215...19P}
provide stellar properties, including surface gravity $\log g$ and
chemical abundances such as $\FeH$, $\aFe$ and $\CN$, which can
be compared with stellar evolution models. The $\CN$ ratio is particularly
important because it is expected to be reduced by the mixing of nuclear-processed
material to the stellar surface during ascent of the red giant branch.
This mixing is due to either first dredge up, caused by a deepening
of the stellar convection zone, or other mechanisms such as thermohaline
mixing, magnetic mixing or rotational mixing, all of which are known
as extra-mixing (e.g.~\citealt{2000A&A...356..181W,2009MNRAS.396.2313S,ChaLag10}).

Galactic thick disc stars, because of their low mass, are not expected
to much change their stellar surface $\CN$ during ascent of the red-giant
branch \citep{Salaris2015}. The convection zone does not mix to great
depth, so little nitrogen is mixed to the surface. Thus it is surprising
that some stars, selected from the $\APO$ sample to be thick-disc
members by virtue of their $\aFe$, have $\CN$ as low as $-0.8\,\mathrm{dex}$
\citep{2017MNRAS.464.3021M}. Even more strangely, a number of the
$\APO$ thick-disc stars have masses, measured from asteroseismological
scaling relations, in excess of the $1.2\hyphen1.3\mathrm{\,M_{\odot}}$
upper limit expected from canonical stellar evolution of stars so
old. These stars challenge both stellar and Galactic evolution models.
Their high masses and alpha enhancements imply that they may have
formed in the Galactic centre and subsequently migrated into the thick
disc \citep{Chiappini2015}. However, we already know from radial-velocity
monitoring that many of the extra-massive thick-disc stars are plausibly
multiple systems \citep{Jofre2016}. In this work we test the possibility
that these stars originate in multiple stellar systems and their properties
arise from binary-star interactions \citep{2017PASA...34....1D}. 

Stars that are more massive than they should be in a population of
fixed age are well known to those who study globular clusters. On
the main sequence these stars are bluer and brighter. They look younger
than stars at the \change{blue} end of the main sequence and so
they are known as blue stragglers\emph{. }Blue stragglers form by
mass transfer or stellar merging in multiple stellar systems \citep[e.g.][]{2002ApJ...568..939L,2011Natur.478..356G}.
In binary stars, mass transfer either by direct Roche-lobe overflow
or wind mass transfer is relatively common \citep{2017PASA...34....1D}.
The increase in mass of a star causes it to look younger than it actually
is \citep{1953AJ.....58...61S,1997MNRAS.291..732T}. There is every
reason to believe that such stellar systems exist in the Galactic
thick disc \citep{Jofre2016}, however they are more difficult to
identify than in stellar clusters because of the lack of a clear turn-off
in the Hertzsprung\textendash Russell or colour\textendash magnitude
diagram.

The stars in the $\APO$ thick-disc sample with mass in excess of
$\Mlimit$ are not blue stragglers. They are red giant stars which
may once have been blue stragglers. Indeed, as we show below, a fraction
of the extra-massive thick-disc stars were probably once blue stragglers.
However mass transfer in a binary system is most likely after the
main sequence as the more massive star in the binary ascends the giant
branch. This can lead to an increase in mass of a companion star or
to common-envelope evolution in which the core of the giant and its
companion orbit each other inside a shared stellar envelope \citep{Izzard2012,2013A&ARv..21...59I}.
Often this leads to envelope ejection but alternatively the stars
merge to form a new giant. These stars are quite possibly those that
are seen in the $\APO$ sample.

In the following we assess the possibility that binary stars make
up at least some of the thick-disc stars with mass in excess of $\Mlimit$.
In section~\ref{sec:Stellar-modelling} we combine detailed stellar
evolution models of first dredge up in red giants with a stellar population\textendash nucleosynthesis
model to construct populations of red-giant stars suitable for comparison
with observations of thick-disc stars. We select a thick disc sample
from $\APO$ in section~\ref{sec:APOKASC}. In section~\ref{sec:Results}
we examine the properties of our model stars while varying model parameters
and initial distributions of masses and periods within reasonable
uncertainties. We discuss the implications of our results and suggestions
for future modelling in section~\ref{sec:Discussion}. Our conclusions
then follow in section~\ref{sec:Conclusions}.

\section{Stellar modelling}

\label{sec:Stellar-modelling}We employ two stellar evolution codes
to model Galactic thick-disc stars. First, in section~\ref{subsec:Detailed-modelling},
we calculate a set of detailed stellar evolution models with the Cambridge
\emph{$\STARS$} code to model the nucleosynthesis of carbon and nitrogen
from the pre-main sequence to helium ignition in stars of mass $0.8$
to $20\,\msun$ with $10^{-4}\leq Z\leq0.03$. Second, in sections~\ref{subsec:Nucleosynthesis-with-binary_c}\textendash \ref{subsec:Stellar-mass-loss}
we describe how we embed our \emph{$\STARS$} data into our \emph{$\binaryc$
}population-nucleosynthesis code. \emph{$\Binaryc$} models millions
of single and binary stars quickly, so it is ideal to explore the
large parameter space associated with binary stars. section~\ref{subsec:Stellar-populations-with-binary-c}
describes how we model the thick-disc stellar population.

\subsection{Detailed stellar modelling with \emph{$\protect\STARS$}}

\label{subsec:Detailed-modelling}We construct a grid of 420 detailed
stellar models with logarithmically distributed masses, $M$, in the
range $0.8$ to $20\mathrm{\,M_{\odot}}$ and metallicities $Z=10^{-4}$,
$3\times10^{-4}$, $0.001$, $0.004$, $0.02$ and $0.03$ using the
\emph{$\STARS$} stellar evolution code \citep{Eggleton1971,1995MNRAS.274..964P}.
Each metallicity has a corresponding opacity table which accommodates
changes in carbon and oxygen compared to the solar mixture \citep{2004MNRAS.348..201E}.
Initial abundances, by mass fraction, are scaled by $Z/0.02$ with
the mix of \citet{Anders1989}\change{. The hydrogen abundance $X=0.76-3Z$}
and helium abundance $Y=0.24+2Z$. Our initial models start on the
pre-main sequence so that any chemical abundance profile set up during
that phase is preserved when the star begins central hydrogen burning.
\change{The models each have 999~mesh points to minimise numerical
diffusion.} We set the convective mixing length parameter \change{$\alpha_{\mathrm{MLT}}=2.0$},
the overshoot parameter \change{$\delta_{\mathrm{ov}}=0.12$ \citep{1997MNRAS.285..696S}}
and \change{define the convectively-mixed stellar envelope by $\nabla_{\mathrm{r}}-\nabla_{\mathrm{a}}\geq0.01$},
where $\nabla_{\mathrm{r}}$ and $\nabla_{\mathrm{a}}$ are the radiative
and adiabatic logarithmic gradients of temperature\change{, to avoid
numerical noise at its lower boundary}. We disable mass loss. Our
input models and their parameters, such as radius and luminosity,
\change{are calibrated to those measured for the Sun} \citep{2009MNRAS.396.1699S}.

All these models all pass through the terminal-age main sequence (TAMS)
and evolve \change{to} helium ignition\change{. First} dredge
up is modelled in all our stars. \change{The initial time step is
chosen to be shorter than $100\,\mathrm{yr}$ and is then allowed
to vary to keep the mean modulus of the step-to-step variation in
the independent variables of the code close to constant.} \textbf{}
We define the TAMS to be the time when the central hydrogen mass fraction
drops below $10^{-5}$. Profiles of elemental abundances of hydrogen,
helium, carbon, nitrogen and oxygen are sampled at $100$~\change{Lagrangian
mass co-ordinates $m=m(r)$, equally-spaced between the centre and
surface,} in each star at the TAMS. These profiles are exported to
a lookup table as a function of $M$, $m/M$ and $Z$ \change{with}
$0\leq m/M\leq1$ and $0\leq r/R\leq1$, where $R$ is the stellar
radius. In addition, the depth of first dredge up, the post dredge
up surface carbon and nitrogen abundances and the surface gravity
when the convective envelope is at its deepest are tabulated as a
function of $M$ and $Z$.

\subsection{Nucleosynthesis with \emph{$\protect\binaryc$}}

\emph{\label{subsec:Nucleosynthesis-with-binary_c}} We use the stellar
population nucleosynthesis code $\binaryc$ to model populations of
single and binary stars. This is a C\emph{ }port of the binary star
evolution ($\BSE$) algorithm developed by \citet{2002MNRAS_329_897H}
extended to include nucleosynthesis \citep{Izzard_et_al_2003b_AGBs,2006A&A...460..565I,2009A&A...508.1359I}.
Recent relevant updates include Wind Roche lobe overflow \citep[WRLOF,][]{2013A&A...552A..26A,2015A&A...581A..62A},
an improved treatment of stellar rotation \citep{2013ApJ...764..166D},
updated stellar lifetimes \citep{2014ApJ...780..117S} and an improved
algorithm describing the rate of Roche-lobe overflow \citep{2014A&A...563A..83C}.

Previous versions of $\binaryc$ treated first dredge up as an instantaneous
event on the giant branch. Stellar surface abundances were shifted
by amounts tabulated as a function of total stellar mass based on
detailed stellar evolution models \citep{Karakas2002}. This approach
is insufficiently accurate for this work so we improve the model of
first dredge up in \emph{$\binaryc$} in stars with $0.8\leq M/\mathrm{M_{\odot}}\leq20$
by modelling the stellar envelope as a set of shells, as would a detailed
stellar evolution code. Each star initially has 200 shells equally
spaced in mass\emph{.}

Initial chemical abundance profiles are linearly interpolated in mass
and metallicity based on the TAMS abundance tables of section~\ref{subsec:Detailed-modelling}.
The chemical profile in the interior at the TAMS arises from pre-main
sequence (PMS) CN burning and main-sequence pp and CN burning. The
CN profile created during the PMS is unchanged during the main sequence
except in the stellar core, so the TAMS abundance profile well represents
the envelope abundance profile at any stage of the main sequence.
In the core the abundance profile changes with time as hydrogen is
converted to helium, but it is difficult to expose core material unless
there is extreme mass loss so we assume the TAMS abundances apply.
Accreted mass is unlikely to mix into the core (section~\ref{subsec:Stellar-mass-loss};
also \citealp{2007A&A...464L..57S}). Wind mass loss is negligible
on the main sequence except in the most massive of our stars. Even
then, it occurs near the end of the main sequence so the TAMS profile
is a reasonable approximation to the true chemical profile.

Hertzsprung gap evolution is too fast for significant nuclear burning
so again the TAMS abundances well represent the stellar interior.
When the star ascends the giant branch its helium core grows and its
convective envelope deepens during first dredge up. This deepening
is relevant to our stellar population because we select such giants
for comparison with observations. Chemical changes during canonical
first dredge up are a result only of mixing, not nuclear burning,
so the TAMS profile is suitable for describing this process (section~\ref{subsec:First-dredge-up}).

After helium ignition at the tip of the red giant branch, second and
third dredge up in $\binaryc$ are treated as described by \citet{2006A&A...460..565I,2009A&A...508.1359I},
although neither process is important to our main conclusions. Our
third dredge up efficiency is that of \citet{Karakas2002} without
the enhancement required to match the number of carbon-enhanced metal-poor
stars \citep{2009A&A...508.1359I} or Magellanic-cloud carbon-star
luminosity functions \citep{Binary_Origin_low_L_C_Stars}.

Our introduction of a shell structure increases the run time of \emph{$\binaryc$}
from about $0.1$ to $1\,\mathrm{s}$ per system. While this slows
our computations significantly, we take advantage of modern multi-core
CPUs, \emph{$\HTCondor$} and the latest \emph{$\binaryc$} support
tools to offset the extra cost.

\subsection{First dredge up in $\protect\binaryc$}

\label{subsec:First-dredge-up}
\begin{figure*}
\begin{tabular}{cc}
\includegraphics[width=0.5\textwidth]{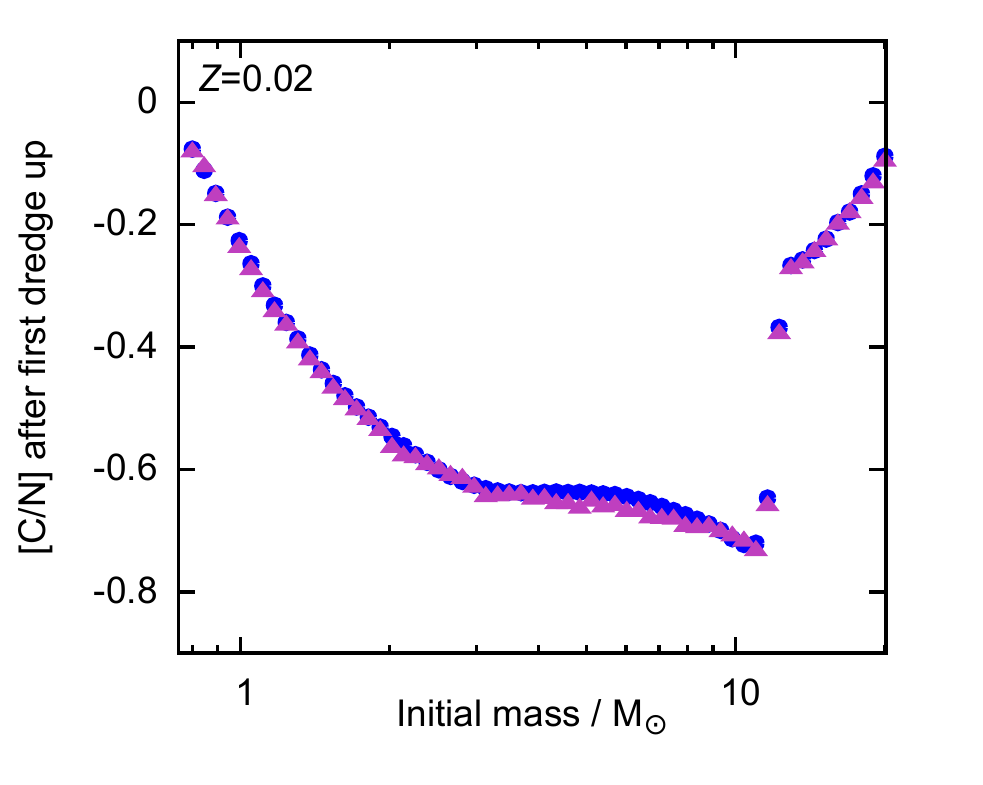} & \includegraphics[width=0.5\textwidth]{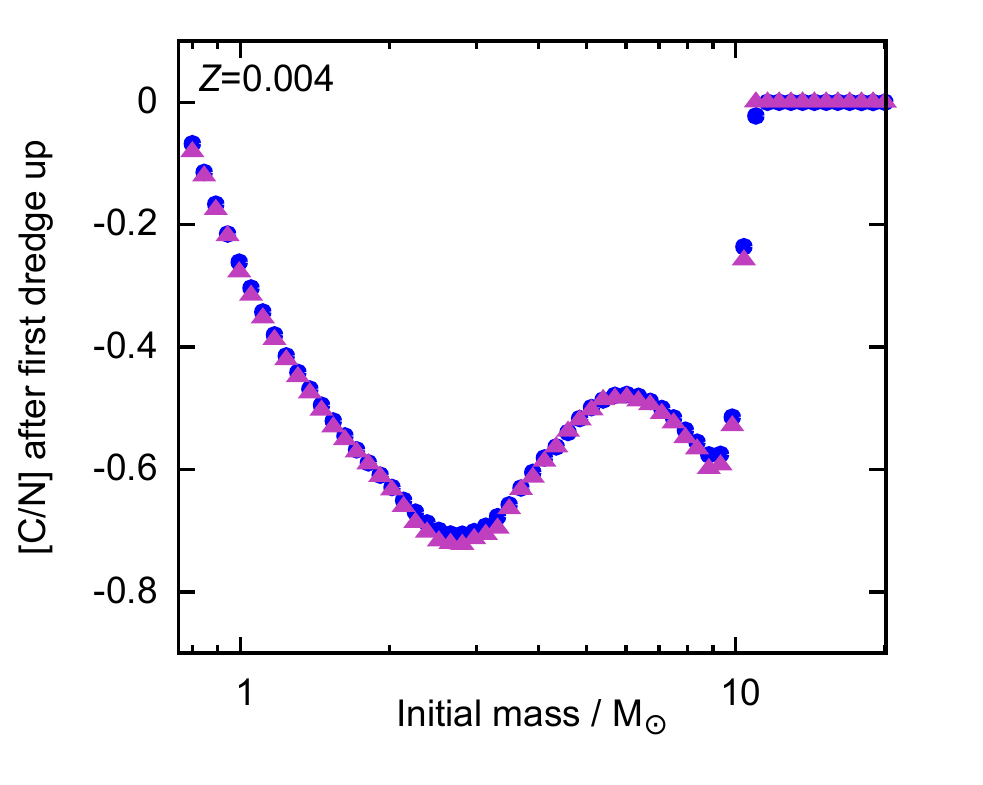}\tabularnewline
\includegraphics[bb=0bp 0bp 283bp 226bp,width=0.5\textwidth]{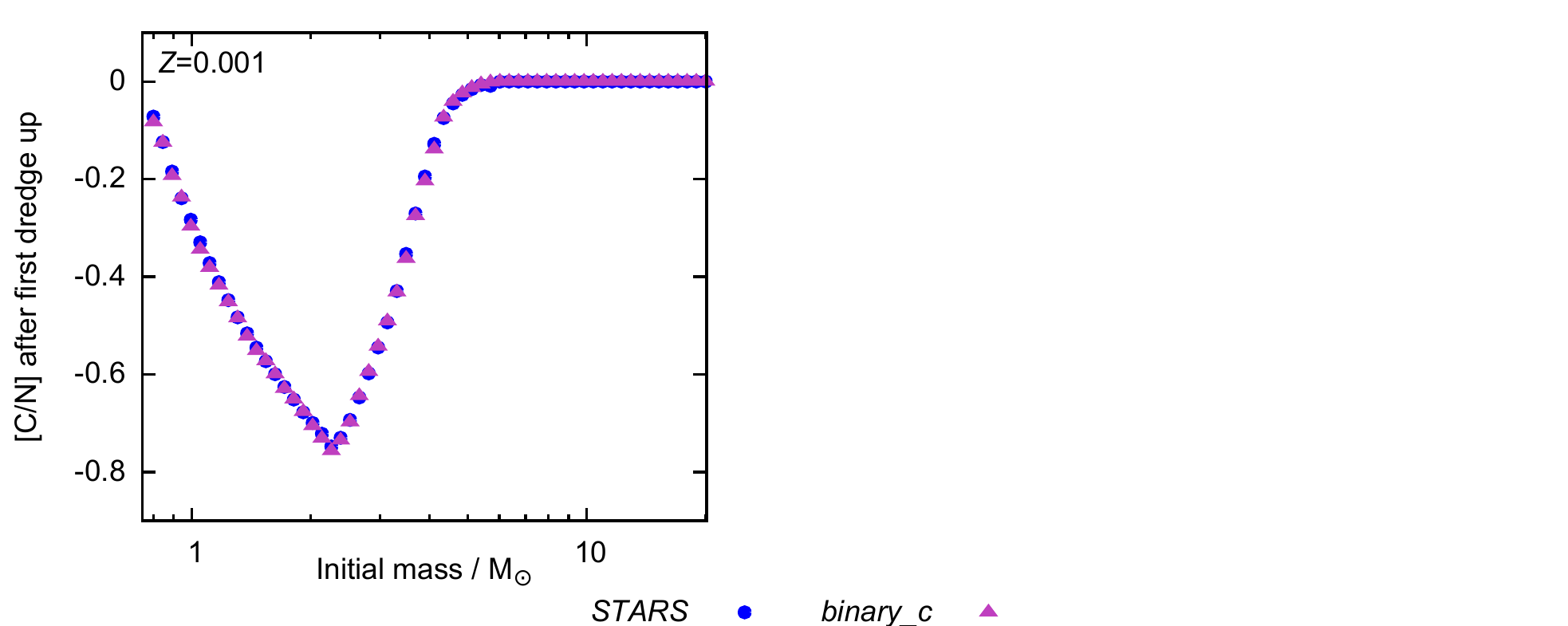} & \includegraphics[width=0.5\textwidth]{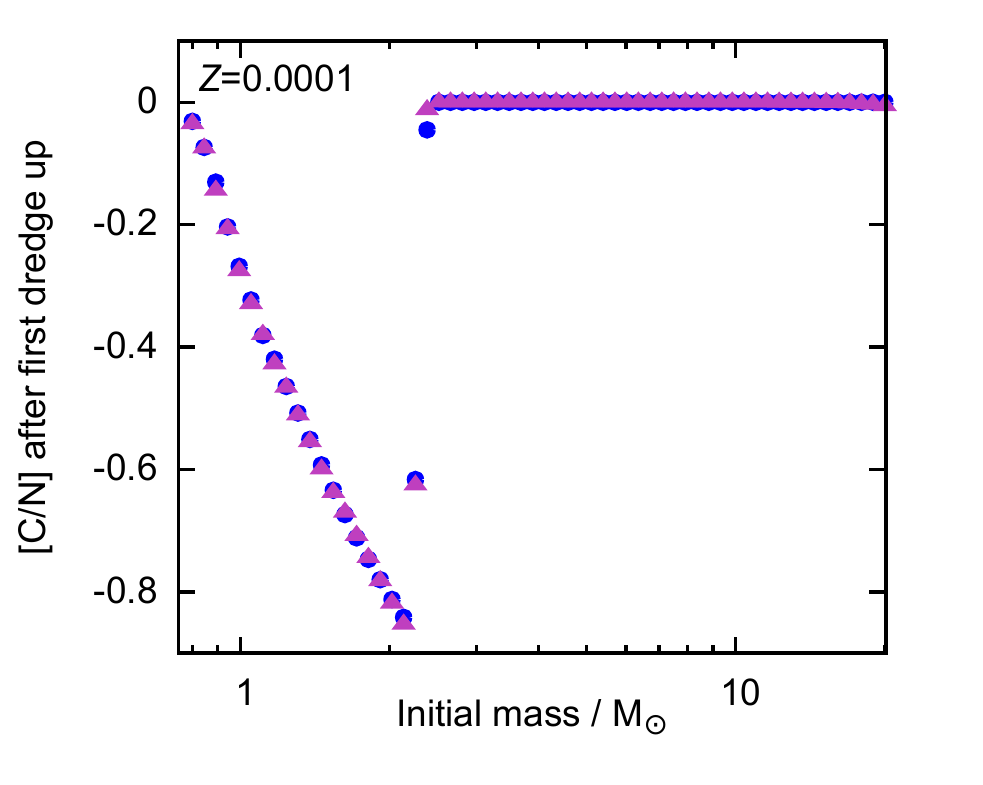}\tabularnewline
\end{tabular}

\caption{\label{fig:CNdredge}The ratio $\protect\CN=\protect\CNdef$ after
first dredge up in our detailed stellar evolution models calculated
with the \emph{$\protect\STARS$} code (blue circles) and our rapid
stellar evolution models calculated with \emph{$\protect\binaryc$}
(purple triangles). The \emph{$\protect\binaryc$} models each run
in less than a second yet reproduce the \emph{$\protect\STARS$} results
to within fractions of a dex over the full mass and metallicity range. }
\end{figure*}

First dredge up is modelled by homogenizing
the stellar convective envelope to a depth $M_{\mathrm{DUP}}$ given
by that calculated in \emph{$\binaryc$} based on the algorithm of
\emph{$\BSE$. }We also limit $M_{\mathrm{DUP}}$ to the dredge up
depth calculated from our $\STARS$ models as a tabulated function
of the mass, $M_{0}$ , of the star at the base of the giant branch
(the notation of \citealp{2000MNRAS.315..543H}). This more accurate
limit is required to match the \emph{$\binaryc$} and $\STARS$ surface
abundances, namely $\CN=\log_{10}\left(N_{\mathrm{C}}/N_{\mathrm{C},\odot}\right)-\left(N_{\mathrm{N}}/N_{\mathrm{N},\odot}\right)$,
after first dredge up to within $0.1\,\mathrm{dex}$. A comparison
between the post-dredge up surface $\CN$ in our \emph{$\binaryc$}
and $\STARS$ models at various metallicities is shown in Fig.~\ref{fig:CNdredge}.
The logarithmic abundances at the stellar surface calculated with
the two codes match to within a few hundredths. This is more accurate
than the observations to which we compare and hence sufficient for
our purposes, especially given that $\binaryc$ remains many orders
of magnitude faster than $\STARS$ yet gives the same result.

\subsection{Stellar mass loss and gain in $\protect\binaryc$ }

\label{subsec:Stellar-mass-loss} In our single-star models, mass
is lost only by stellar winds, the rates of which are described in
Appendix~\ref{sec:stellar-winds}. In binary stars, mass is lost
by both stellar winds and Roche-lobe overflow (RLOF). Our default
RLOF prescription follows \citet{2014A&A...563A..83C}. This modifies
the algorithm of \emph{$\BSE$ }by enhancing the mass-transfer rate
\change{so} that the stellar radius $R$ remains close to the Roche
lobe radius $R_{\mathrm{L}}$ while maintaining numerical stability.
This algorithm compares well to detailed binary-evolution models \citep*{2001A&A...369..939W}
which enforce $R=R_{\mathrm{L}}$ during both fast and slow case-A
mass transfer \citep{2014ApJ...780..117S}.

We model chemical changes that result from mass loss in $\binaryc$
by removing material from successive shells at the surface of the
star, so eventually exposing nuclear-processed material at depth.
Mass loss is most likely to occur at the end of the main sequence
or during subsequent evolution, so our assumption that the abundance
profile in the star is that at the TAMS is justified. We assume that
first dredge up reaches the same depth as in a star with the same
initial mass but without mass loss (see also section~\ref{subsec:Model-uncertainties:-binary-nucsyn}).

Stars gain material either by Roche-lobe overflow or wind mass transfer.
Accretion is modelled in \emph{$\binaryc$} by adding mass into shells
at the surface with the chemical abundance of the surface of the companion
star. Each shell is allowed to increase its mass up to 1/200 the initial
stellar mass, at which point a new shell is added. If the number of
shells exceeds 750, neighbouring shells are merged in pairs to halve
the number of shells. This limit is chosen to keep code run time short
while maintaining accuracy.

Wind mass transfer follows the formalism of \citet{2013A&A...552A..26A}
which is based on the smoothed-particle hydrodynamic simulations of
\citet{2007ASPC..372..397M}. This algorithm increases the wind accretion
rate relative to the Bondi-Hoyle prescription \citep{Bondi1944} in
binaries with periods between $10^{3}$ and $10^{5}\,\mathrm{d}$.
In such systems slow winds from red-giant stars are channelled inside
the Roche lobe\change{ of the accretor so that} accretion is \change{very}
efficient. The numbers of barium, CH and carbon-enhanced metal poor
(CEMP) stars are affected by the wind accretion rate. \citet{2015A&A...581A..62A}
found that extra mass accretion helps to reconcile the paucity of
CEMP stars in our population models to the number observed \citep{2014ApJ...788..131L}.

When a star accretes, either by RLOF or from a wind, the material
it gains may have a greater molecular weight than the stellar envelope
below it. If so, thermohaline mixing acts to homogenize the upper
part of the envelope until its molecular weight matches the molecular
weight of material immediately below the mixed zone. \changemore{Our
new} approach leads to less dilution of accreted material than in
our previous study of CEMP stars which assumed full mixing of the
star \citep{2009A&A...508.1359I}. \change{The detailed models of
\citet{2007A&A...464L..57S} show that even a small amount of accretion
can cause most of the star to mix. They accrete $0.09\mathrm{\,M_{\odot}}$
of helium-enhanced material, typical of $2\mathrm{\,M_{\odot}}$ AGB
ejecta with molecular weight $0.657$, on to a $0.74\mathrm{\,M_{\odot}}$,
$Z=10^{-4}$ star with molecular weight $0.593$. The resulting $0.83\mathrm{\,M_{\odot}}$
star mixes by the thermohaline instability to within about $0.1\mathrm{\,M_{\odot}}$
of its centre in one tenth of the star's main-sequence life time.
\changemore{We assume thermohaline mixing is instantaneous even though
it is} expected to act on the stellar thermal time-scale. In all
our stars this is much shorter than, typically $10\pc$ of, the main-sequence
life time (\citealp{2008ApJ...679.1541D} argue that the mixing time-scale
should perhaps be much longer than the star's thermal time-scale).
We do not include gravitational settling, radiative levitation \changemore{or
rotational mixing} which could also \changemore{alter} the surface
abundances \changemore{of} old stars \changemore{\citep{2016A&A...592A..29M,2017arXiv170709434M}}}.

When two dwarf stars merge their combined envelope is sorted by molecular
weight. This simulates the results of detailed stellar evolution models
of low-mass merged stars by the use of molecular weight as a proxy
for entropy \citep{2008MNRAS.383L...5G,2010MNRAS.407..277S,2013A&ARv..21...59I}.\textbf{
}When a binary \change{star} enters a common envelope phase both
\change{stellar envelopes} are homogenized in the process under
the assumption that the orbital energy and angular momentum deposited
in the common envelope mixes it completely prior to ejection or merging.
The relatively compact cores in the common envelope are not mixed
with the envelope.

\subsection{Stellar populations with \emph{$\protect\binaryc$}}

\emph{\label{subsec:Stellar-populations-with-binary-c}} Our stellar
population models each contain $10^{4}$ single stars (model sets
$\mathrm{S}n$) or $100^{3}$ binary stars (model sets $\mathrm{B}n$),
evolved up to $13.7\,\mathrm{Gyr}$. Model sets X and Y are $50:50$
mixtures of model sets S1 and B1, i.e.~they have an initial binary
fraction similar to that of solar-neighbourhood stars of mass around
$1\,\msun$ \citep{2010ApJS..190....1R}. Primary masses, $M_{1}$,
are distributed between $0.1$ and $6\,\msun$ according to the initial
mass function of \citet*{KTG1993MNRAS-262-545K}. Secondary masses
follow a flat distribution in the mass ratio $q=M_{2}/M_{1}$ such
that $0.1\msun/M_{1}\leq q\leq1$.  Orbital parameters are distributed
either as a hybrid orbital-period distribution (Appendix~\ref{subsec:hybrid-period-distribution})
which interpolates between the log-normal distribution of solar-neighbourhood
G/K dwarfs at around $1\mathrm{\,M_{\odot}}$ \citep{1991A&A...248..485D}
and a distribution appropriate to O-type stars at high mass \citep{2012Sci...337..444S},
or orbital separations, $a$, are distributed according to a flat-$\ln a$
distribution between $3$ and $10^{4}\,\rsun$ (Appendix~\ref{subsec:opikdist}).

Our stars have a default metallicity $Z=\defaultZ$ (section~\ref{sec:APOKASC}
and Fig.~\ref{fig:Metallicity-distribution-of-thick-disc-stars})
but we also evolve populations with $Z=10^{-4}$, $0.001$ and $0.02$.
The chemical composition of our stars is a solar-scaled mixture based
on \citet*{Anders1989} with $\CFe$ enhanced by $+0.2\,\mathrm{dex}$
to match the chemical evolution of the thick disc \citep{2006MNRAS.367.1181B,Masseron2015}.
We do not enhance the $\alpha$-element abundances because our \emph{$\binaryc$}
stellar evolution models assume solar-scaled abundances.

Our single-star evolution defaults to the $\SSE$ and $\BSE$ standard
models of \citet{2000MNRAS.315..543H} and \citet{2002MNRAS_329_897H}
respectively. Nucleosynthesis is described in section~\ref{subsec:Nucleosynthesis-with-binary_c}.
Our binary-star evolution model sets include the following physics
as defined in Table~\ref{tab:Model-set-parameters}.
\begin{itemize}
\item Wind-Roche-lobe overflow (WRLOF) follows the description of \citet{2013A&A...552A..26A}
using their Eqs.~5 or 9, based on the detailed hydrodynamic models
of \citet{2007ASPC..372..397M}, or the Bondi-Hoyle prescription described
in \citet{2002MNRAS_329_897H}.
\item Common envelopes are treated according to the energy-balance algorithm
described by \citet{2002MNRAS_329_897H} with parameters $\alpha_{\mathrm{CE}}=0.2$
and $\lambda_{\mathrm{CE}}$ fitted to the models of \citet{Dewi2000}.
We also test $\alpha_{\mathrm{CE}}=0.5$ and $1.0$.
\item The Companion Reinforced Attrition Process (CRAP) of \citet{1988MNRAS.231..823T}
is applied with a parameter $B_{\mathrm{C}}=0$ (i.e.~disabled by
default, as by \citealp{2002MNRAS_329_897H}), $10^{3}$ or $10^{4}$.
This process enhances the stellar wind mass loss rate, $\dot{M}_{\mathrm{wind}}$,
because of the tidal influence of a companion star according to 
\begin{alignat}{1}
\dot{M}_{\mathrm{wind}} & =\dot{M}_{\mathrm{wind}}\left(B_{\mathrm{C}}=0\right)\times\left(1+B_{c}\max\left[\frac{1}{2},\frac{R}{R_{\mathrm{L}}}\right]^{6}\right)\label{eq:CRAP}
\end{alignat}
where $R$ and $R_{L}$ are the stellar radius and Roche radius, respectively,
and $\dot{M}_{\mathrm{wind}}\left(B_{\mathrm{C}}=0\right)$ is the
mass-loss rate in the absence of CRAP.
\item Material lost from the system during non-conservative RLOF carries
the specific angular momentum of the accretor (our default model,
$\gamma_{\mathrm{RLOF}}=-2$), the donor ($\gamma_{\mathrm{RLOF}}=-1$),
or a fraction $\gamma_{\mathrm{RLOF}}\geq0$ of the specific orbital
angular momentum, where $\gamma_{\mathrm{RLOF}}=0$, $1$ or $2$.
\item RLOF is based on the formalism of \citet{2014A&A...563A..83C} which
defines the mass-transfer rate as a steep function of the ratio $R/R_{L}$.
Alternatives include the original $\BSE$ prescription \citep{2002MNRAS_329_897H}
and the adaptive-RLOF of \citet{2014ApJ...780..117S} who compute
the mass-transfer rate such that $R=R_{L}$.
\item \change{Tides are based on \citet{1981A+A....99..126H} as prescribed
by \citet{2002MNRAS_329_897H} with time-scales from \citet{1977A+A57-383Z}.
The parameter $E_{2}$ is as fitted by \citet{2013A&A...550A.100S}.}
\end{itemize}
We form stars at a constant rate between $\agemin$~and~$\agemax\,\mathrm{Gyr}$
ago such that our thick-disc model stars have masses between about
$0.95$ and $1.3\,\msun$ to match the bulk population in our thick-disc
observational sample (section~\ref{sec:APOKASC}). Either star in
a binary can contribute to stellar number counts and if both stars
concurrently satisfy our thick-disc criteria then both are counted
separately. The number of such systems is very small.

We also make model sets containing stars with ages between $8$ and
$13\,\mathrm{Gyr}$ which are more typical for the thick disc \citep{2009IAUS..258...23F,Haywood2013}.
We also consider, in model set Y, a limited range of surface gravity
to match selection effects of the $\APO$ sample. 

We count stars with a radial velocity amplitude exceeding $1\,\mathrm{km}\,\mathrm{s}^{-1}$
as binaries. This limit is comparable to that in the observations
to which we compare (section~\ref{sec:APOKASC}). We take into account
the fact that binary star systems are randomly inclined when calculating
our modelled number counts (Appendix~\ref{sec:orbitalinclination}).
We also count blue stragglers which are defined as main-sequence stars
which have accreted mass and are older than the main-sequence lifetime
appropriate to their mass \citep[cf.][]{2001MNRAS.323..630H}.

\begin{table*}
\begin{tabular}{ccc>{\centering}p{10cm}}
\hline 
Model set & Single/binary & Metallicity & Parameters \tabularnewline
\hline 
S1 & Single & $\defaultZ$ & As $\SSE$/$\BSE$ (see text).\tabularnewline
S2 & Single & $0.0001$ & As S1 with $Z=0.0001$\tabularnewline
S3 & Single & $0.001$ & As S1 with $Z=0.001$\tabularnewline
S4 & Single  & $0.02$ & As S1 with $Z=0.02$\tabularnewline
S5 & Single & $\defaultZ$ & As S1 with star formation between $8$ and $13\,\mathrm{Gyr}$\tabularnewline
S6 & Single & $\defaultZ$ & As S1 with $2\leq\logg\leq3$\tabularnewline
B1 & Binary & $\defaultZ$ & $Z=\defaultZ$,$\alpha_{\mathrm{CE}}=0.2$, $B_{\mathrm{C}}=0$, $q_{\mathrm{crit}}=1.6$,
$\gamma_{\mathrm{RLOF}}=-2$, RLOF Claeys (2014), WRLOF Abate (2013)
eq.~(5), hybrid initial-period distribution \tabularnewline
B2 & Binary & $0.0001$ & As B1 with $Z=0.0001$\tabularnewline
B3 & Binary & $0.001$ & As B1 with $Z=0.001$\tabularnewline
B4 & Binary & $0.02$ & As B1 with $Z=0.02$\tabularnewline
B5 & Binary & $\defaultZ$ & As B1 with $q_{\mathrm{crit}}=1.8$\tabularnewline
B6 & Binary & $\defaultZ$ & As B1 with $q_{\mathrm{crit}}=3$\tabularnewline
B7 & Binary & $\defaultZ$ & As B1 with $\alpha_{\mathrm{CE}}=0.5$\tabularnewline
B8 & Binary & $\defaultZ$ & As B1 with $\alpha_{\mathrm{CE}}=1$\tabularnewline
B9 & Binary & $\defaultZ$ & As B1 with no WRLOF\tabularnewline
B10  & Binary & $\defaultZ$ & As B1 with WRLOF Abate 2013 eq.~(9)\tabularnewline
B11 & Binary & $\defaultZ$ & As B1 with $\BSE$ RLOF\tabularnewline
B12 & Binary & $\defaultZ$ & As B1 with adaptive RLOF of Schneider (2014)\tabularnewline
B13 & Binary & $\defaultZ$ & As B1 with $B_{C}=10^{3}$\tabularnewline
B14 & Binary & $\defaultZ$ & As B1 with $B_{C}=10^{4}$\tabularnewline
B15 & Binary & $\defaultZ$ & As B1 with $\gamma_{\mathrm{RLOF}}=-1$ (from donor)\tabularnewline
B16 & Binary & $\defaultZ$ & As B1 with $\gamma_{\mathrm{RLOF}}=0$ \tabularnewline
B17 & Binary & $\defaultZ$ & As B1 with $\gamma_{\mathrm{RLOF}}=1$ \tabularnewline
B18 & Binary & $\defaultZ$ & As B1 with $\gamma_{\mathrm{RLOF}}=2$ \tabularnewline
B19 & Binary & $\defaultZ$ & As B1 with logarithmically-flat initial-separation distribution\tabularnewline
B20 & Binary & $\defaultZ$ & As B1 with star formation between $8$ and $13\,\mathrm{Gyr}$\tabularnewline
B21 & Binary & $\defaultZ$ & As B1 with $2\leq\logg\leq3$\tabularnewline
X & Mixed & $\defaultZ$ & $\frac{1}{2}\,\mathrm{S1}+\frac{1}{2}\,\mathrm{B1}$\tabularnewline
Y & Mixed & $\defaultZ$ & $\frac{1}{2}\,\mathrm{S6}+\frac{1}{2}\,\mathrm{B21}$, i.e.~as S6
with $2\leq\logg\leq3$\vspace{1mm}\tabularnewline
\hline 
\end{tabular}

\caption{\label{tab:Model-set-parameters}Parameters of our $\protect\binaryc$
stellar-population model sets. Column one labels the model set, column
two shows the multiplicity, column three the metallicity and column
four any other parameters which are changed. S1 and B1 are our default
model sets, X best represents the Galactic thick disc and Y our $\protect\APO$
thick-disc sample.}
\end{table*}

\section{Our observational sample}

\label{sec:APOKASC}We extract a sample of thick-disc stars from $\APO$
\citep{2014ApJS..215...19P}. Abundance data, i.e.~$\FeH$, $\CN$
and $\aFe$, are from the APOGEE data release 12 \citet{2015AJ....150..148H}.
Asteroseismological masses and stellar parameters ($T_{\mathrm{eff}}$,
$\log g$) are from \citet{2014ApJS..215...19P}. Nuclear-burning
stage identifications, such as hydrogen-shell burning or core-helium
burning (red clump), are from \citet{2017MNRAS.466.3344E}. Most of
these stars are relatively unevolved red giants ($\loggsquare\gtrsim2.1$)
or are helium-burning in the red clump.

\begin{figure*}
\begin{centering}
\includegraphics[scale=0.8]{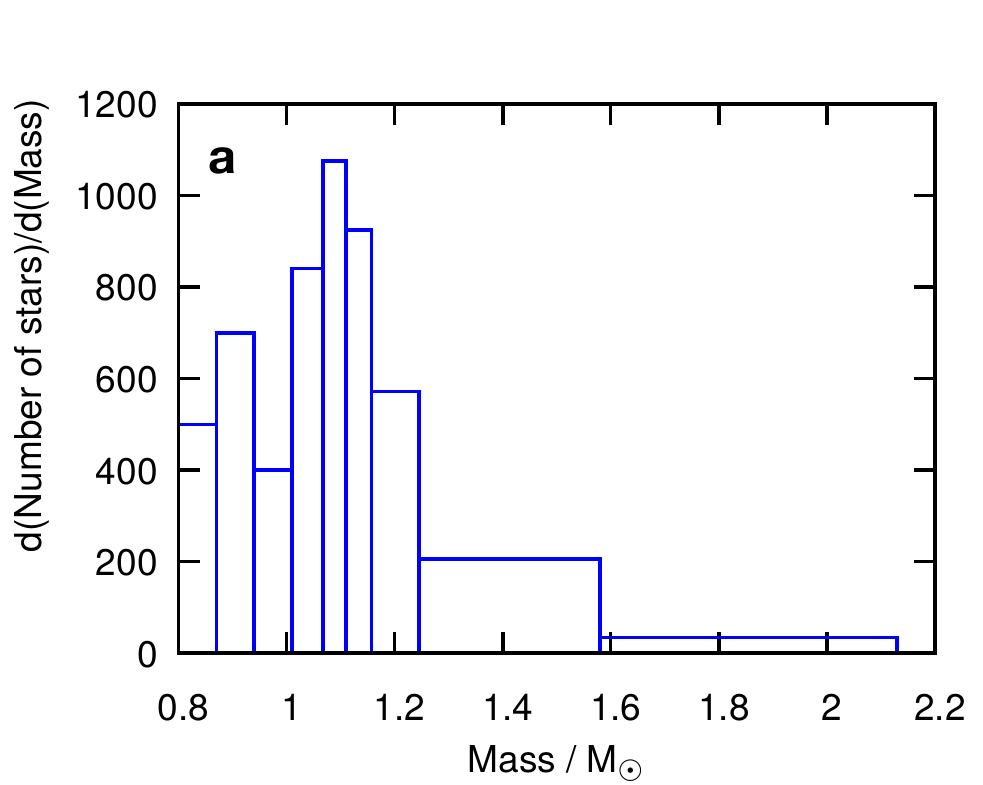}\includegraphics[scale=0.8]{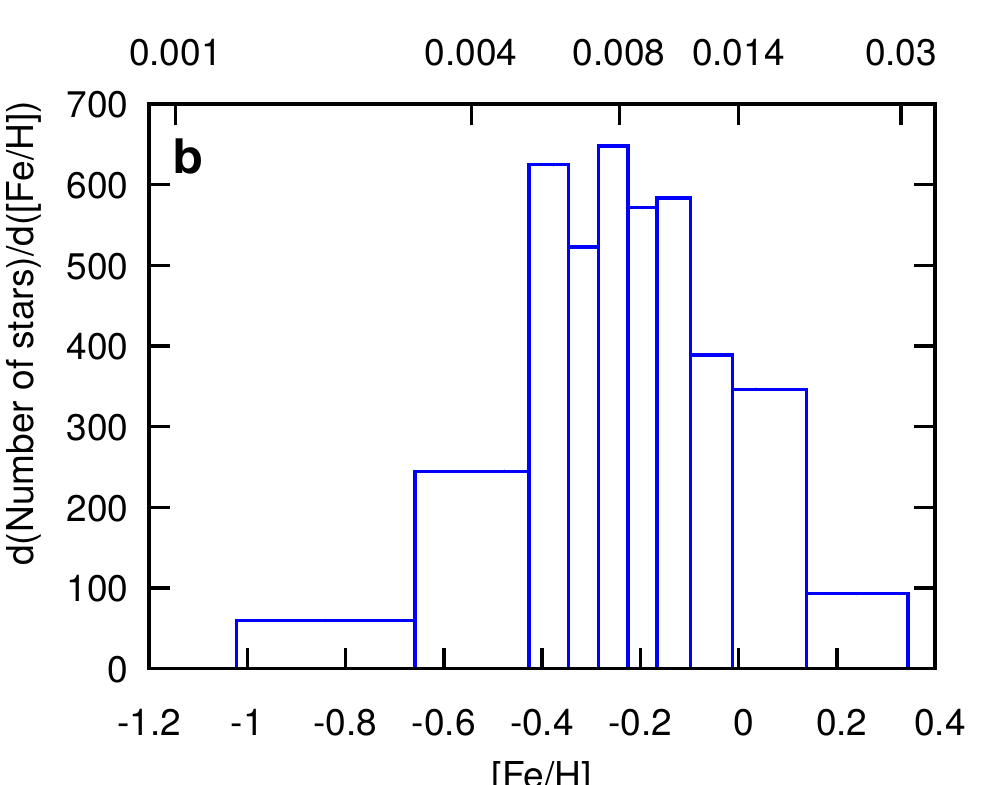}
\par\end{centering}
\caption{\label{fig:Metallicity-distribution-of-thick-disc-stars}Mass (left,
\textbf{a}) and metallicity (right, \textbf{b})\textbf{ }distributions
in our selection of thick-disc giant stars from $\protect\APO$. In
\textbf{b }the lower abscissa shows $\protect\FeH=\log_{10}\left(N_{\mathrm{Fe}}/N_{\mathrm{\mathrm{Fe},\protect\sun}}\right)-\log_{10}\left(N_{\mathrm{H}}/N_{\mathrm{H},\protect\sun}\right)$
while the upper shows $Z\approx Z_{\protect\sun}10^{\protect\FeH}$
where $Z_{\protect\sun}=0.014$ and $N_{i}$ are number densities
of species $i$ at the stellar surface.}
\end{figure*}

From $\APO$ we have 1989 giant stars. We select thick-disc stars
based on the abundance of $\alpha$ elements, $\aFe$, such that thick-disc
stars satisfy,
\begin{alignat}{1}
\aFe & >-0.06\times\FeH+0.1\,.\label{eq:thick-disc-condition}
\end{alignat}
This leaves us with $345$ stars. Of these, 75 ($21\pc)$, 47 ($\APOpc$)
and 31 ($9\pc$) have masses exceeding $1.2$, $1.3$ and $1.4\mathrm{\,M_{\odot}}$
respectively (Fig.~\ref{fig:Metallicity-distribution-of-thick-disc-stars}a).
If we select, in addition to Eq.~(\ref{eq:thick-disc-condition}),
only stars with $\FeH<-0.2$, the fraction of the remaining 189 stars
with mass exceeding $1.2$, $1.3$ and $1.4\mathrm{\,M_{\odot}}$
is $17\pc$, $11\pc$ and $6\pc$ respectively. There is a significant
population of thick-disc stars in excess of $1.3\mathrm{\,M_{\odot}}$
regardless of metallicity. Mass estimates have a typical associated
error of $0.15\mathrm{\,M_{\odot}}$ although some stars have mass
uncertainties of $0.3\mathrm{\,M_{\odot}}$ \citep{Miglio2012,2016AN....337..793B,2016AN....337..774D,Miglio2016}. 

The metallicity distribution of our thick-disc selection is shown
in Fig.~\ref{fig:Metallicity-distribution-of-thick-disc-stars}b.
The distribution peaks at a metallicity of about $0.008$ assuming
that abundances are solar-scaled and that $\FeH=0$ corresponds to
$Z_{\sun}=\Zsolar$. We thus use $Z=0.008$ in our population models.
The tail of the distribution, with metallicities in excess of $Z_{\sun}$,
may contain thin-disc stars which contaminate our thick-disc sample
\citep[cf.][]{2011MNRAS.412.1203N}. Eq.~(\ref{eq:thick-disc-condition})
demands that $\aFe\lesssim0.1$ when $\FeH=0.2$. At this and higher
metallicity, the measurement error on $\aFe$ is similar to the lower
limit of Eq.~(\ref{eq:thick-disc-condition}) and hence $\aFe$ cannot
be used reliably to determine thick- or thin-disc membership. Despite
this uncertainty, most of our stars with mass in excess of $\Mlimit$
have $\FeH<-0.2$ can be attributed to the thick-disc.

\section{Results}

\label{sec:Results}
\begin{table*}
\global\long\def\pc{\%}
\input{results}\global\long\def\pc{\,\mathrm{per\,cent}}

\caption{\label{tab:Results}Results of our simulations of thick-disc giant
populations of single and binary stars. The physical parameters corresponding
to the datasets specified in column one are defined in Table~\ref{tab:Model-set-parameters}.
Column two shows the fraction of giant stars in a population which
have masses above $\protect\Mlimit$. The remaining columns show fractions
of these stars with $M>\protect\Mlimit$ with various properties:
abundance as measured by $\protect\CN$, the fractions which would
be observed as single and binary, and the fractions that were and
were not a blue straggler star (BSS) prior to ascent of the giant
branch, respectively.}
\end{table*}
In this section we compare our thick-disc stellar population models
to our thick-disc stellar sample extracted from $\APO$. We focus
on $\CN$ vs.~mass, core-mass or $\log g$. In the figures that follow,
the number of stars in each bin is represented by the depth of shading
which is proportional to the log of the number of stars in each bin.
The colour gives the binary fraction with red single and blue binary.
We provide the number of stars in excess of $\Mlimit$ and their properties
because such stars can only form in our models by binary-star interaction.
Our model set data are in Table~\ref{tab:Results}.

\subsection{Single stars }

\label{subsec:Single-stars-vs-APOKASC}
\begin{figure*}
\hspace*{-1.0cm} 

\begin{tabular}[t]{>{\centering}m{9cm}>{\centering}m{9cm}}
Single stars, model set S1\vspace*{-0.5cm} & Binary stars, model set B1\vspace*{-0.5cm}\tabularnewline
\includegraphics[width=9.5cm,valign=t]{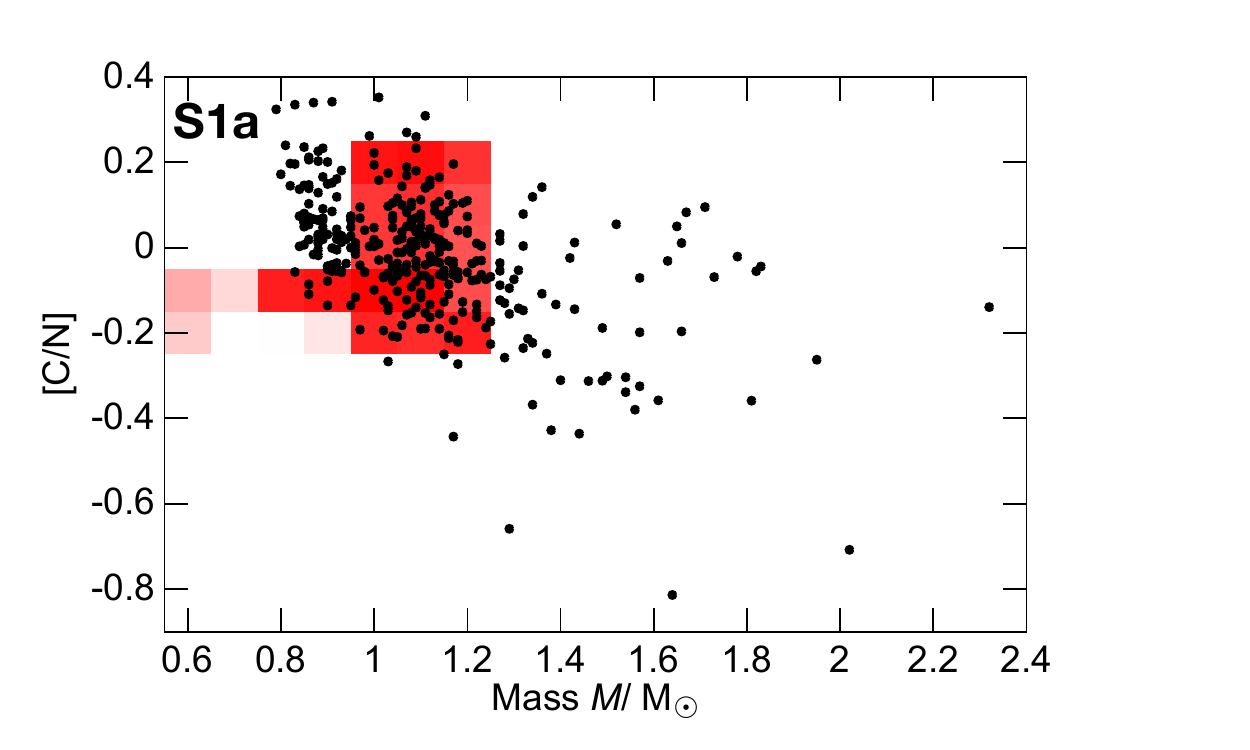} & \includegraphics[bb=0bp 0bp 360bp 216bp,width=9.5cm]{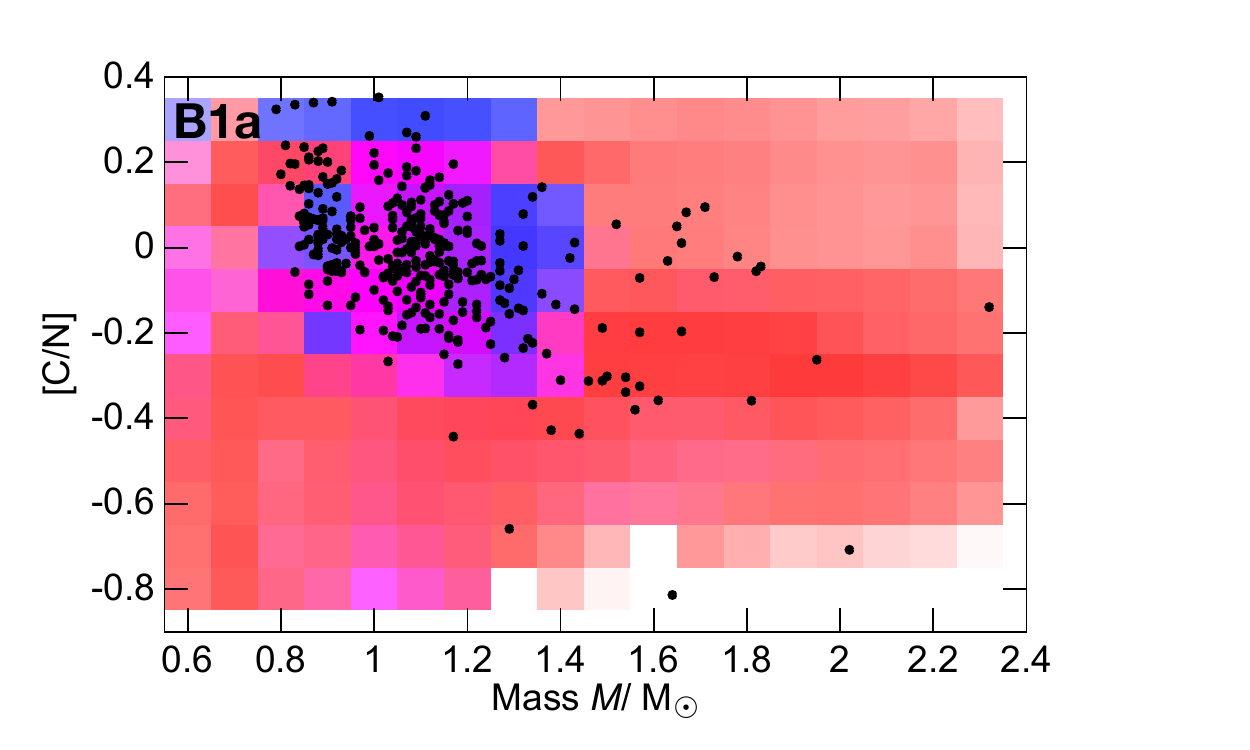}\tabularnewline
\includegraphics[bb=0bp 0bp 360bp 216bp,width=9.5cm]{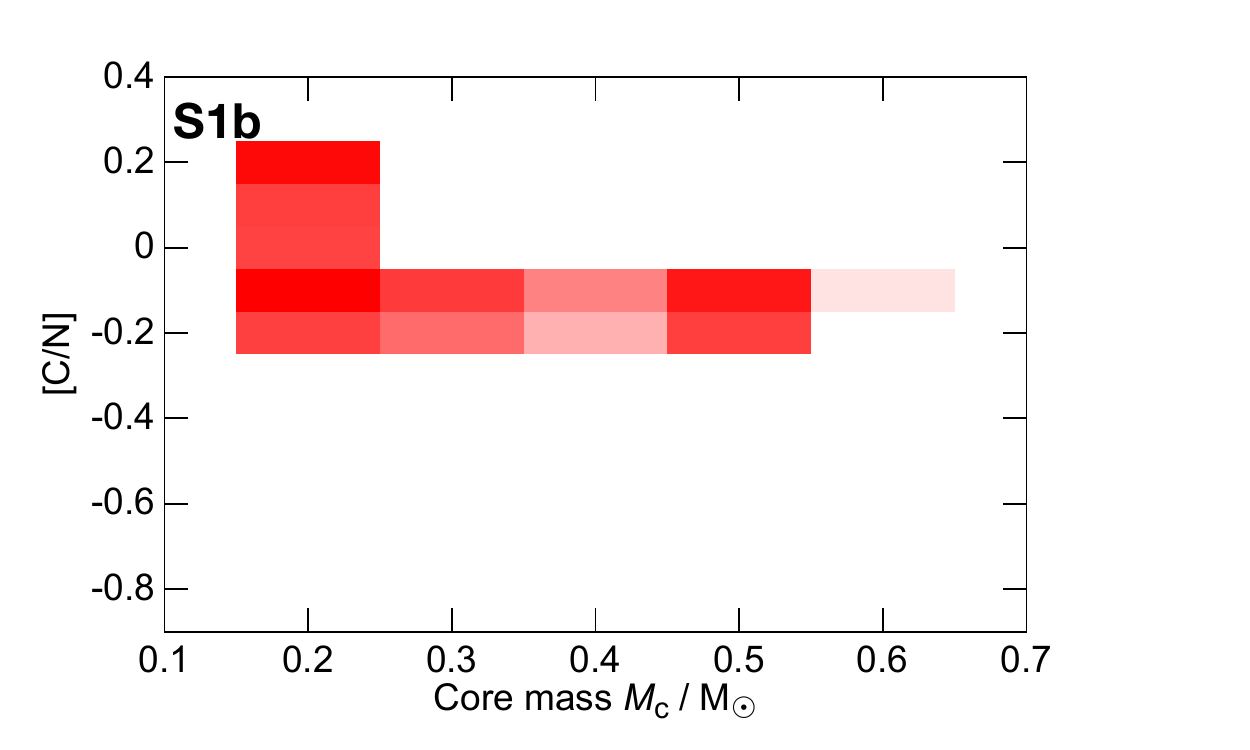} & \includegraphics[bb=0bp 0bp 360bp 216bp,width=9.5cm]{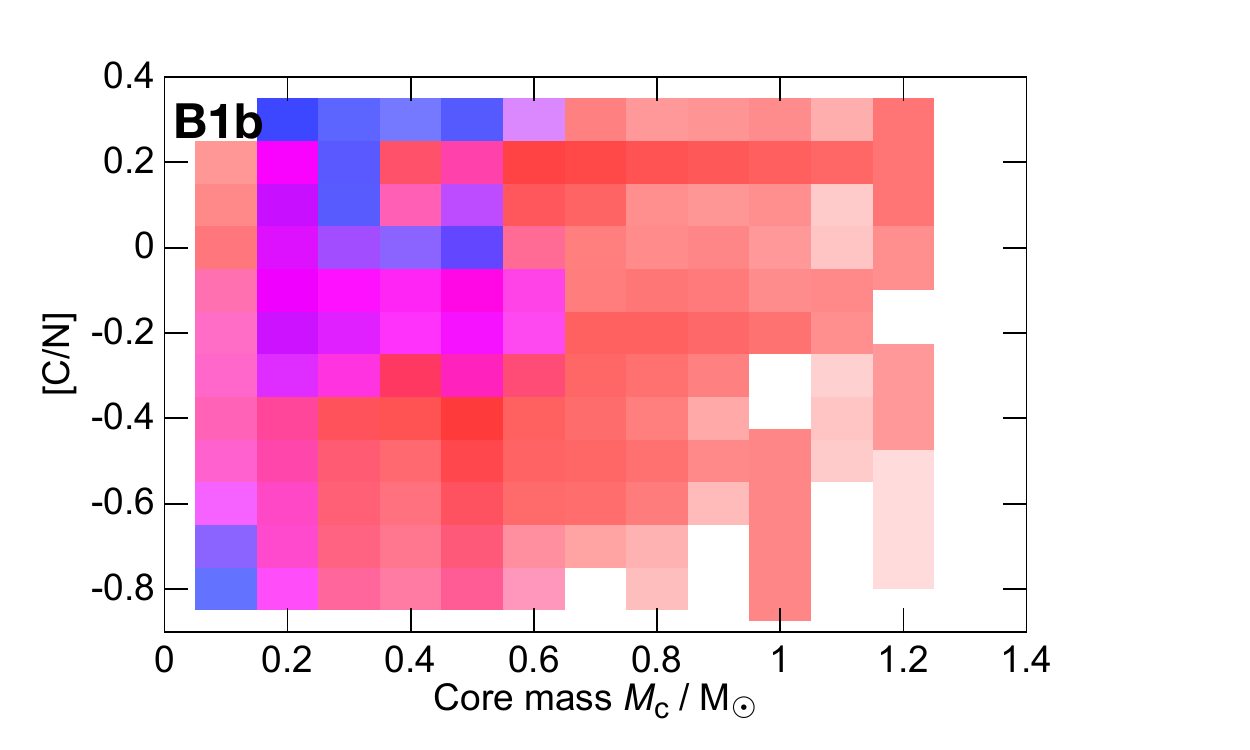}\tabularnewline
\includegraphics[bb=0bp 0bp 360bp 216bp,width=9.5cm]{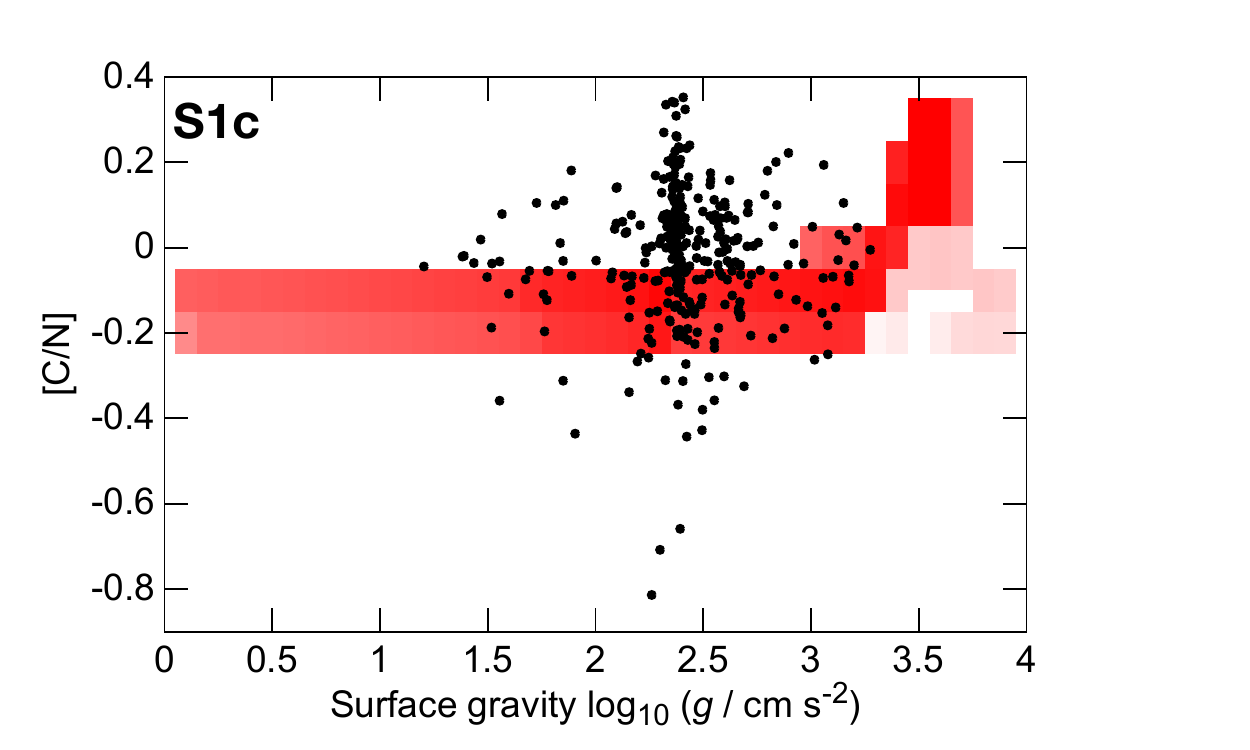} & \includegraphics[bb=0bp 0bp 360bp 216bp,width=9.5cm]{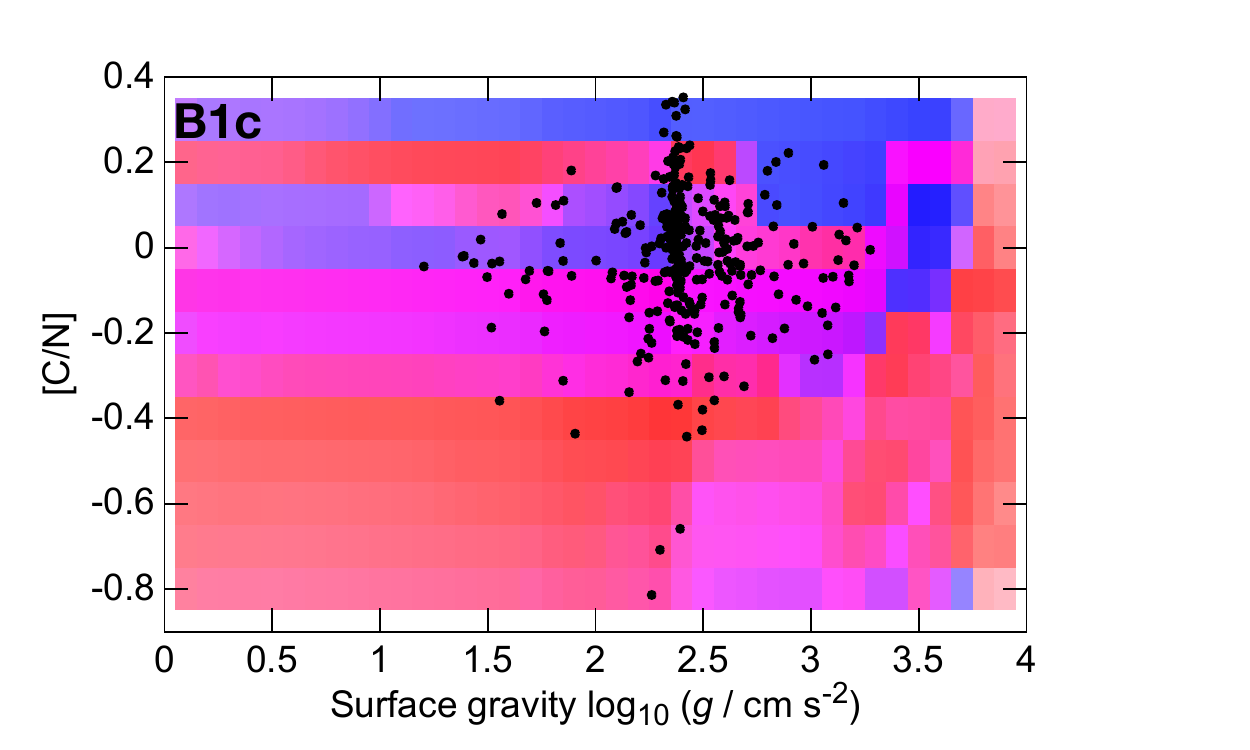}\tabularnewline
\multicolumn{2}{c}{\includegraphics[bb=0bp 55bp 360bp 150bp,width=9.5cm]{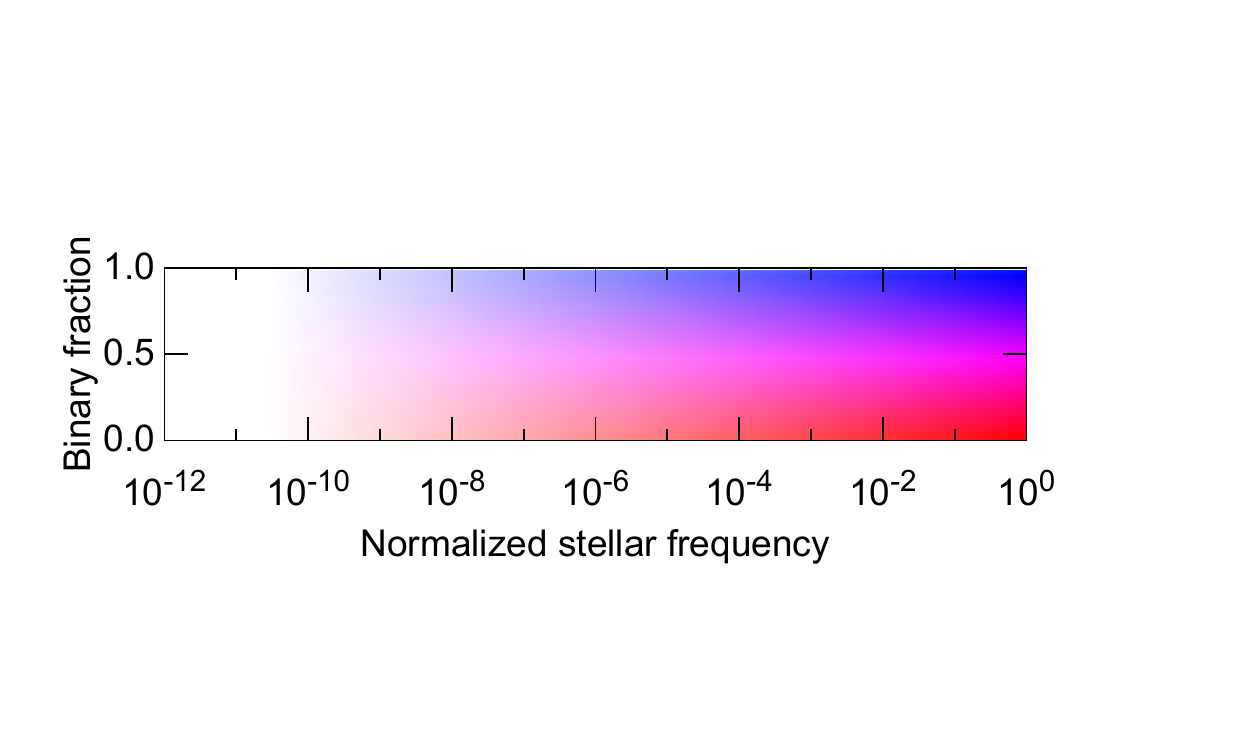}}\tabularnewline
\end{tabular}

\caption{\label{fig:S1-B1}Properties of our thick-disc stellar population
models sets S1 (all single stars) and B1 (all binary stars) of the
thick disc made with \emph{$\protect\binaryc$} vs.~our $\protect\APO$
thick-disc stellar sample (black points). The top row\textbf{ (a})
shows $\protect\CN$ vs.~mass, middle row (\textbf{b}) shows $\protect\CN$
vs.~core mass and the bottom row (\textbf{c}) shows $\left[\mathrm{C}/\mathrm{N}\right]$
vs.~$\log g$. The depth of shading represents the logarithm of the
number of stars in each bin {\change{relative to the maximum in each
panel}}. The colour is the binary fraction: single stars are red
while binary stars are blue. We show $-0.9\leq\mathrm{\protect\CN}\leq0.4$
to match the range of the $\protect\APO$ data, excluding its one
star with $\protect\CN\approx+1$. }
\end{figure*}
Our single-star, $Z=\defaultZ$ thick-disc population is compared
with the $\APO$ observational sample in the left column of Fig.~\ref{fig:S1-B1}.
The peak of the distribution of $\left[\mathrm{C}/\mathrm{N}\right]$
vs.~mass (Fig.~\ref{fig:S1-B1} S1a), at around $M=1.0\pm0.1\mathrm{\,M_{\odot}}$
and $\CN=0.0\pm0.2$, and the spread of masses in our single-star
models, from about $0.8\mathrm{\,M_{\odot}}$ to $1.3\mathrm{\,M_{\odot}}$
agrees well with the bulk of the observed stars. This agreement is
by design because we choose our stellar ages, $\agemin$ to $\agemax\,\mathrm{Gyr}$,
to match the $\APO$ asteroseismological masses (section~\ref{sec:APOKASC}).
We also boost our initial $\CN$ by $+0.2\,\mathrm{dex}$ to mimic
chemical evolution in the thick disc.  Importantly, none of our single-star
models has a mass exceeding $\Mlimit$.

The distribution of core masses is as expected from single-star evolution
(Fig.~\ref{fig:S1-B1} S1b). Red giants contribute to the peak at
$0.2-0.3\mathrm{\,M_{\odot}}$ because this is when their evolution,
and hence core mass growth rate, is slowest. Core helium-burning and
asymptotic giant stars give the peak at around $0.5\mathrm{\,M_{\odot}}$.

Our model giants are in all stages of post-main sequence, giant-star
evolution, Hertzsprung gap, giant branch, core helium burning (red
clump) and the asymptotic giant branch so these stars have surface
gravities, $\logg$, from $+4$, typical of the main-sequence turn-off,
to $0$, typical of asymptotic giant branch (AGB) stars (Fig.~\ref{fig:S1-B1}
S1c). Model sets S6, B6 and Y select only stars with $+2\leq\logg\leq+3$
to better match $\APO$ and are discussed in section~\ref{subsec:Mixed-populations}
below. 

\subsection{Binary stars}

\label{subsec:Binary-stars-vs-APOKASC}

In the right panels of Fig.~\ref{fig:S1-B1} we show the results
of repeating the above analysis with model set B1, a population of,
initially, only binary stars. The bulk of our $\APO$ sample stars
coincide with the bulk of our model B1 stars, just as with our single-star
model set S1 (section~\ref{subsec:Single-stars-vs-APOKASC}). The
predicted observed binary fraction in the bulk of the stars, around
$M=1.0\pm0.1\mathrm{\,M_{\odot}}$ and $\CN=0.0\pm0.2$, is about
$65\pc$. While all stars in this model set B1 are born binary, evolution
and inclination reduce the observed binary fraction to less than $100\pc$.
\changemore{Our binary stars have their $\CN$ boosted by $+0.2\,\mathrm{dex}$
identically to our single stars.} 

Outside the bulk population, the products of binary evolution are
scattered throughout the $M\,\mathrm{vs}\,\CN$ plane. In particular,
stars with masses in excess of $\Mlimit$, which can only be made
in binaries, make up \data{B1.1} of the giants. Most of these, \data{B1.5},
and almost all stars with masses in excess of $1.5\mathrm{\,M_{\odot}}$,
are single, meaning they are merged binaries. Of these merged stars,
\data{B1.7} are blue-stragglers during the main sequence. The rest,
\data{B1.8}, merge as giants during common envelope evolution.

Mass accretion drives stars to the right in Fig.~\ref{fig:S1-B1}
B1a, while mixing tends to drive stars downwards by decreasing $\CN$.
The mixing is caused either by common envelope evolution or subsequent
first dredge up in the star. We assume that merged stars have first
dredge up equally as deep as in a single star of the same mass. This
is likely a simplification but our model stars still cover a similar
parameter space to the $\APO$ sample (see also section~\ref{subsec:Model-uncertainties:-binary-nucsyn}).

We also see a population of stars which gain mass by wind accretion
and remain observable as binaries. These accrete up to $0.3\mathrm{\,M_{\odot}}$
from the wind of their companion while on the main sequence \citep{2013A&A...552A..26A}
and remain polluted as they ascend the giant branch. As for the merged
stars, after accretion these stars subsequently reduce their $\CN$
by first dredge up so $0.3<\CN<-0.5$. Stars that accrete from an
AGB companion are often rich in carbon with $\CN\gtrsim0.2$. These
would likely be rejected by $\APO$ and are discussed further in section~\ref{subsec:The-extended-binary-population}. 

The distributions of $\CN$ vs.~core mass (Fig.~\ref{fig:S1-B1}
B1b) and $\CN$ vs.~$\logg$ (Fig.~\ref{fig:S1-B1} B1c) peak similarly
to the single stars of model set S1, but are smeared out by binary
interactions. Stars with core masses above $0.6\mathrm{\,M_{\odot}}$,
and those with $\CN<-0.4$, are almost all merged binaries. This prediction
is being tested by observations, the first results of which are reported
in \citet{Jofre2016}.

\subsection{A thick-disc stellar population}

\label{subsec:Mixed-populations}

\begin{figure*}
\hspace*{-1.0cm}%
\begin{tabular}[t]{c>{\centering}p{9cm}}
Model set X\vspace*{-0.5cm} & Model set Y\vspace*{-0.5cm}\tabularnewline
\includegraphics[width=0.55\textwidth]{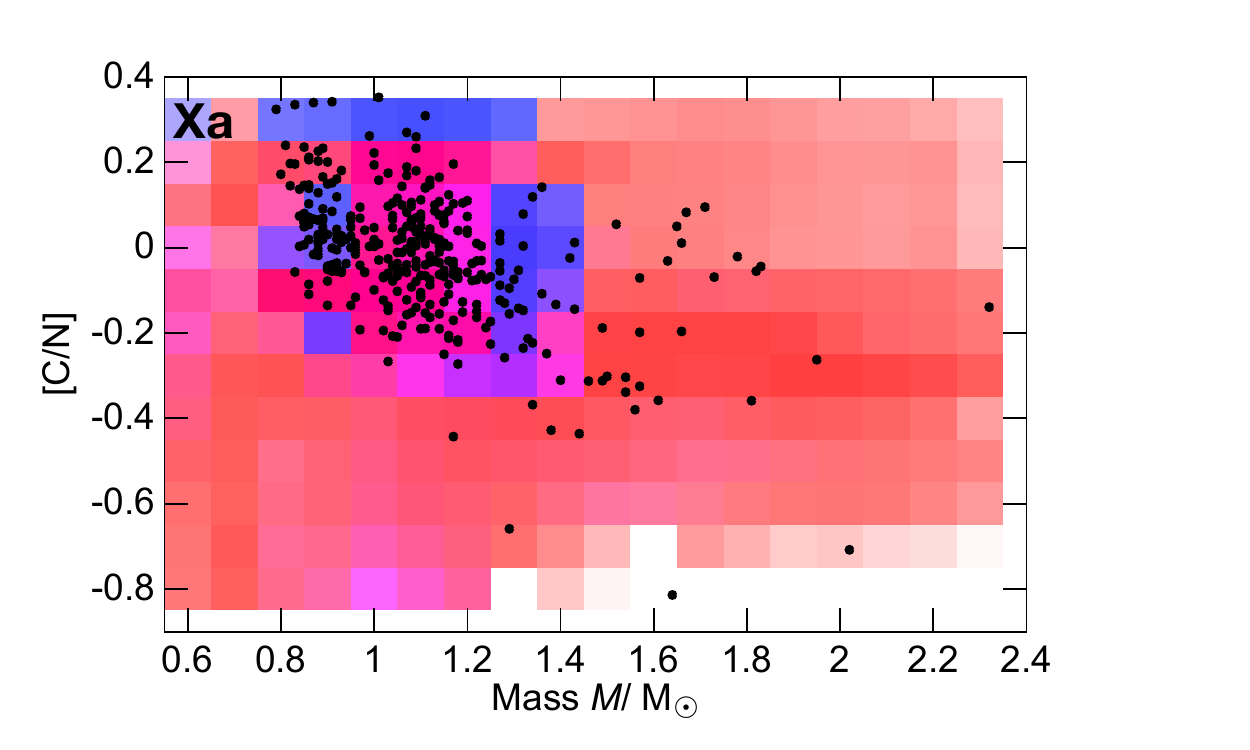} & \includegraphics[width=0.55\textwidth]{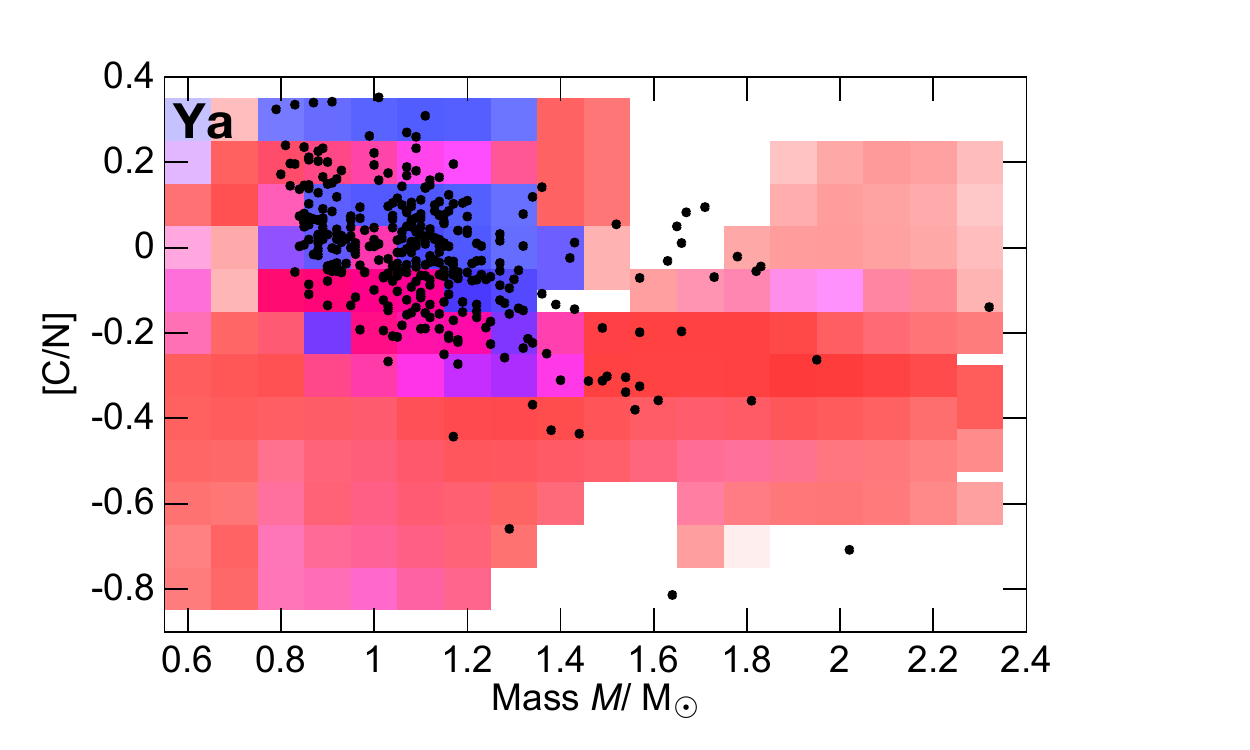}\tabularnewline
\includegraphics[width=0.55\textwidth]{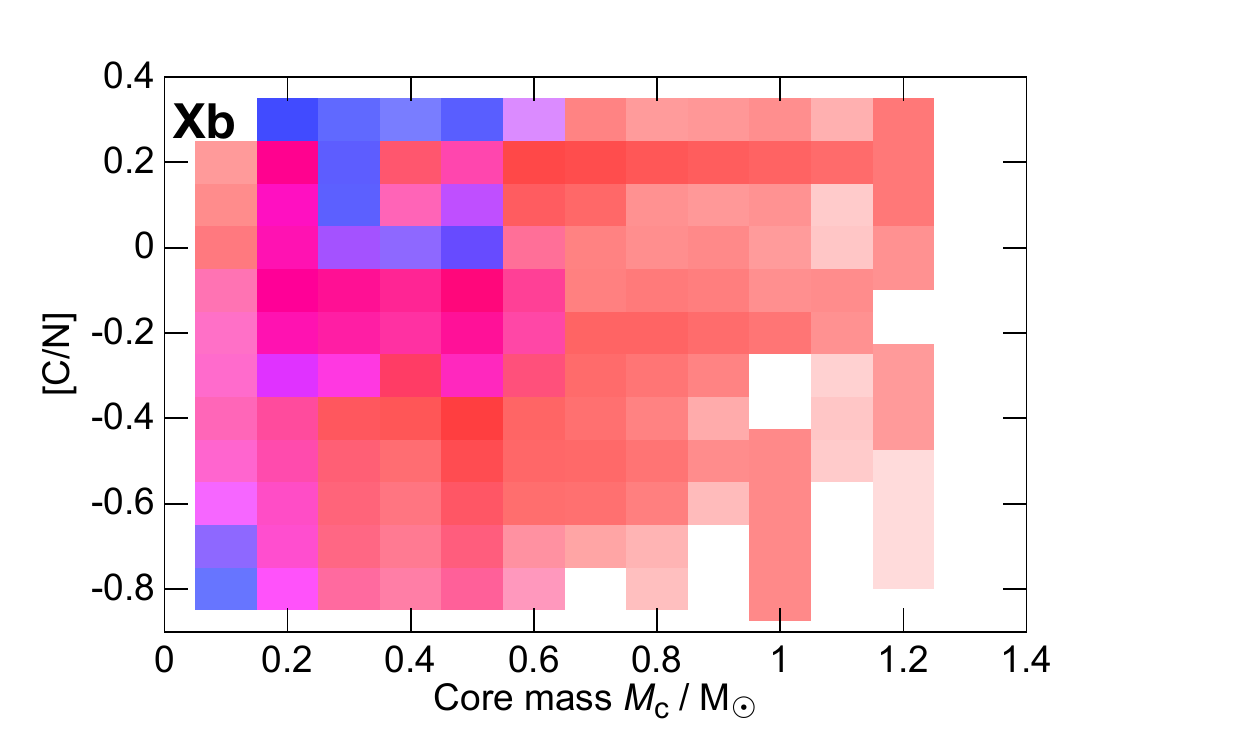} & \includegraphics[width=0.55\textwidth]{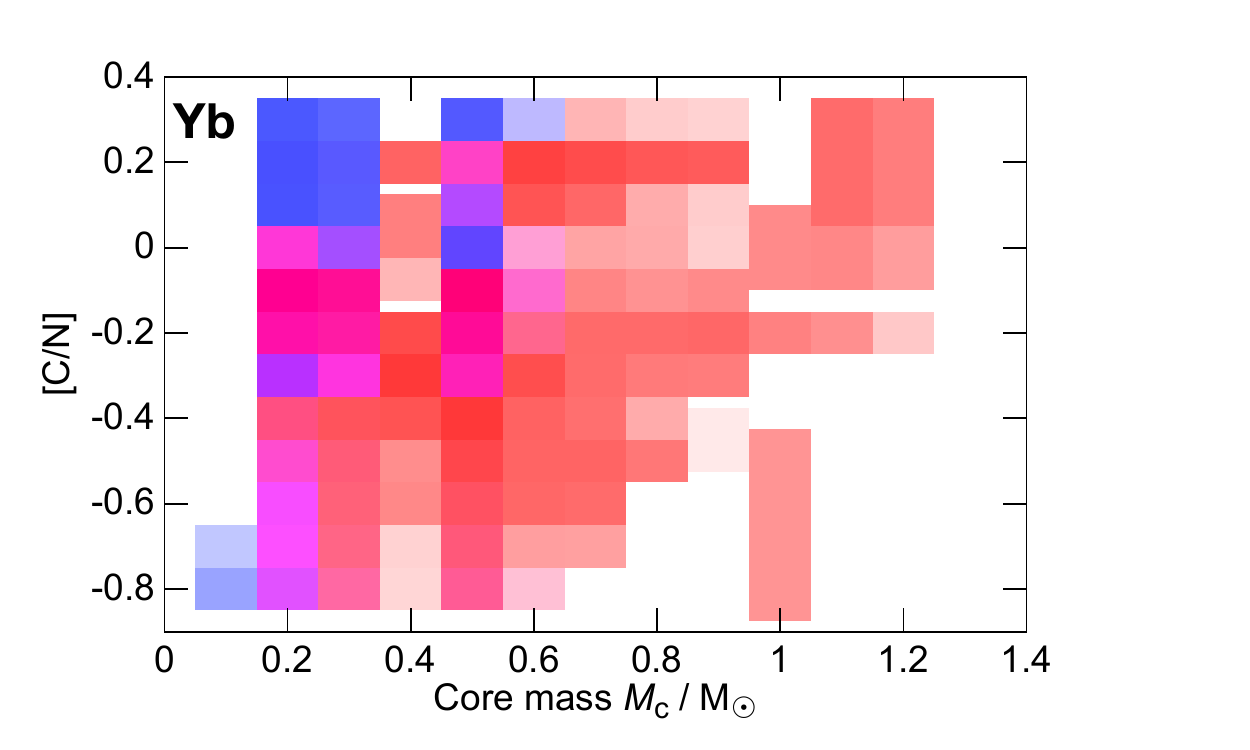}\tabularnewline
\includegraphics[width=0.55\textwidth]{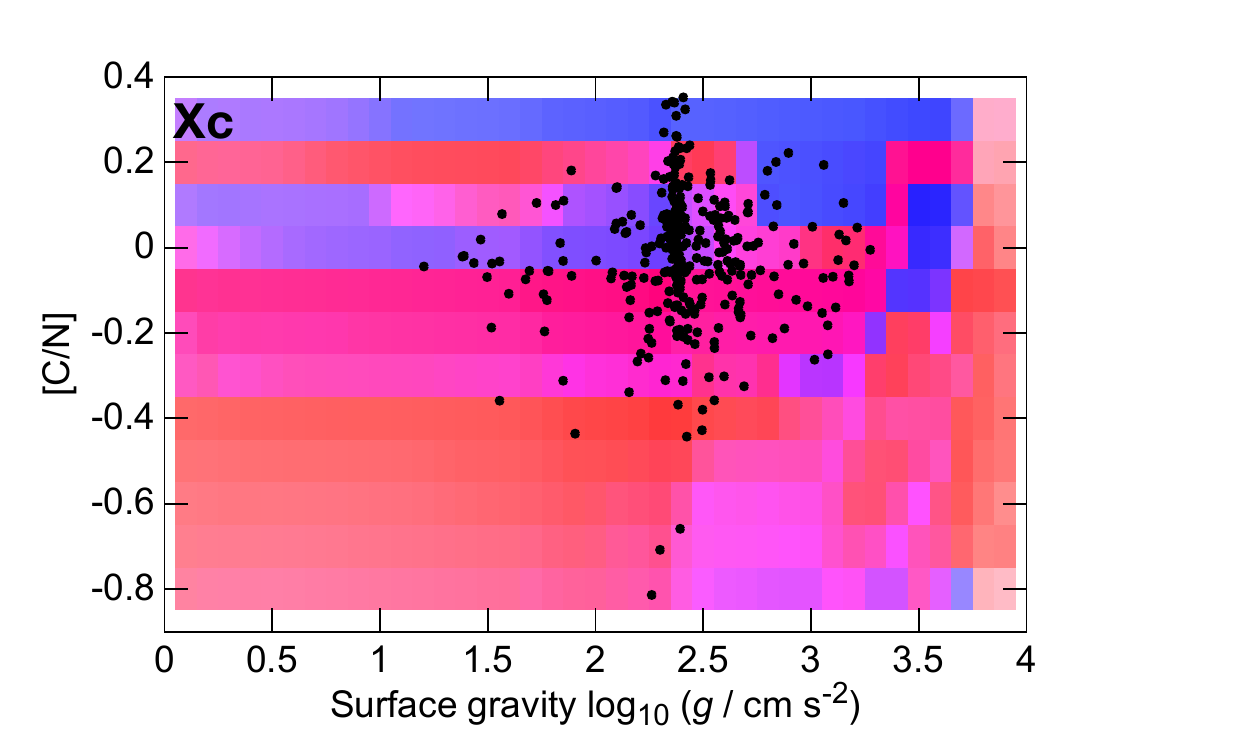} & \includegraphics[width=0.55\textwidth]{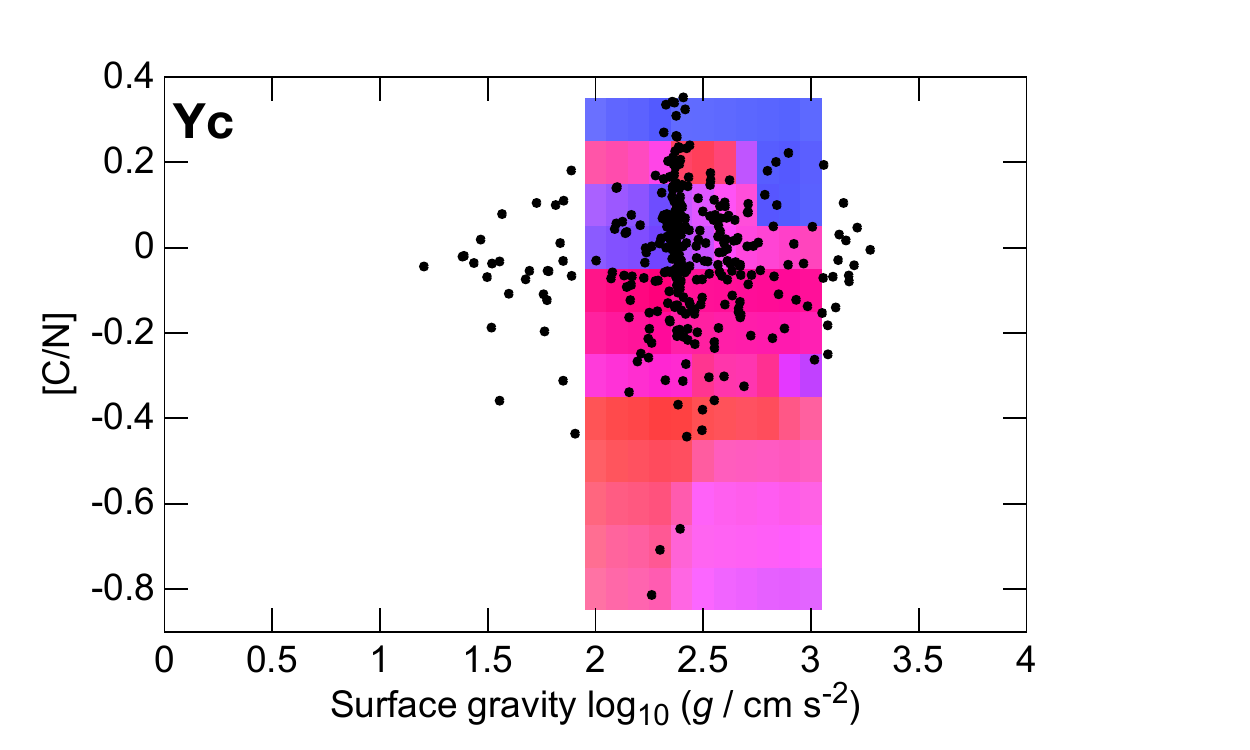}\tabularnewline
\multicolumn{2}{c}{\includegraphics[bb=0bp 55bp 360bp 150bp,width=9.5cm]{images/Fig3_colorbox}}\tabularnewline
\end{tabular}

\caption{\label{fig:X-Y}As Fig.~\ref{fig:S1-B1} with populations of $50\protect\pc$
single and $50\protect\pc$ binary stars. In the left column $\log g$
is unconstrained (model set X) while in the right column we select
only stars with $2\leq\protect\logg\leq3$ (model set Y) to match
our thick-disc sample from $\protect\APO$.}
\end{figure*}
In Fig.~\ref{fig:X-Y} we show two populations which contain a mix
of initially single and initially binary stars. The first, model set
X, is a 50:50 mix of model sets S1 and B1, our default single- and
binary-star populations. A $50\pc$ binary fraction represents stars
of around $1$ to$1.2\mathrm{\,M_{\odot}}$ in the solar neighbourhood
\citep{2010ApJS..190....1R} and we assume this also represents the
thick disc, as \citet{2015ApJ...799..135Y} suggests. Model set X
is thus the population of red giants in the thick disc with no selection
criterion except our $1\,\mathrm{km}\,\mathrm{s}^{-1}$ radial velocity
cut (section~\ref{subsec:Stellar-populations-with-binary-c}). Model
set Y is identical to set X except that it contains only stars with
$2\leq\logg\leq3$ to better match the $\APO$ sample.

The fractions of stars more massive than $\Mlimit$ are \data{X.1}
and \data{Y.1} in model sets X and Y, with binary fractions \data{X.6}
and \data{Y.6}, respectively. In this respect the data sets are quite
similar but they differ because the $\log g$ restriction in model
set Y preferentially selects red clump (helium burning) stars. These
have core masses around the helium ignition core mass, $0.48\mathrm{\,M_{\odot}}$
at $Z=0.008$, so model set Y has an excess of stars with this core
mass (section~\ref{subsec:Masses-and-core-masses}).

The distributions of binary properties are also similar in model sets
X and Y (Table~\ref{tab:Results}). In the following sections we
discuss the thick-disc giants in general, our model set X, rather
than the $\APO$-specific model set Y. Our conclusions are essentially
the same except for the core-mass distribution. Model set X, without
the $\log g$ restriction, is likely more applicable to future surveys.

\subsection{Extended parameter space}

\label{subsec:The-extended-binary-population}
\begin{figure*}
\hspace*{-1.0cm}%
\begin{tabular}{cc}
\includegraphics[width=17cm]{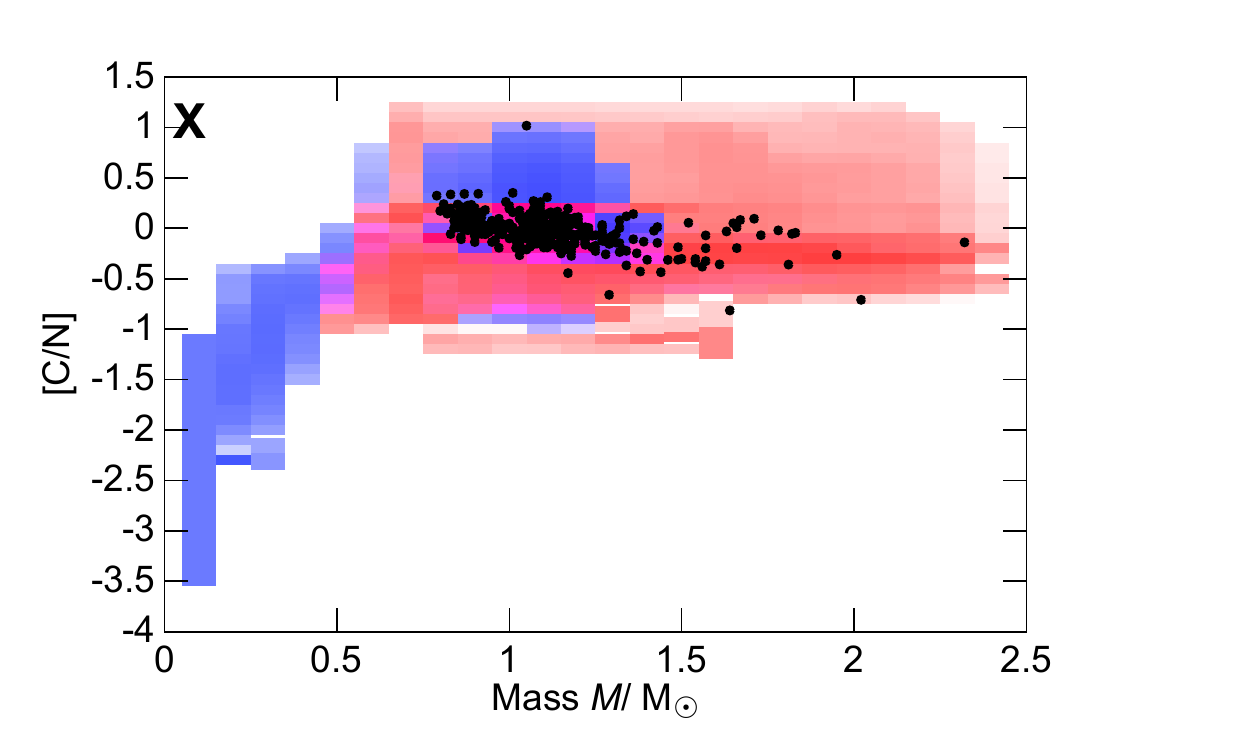} & \hspace*{-6.5cm}\includegraphics[bb=0bp -4bp 360bp 216bp]{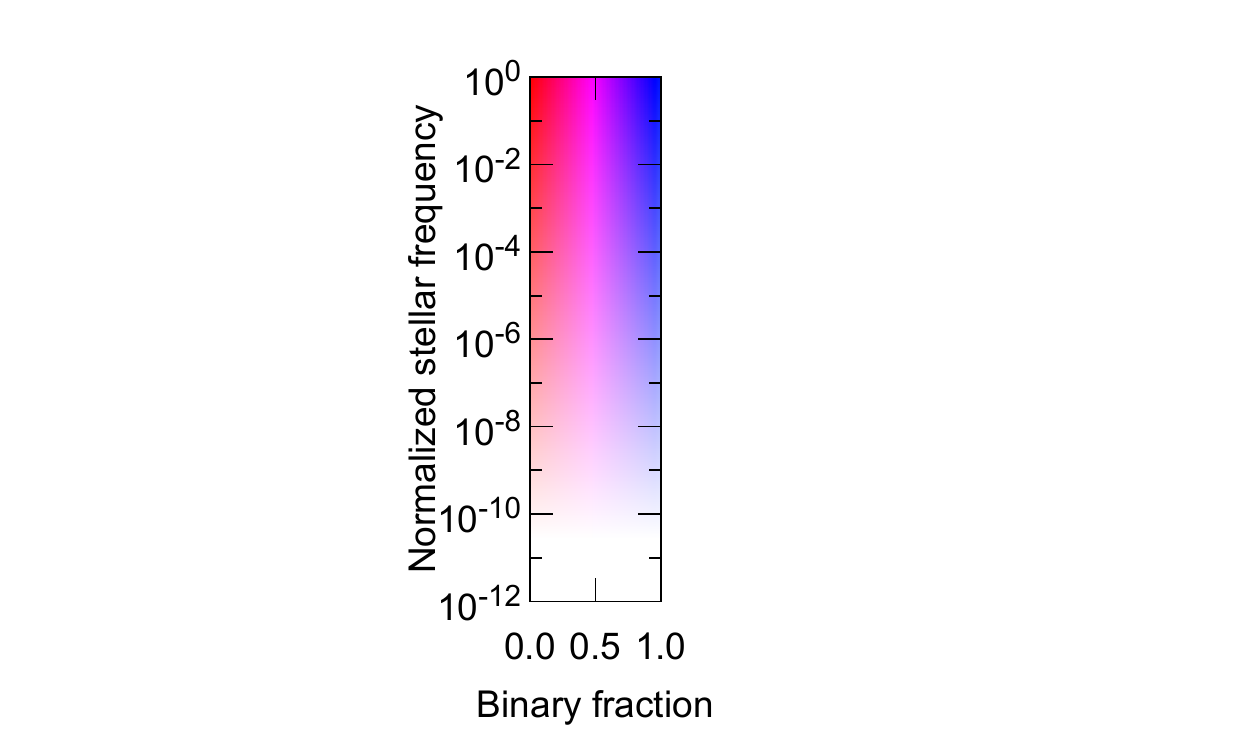}\tabularnewline
\end{tabular}

\caption{\label{fig:X-CNunlimited}$\protect\CN$ vs mass in our model set
X, which best represents an unbiased thick-disc stellar population,
with the $\protect\CN$ range chosen to include all giant stars in
our model including stripped giants (Algols), merged giants and mass-transfer
products (AGB and CH stars). The depth of shading represents the (logarithm
of) the number of stars in each bin relative to the maximum. The colour
is the binary fraction: single stars are red while binary stars are
blue. }
\end{figure*}
Most stars in our thick-disc $\APO$ selection have $-0.8\leq\CN\leq0.4$
and one has $\CN=+1.0$. Fig.~\ref{fig:X-CNunlimited} shows our
model set X predictions over the full range of $\CN$ against mass.
Almost all our model stars with mass exceeding $1.5\mathrm{\,M_{\odot}}$
are single, merged binaries. While most of these are first giant branch
stars, with $-1\lesssim\CN\lesssim0$ corresponding to first dredge
up appropriate to their mass, we predict a number of stars with significant
carbon excess, $\CN>0$ (\data{X.2} of all giants in model set X),
some of which have mass in excess of $2\mathrm{\,M_{\odot}}$.

Many of these carbon-rich stars and all those with $M>1.5\mathrm{\,M_{\odot}}$
are single, asymptotic giants (AGB stars). They accrete sufficient
mass to exceed the minimum mass for third dredge up (around $1.2\mathrm{\,M_{\odot}}$,
\citealp{Binary_Origin_low_L_C_Stars}) and hence enhance their surface
carbon abundance. The one $\APO$ star with $\CN=+1$ and $M\approx1\mathrm{\,M_{\odot}}$
could be a thermally-pulsing AGB star that was previously more massive
but has since lost material in its stellar wind. Few of these stars
are expected to be seen in $\APO$ because the thermally-pulsing AGB
(TPAGB) phase is short in duration relative to both the giant branch
and red clump. However, they should be anomalously luminous compared
to first giant branch stars. \emph{Gaia }distances may help here.

Related to this population are the binary stars with mass around $1.1\mathrm{\,M_{\odot}}$
which also have enhanced surface carbon. These are the equivalent
of barium, CH and CEMP stars. Carbon-rich material is accreted from
a more massive asymptotic giant companion while these stars are on
the main sequence. Despite thermohaline dilution and the effects of
first dredge up, the stars remain carbon rich as they ascends the
giant branch. In model set B2, with $Z=10^{-4}$, \data{B2.4} of
stars have $\CN>0.5$, a similar fraction to the approximately $1\pc$
of halo stars which are CH stars and the $1\pc$ of G/K giants which
are barium stars \citep{1998A&A...332..877J}. We discuss the implications
of wind mass transfer further in section~\ref{subsec:Comparison-with-CEMPs}.

The binary stars in model set X with masses below $0.6\mathrm{\,M_{\odot}}$
and decreased $\CN$, so enhanced surface nitrogen, are stripped red
giants which are similar to low-mass Algol systems. Mass transfer
exposes CN-cycled material in their cores. Because of their low mass,
reduced luminosity and anomalous $\CN$, these stars are likely not
selected in $\APO$. These Algols number $0.1-0.3\pc$ of all giant
stars in all our data sets except B19 with \data{B19.algol} which
is enhanced because of the logarithmically-flat initial-separation
distribution (Appendix~\ref{subsec:opikdist}). If the number of
Algol systems could be reliably measured as a fraction of the number
of red giants, it would provide a powerful diagnostic of the initial
binary period distribution. These stars may weak G-band stars which
are carbon poor and nitrogen rich \citep{2013ApJ...765..155A,2016A&A...587A..42P}
for which monitoring of duplicity is rather incomplete.

\subsection{Variation of parameters}

\label{subsec:model-parameters}

Our main conclusions are relatively robust to uncertainties in our
model parameters. The fraction of giants with mass in excess of $\Mlimit$
is $0.8\,\mathrm{to}\,3\pc$ in all our binary-star model sets except
B19. It is \data{B1.1} in our initially $100\pc$ binary population
(model set B1), \data{X.1} in our 50:50 mix of single and binary
stars (model set X), and is \data{Y.1} in our $\log g$-selected
model set (Y) which should match the $\APO$ sample. The finer details,
such as binary fraction among the stars more massive than $\Mlimit$
and the number of carbon-rich stars, vary somewhat from set to set.

Model sets B5 and B6 test whether changing the critical mass ratio
for mass transfer on the main sequence is important to our results.
It is not. Varying the parameter only slightly changes the number
of blue straggler stars. Model sets B7 and B8 test variations of the
common-envelope ejection efficiency parameter with $\alpha_{\mathrm{CE}}=0.5$
and $1$ respectively. Increasing the efficiency of common-envelope
ejection reduces the number of merged stars but, even with $\alpha_{\mathrm{CE}}=1$
(model set B8), \data{B8.5} of stars more massive than $\Mlimit$
form by merging inside a common-envelope.

Model set B9, with no wind-RLOF, has only \data{B9.6} giants more
massive than $\Mlimit$ that are binary. Canonical Bondi-Hoyle wind
accretion is not efficient enough to make many stars with $M>\Mlimit$
because typically only about $0.1\mathrm{\,M_{\odot}}$ accretes \citep{2013A&A...552A..26A}.
Only the merging channel makes stars more massive than $\Mlimit$
when there is no wind-RLOF and, likewise, only \data{B9.7} of giants
more massive than $\Mlimit$ are blue stragglers on the main sequence.
In our low-metallicity model sets, B2 with $Z=10^{-3}$ and B3 with
$Z=10^{-4}$, many more carbon-rich stars are made because carbon
production in AGB stars is more efficient in low-mass stars at low
metallicity \citep{Karakas2002,Binary_Origin_low_L_C_Stars}. Model
set B3 has fewer carbon-rich, massive giants because there are few
stars with $M>\Mlimit$ that satisfy our age criteria at $Z=10^{-4}$.

Using our alternative Roche-lobe overflow rate calculation schemes
(B11, B12) has little effect. Changing our prescription of angular
momentum loss during non-conservative RLOF (B15\textendash 18), $\gamma_{\mathrm{RLOF}}$,
has no effect because mass transfer is either conservative, so $\gamma_{\mathrm{RLOF}}$
is irrelevant, or proceeds through common-envelope evolution. Allowing
for the companion-reinforced attrition process (CRAP) with $B_{\mathrm{C}}=10^{3}$
and $10^{4}$ in model sets B13 and B14 changes the fraction of giant
stars in excess of $\Mlimit$ to \data{B13.1} and \data{B14.1} respectively.
Increasing $B_{\mathrm{C}}$ is the only way, other than increasing
the metallicity, to increase the number of binaries among our simulated
red giants with $M>\Mlimit$. 

One model set stands out from the others in terms of number of giants
with $M\gtrsim\Mlimit$. In model set B19 we use the a logarithmically-flat
separation distribution (Appendix~\ref{subsec:opikdist}), $dN\propto d\log a$
where $a$ is the orbital separation, rather than the log-normal distribution
of \citet{1991A&A...248..485D}. This enhances the number of initially
close binaries and hence also the number of stars which transfer mass
or merge. The fraction of giants with $M>\Mlimit$ increases to \data{B19.1},
far closer to the $\APOpc$ of the $\APO$ data. We discuss this further
in the following section.

\section{Discussion}

\label{sec:Discussion}

Our simulated stellar populations match the range of abundances and
masses observed in the $\APO$ red-giant sample, but a number of uncertainties
\textendash{} in the models, observations and interpretation \textendash{}
remain. \change{Alternative evolutionary pathways, such as triple
stars and stellar migration, are possible.} We discuss these below
and suggest avenues for future research that may help resolve the
associated problems.

\subsection{Range of $\protect\CN$ and mass vs.~APOKASC and the number of massive
thick-disc stars\label{subsec:Range-of-CN-and-number-of-massive-thick-disc-stars}}

Interactions in binary-star systems naturally explain the range of
masses in the $\APO$ thick-disc sample of red giants, from $0.9$
to about $2\mathrm{\,M_{\odot}}$ and their range in surface $\CN$.
Many of these binaries are expected to merge and evolve as single
giant stars, which also accounts for the low binary frequency among
these extra-massive stars. However, the number of stars more massive
than $\Mlimit$ in the $\APO$ thick-disc sample is $\APOpc$. It
is difficult to reproduce such a large fraction of stars with our
stellar population models. 

Our model set B19, with a logarithmically-flat orbital separation
 distribution (Appendix~\ref{subsec:opikdist}), contains enough
sufficiently close binary stars that \data{B19.1} of thick-disc stars
are expected to exceed $\Mlimit$, assuming that all stars are born
with companions. In the solar neighbourhood about half of $1\mathrm{\,M_{\odot}}$
stars are binary and a log-normal distribution of periods peaking
at many years \citep{1991A&A...248..485D} is more representative
than a logarithmically-flat initial-separation distribution. If such
distributions also apply to the thick disc, it is difficult to see
how all the $\APOpc$ of these extra-massive stars can be made in
binary systems. Recent suggestions of a high multiplicity fraction,
perhaps in excess of two thirds, among old stars also helps reconcile
our models and the observational data \citep{2017ApJ...836..139F}.

\subsection{Extra mixing\label{subsec:Extra-mixing}}

\change{Our population nucleosynthesis models do not include mixing
except convection and thermohaline mixing of accreted material. They
may then not be able to reproduce observed extra mixing near the tip
of the red giant branch \citep{Lagarde2012} nor any nitrogen depletion
at the helium flash \citep{2017MNRAS.464.3021M}. However, our $\APO$
red giant branch stars are not bright enough to have yet undergone
canonical extra mixing which should occur near the tip of the red
giant branch. So we need not implement extra mixing in our models.

Our $\APO$ red clump stars are more problematic. \citet{2017MNRAS.464.3021M}
show that, at metallicities around $0$ and $-0.55$, $\NFe$ drops
by $0.1$ and $0.2\,\mathrm{dex}$ at the helium flash, while $\CFe$
decreases by at most $0.1\,\mathrm{dex}$. Most of our thick-disc
stars have $0<\FeH<-0.4$, so a change in $\CN$ of about $-0.25\,\mathrm{dex}$
is possible at the tip of the red giant branch. The scatter in $\CN$
among our $\APO$ stars more massive than $\Mlimit$ is several times
this, with $+0.1\lesssim\CN\lesssim-0.8$. So selecting only red giant
branch (hydrogen burning) or red clump (helium burning) stars does
not change our overall conclusion.

Binary interaction is unlikely to solve the question of extra mixing
at helium ignition and associated depletion of nitrogen. This depletion
is seen in the bulk of stars, many of which are presumably single,
and so not among the few per cent which have exchanged mass or merged.
Mass transfer also only increases the likelihood of a deeper first
dredge up, more canonical extra mixing and a further increase in nitrogen.}

\subsection{Binary properties as a function of mass and orbital period\label{subsec:Binary-properties-as-function-of-mass}}

The binary fraction among stars more massive than our threshold mass
of $\Mlimit$ depends little on most of our model parameters. This
reflects the two mechanisms for making stars more massive than $\Mlimit$
star in the thick disc. Wind mass accretion, in which the system is
and remains binary, only increases the stellar mass by a few tenths
of a solar mass at most. In contrast, merging can nearly double the
mass of a star and merged binary-star systems are always single. This
is clear from Fig.~\ref{fig:S1-B1}~B1a: the blue (binary-star)
region extends to around $M=1.5\mathrm{\,M_{\odot}}$ but no further,
while the red region (single stars only) dominates at high mass. This
is a feature typical of all our model sets: the most massive stars
have merged and should all be single or have a wide companion if initially
in a triple system \citep{2016ComAC...3....6T}. Our result agrees
qualitatively with that of \citet{Jofre2016} who conclude that stars
less massive than $1.2\mathrm{\,M_{\odot}}$ are more likely to be
in binaries than those more massive than $1.2\mathrm{\,M_{\odot}}$.
In section~\ref{subsec:Thick-disc-membership} we discuss changes
to the mass threshold.

\begin{figure*}
\begin{centering}
\includegraphics[bb=0bp 0bp 612bp 288bp,scale=0.7]{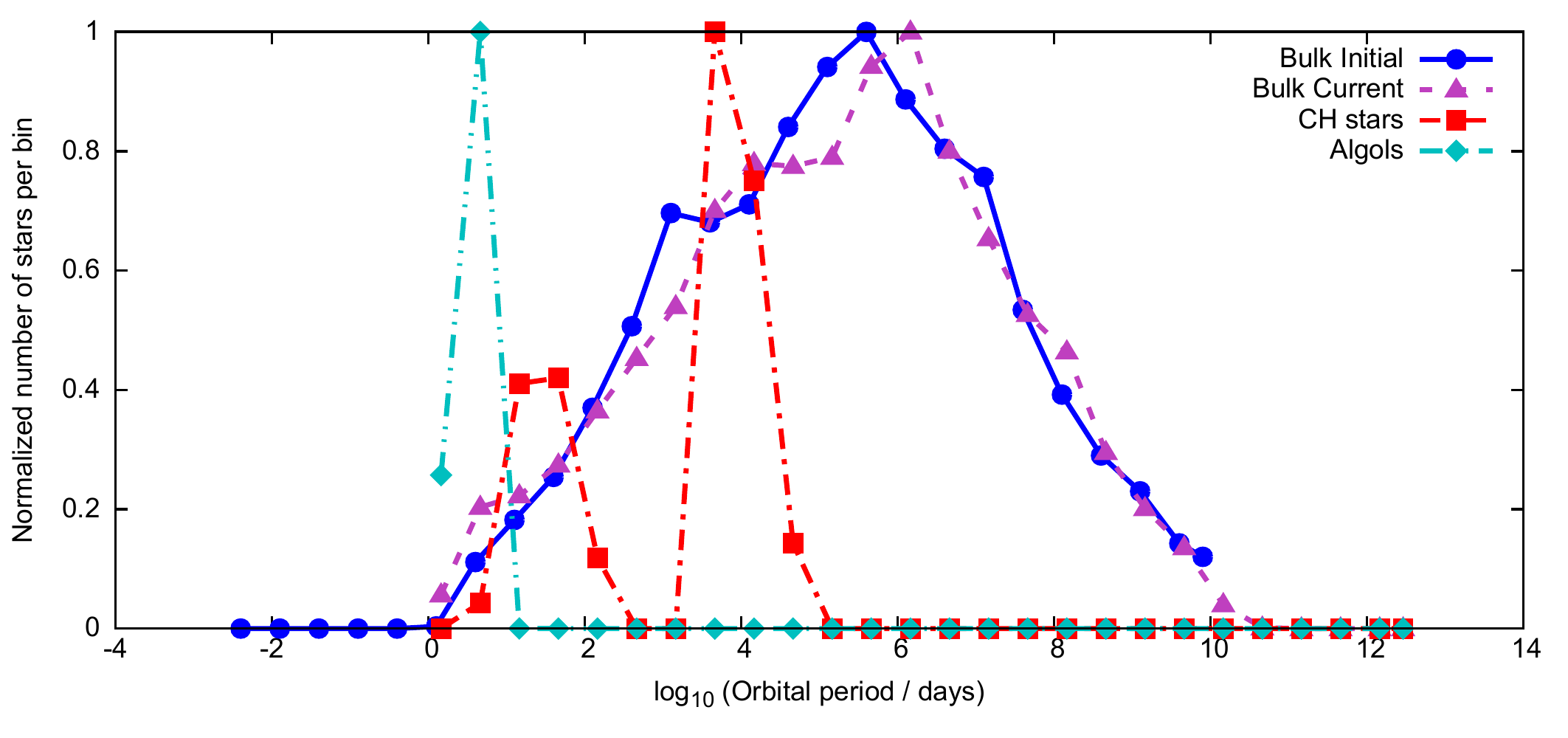}
\par\end{centering}
\caption{\label{fig:orbital-periods}Orbital period distributions in our default
model set of thick-disc binary giants, B1. The blue circles and purple
triangles show the initial and current distributions among the bulk
of our model stars with $M=1.10\pm0.05\mathrm{\,M_{\odot}}$ and $\protect\CN=0\pm0.05$.
The red squares show the period distribution of our CH stars with
$M=1.20\pm0.05\mathrm{\,M_{\odot}}$ and $\protect\CN=0.50\pm0.05$.
The cyan diamonds show the period distribution of the Algol-like systems
with $M=0.20\pm0.05\mathrm{\,M_{\odot}}$ and $\protect\CN=-1.50\pm0.05$.
Logarithmic bins in orbital period are $0.5\,\mathrm{dex}$ wide \change{and}
each distribution has its peak normalized to $1.0$. The bulk initial
distribution is slightly different to our initial binary period distribution
because only some stars enter the range given by $M=1.10\pm0.05\mathrm{\,M_{\odot}}$
and $\protect\CN=0\pm0.05$.}
\end{figure*}
The bulk of stars in model set B1, with $M=1.1\mathrm{\,M_{\odot}}$
and $\CN=0$, have an orbital period distribution which is very similar
to the initial distribution of orbital periods of G/K dwarfs (Fig.~\ref{fig:orbital-periods}).
Our CH stars, with $M=1.2\mathrm{\,M_{\odot}}$ and $\CN=0.5$, have
an orbital-period distribution similar to that predicted for barium
stars \citep{2010A&A...523A..10I}. A period gap between $10$ and
$1000\,\mathrm{d}$, caused by orbital shrinkage during common envelope
evolution, is clearly visible (e.g.~\citealp{2003ASPC..303..290,2013A&A...551A..50D,2015A&A...579A..49V}).
The Algol systems are all in short period binaries, as we expect of
low-mass giants undergoing Roche-lobe overflow.

Our models predict a small number of short-period binaries equivalent
to low-mass Algol systems. Given their low mass these stars may be
rejected by $\APO$ or may be too dim to be seen in great numbers.
Our prediction number of Algols is similar to the $0.1-0.2\pc$ of
Galactic disc stars predicted by \citet{2017arXiv170100746M}. Interaction
is more likely in giant stars than in field dwarfs, which at least
partly explains our greater predicted frequency. If the number of
Algols can be measured accurately it may us allow to constrain the
initial-period distribution. Our lower-metallicity model sets, B2
with $Z=10^{-3}$ and B3 with $Z=10^{-4}$, have Algol fractions of
\data{B2.algol} and \data{B3.algol}, suggesting that any observational
campaign will have to be careful to select stars by metallicity. 

\subsection{Thick disc membership: metallicity and age, threshold mass\label{subsec:Thick-disc-membership}}

Our thick-disc selection is probably biased. \emph{Kepler} observed
only a small fraction of the sky, so does not sample well the whole
thick or thin disc, and the sample is magnitude rather than volume
limited. The metallicity distribution in the $\APO$ sample (Fig.~\ref{fig:Metallicity-distribution-of-thick-disc-stars})
suggests that a fraction of the stars more massive than $1.3\mathrm{\,M_{\odot}}$
are actually high-metallicity interlopers whose origin is not in binary-star
interaction \citep{2011MNRAS.412.1203N}.

Our thick-disc age criteria are likely rather simplistic. We deliberately
choose our age range, $5$ to $\agemax\,\mathrm{Gyr}$, to match the
masses reported by $\APO$, assuming a metallicity of $Z=\defaultZ$
($\FeH=-0.24$). The $\APO$ masses may be systematically \change{overestimated
by $10-15\pc$} because they are derived from scaling relations rather
than more sophisticated techniques \change{\citep{2016ApJ...832..121G,2017MNRAS.tmp..120R}}.

To compensate for this, we can select older stars of lower mass, hence
greater age, when they are giants, as in our model set B20 with ages
$8$ to $13\,\mathrm{Gyr}$. The number of stars with mass in excess
of $\Mlimit$ reduces from \data{B1.1} to \data{B20.1}. However,
this test is unfair. We must also reduce our threshold mass to reflect
the increase in age. In model set S5, the single-star equivalent of
B20, there are no stars of mass exceeding $1.0\mathrm{\,M_{\odot}}$.
The fractions of giant stars with masses exceeding $1.0$, $1.1$
and $1.2\mathrm{\,M_{\odot}}$ in model set B20 are \data{B20.M1.0},
\data{B20.M1.1} and \data{B20.M1.2}, respectively. 

The best solution to the above problem is to construct a full thick-disc
population made up of models with a distribution of metallicity. We
shall do this in the future. Here we concentrate on the effects of
binary stars.

\subsection{Masses and core masses\label{subsec:Masses-and-core-masses}}

Our limited statistical analysis avoids the unpleasant task of taking
into account the sometimes significant errors on the $\APO$ masses.
We assume that, in bulk, these errors have little effect on the relatively
large fraction ($\APOpc$) of stars with masses in excess of $\Mlimit$.
Of 300 stars, and given the high-metallicity interlopers described
above, this seems reasonable.

\begin{figure*}
\begin{centering}
\includegraphics[bb=0bp 0bp 360bp 432bp,scale=0.7]{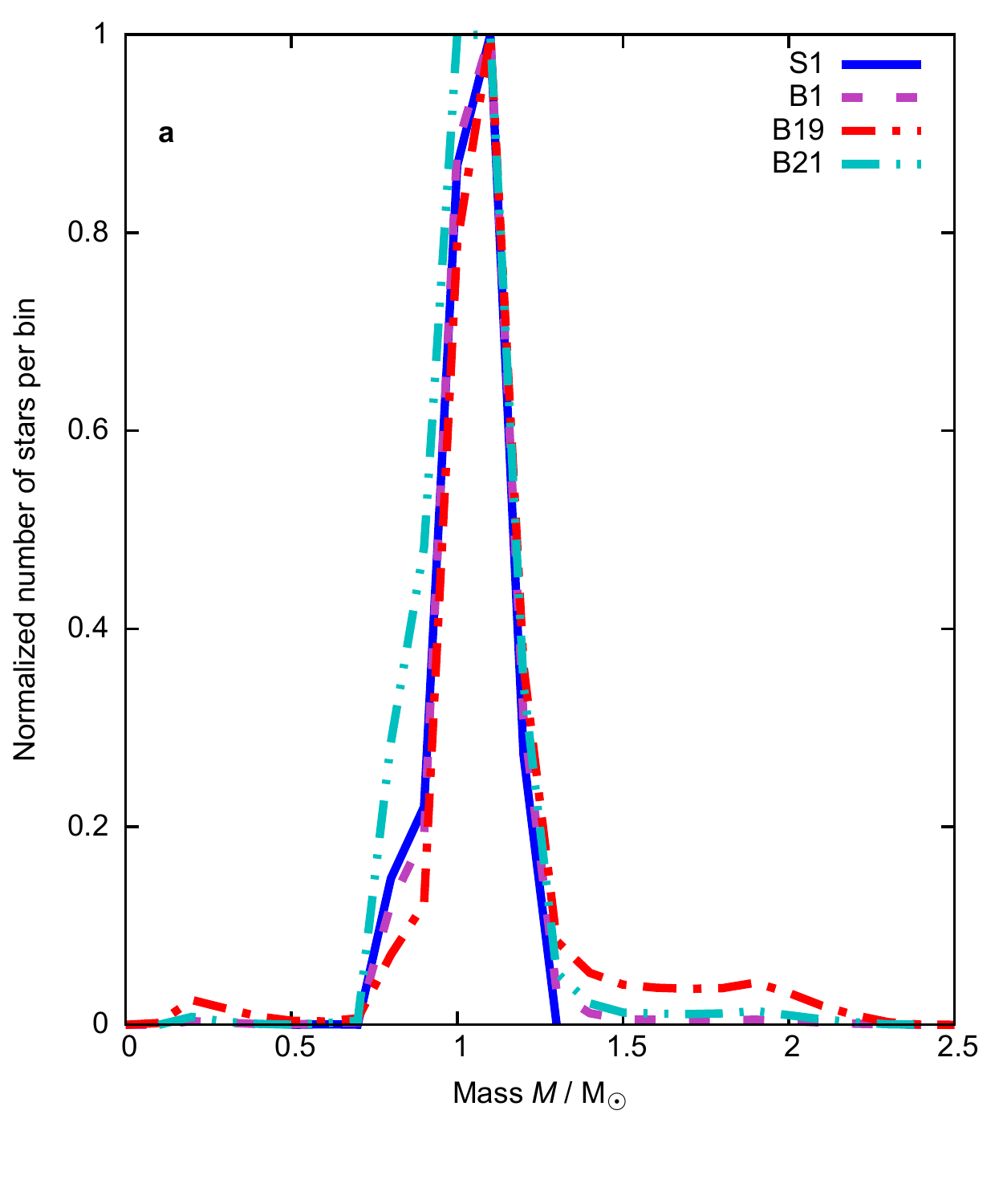}\includegraphics[bb=0bp 0bp 360bp 432bp,scale=0.7]{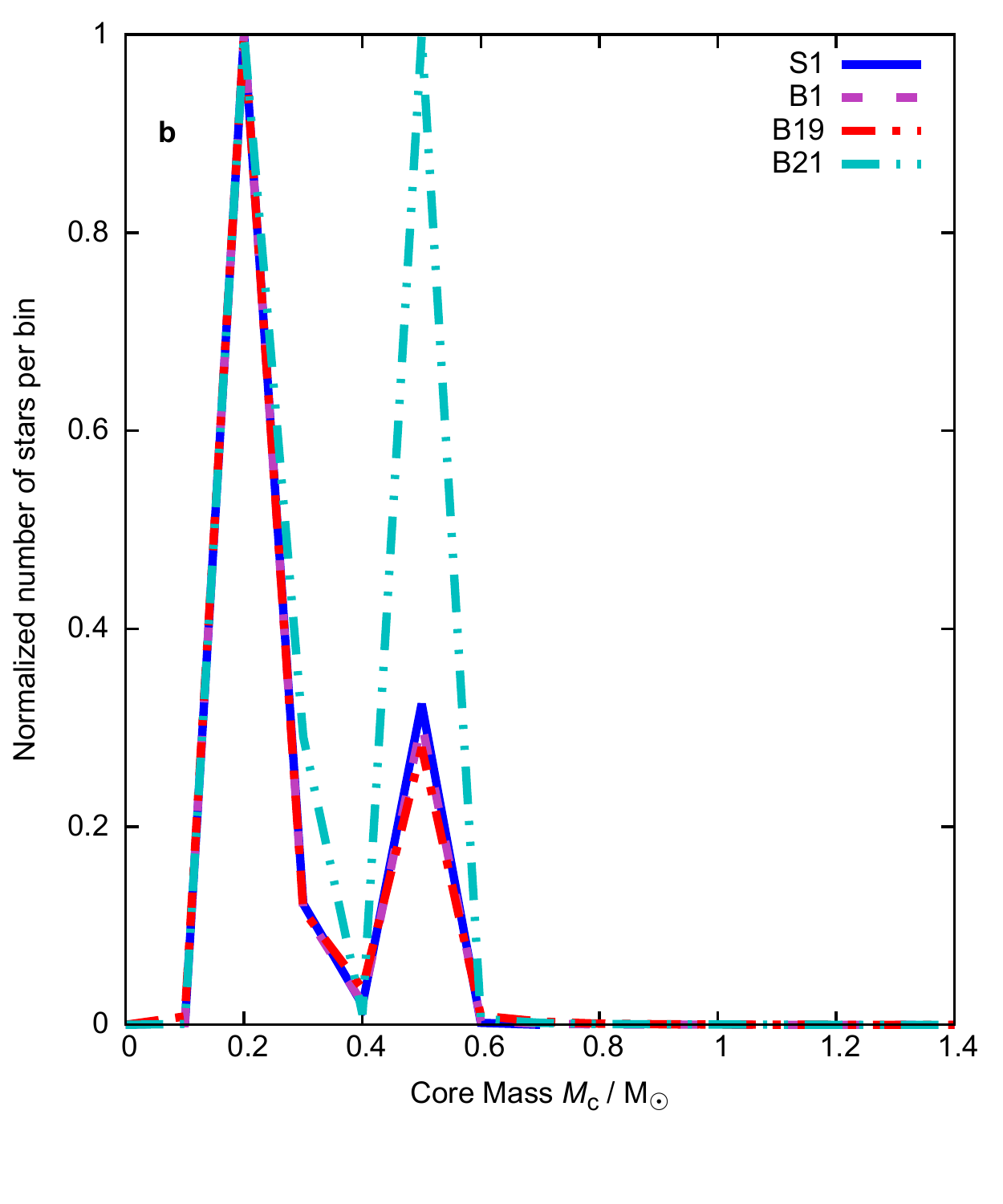}
\par\end{centering}
\caption{\label{fig:masses-and-core-masses}Mass (panel \textbf{a}, left) and
core mass (panel \textbf{b}, right) distributions in our model sets
S1 (single stars with default physics), B1 (initially-binary stars
only with default physics), B19 (as B1 with a logarithmically-flat
initial-separation distribution) and B21 (as B1 with $2\leq\protect\logg\leq3$
selection to mimic the $\protect\APO$ data). Bins have a fixed width
of $0.1\mathrm{\,M_{\odot}}$ and the peak of each distribution is
normalized to $1.0$. The tail of stars more massive than $\protect\Mlimit$
is clear in panel \textbf{a }and a tail of fewer stars with core mass
in excess of $0.6\mathrm{\,M_{\odot}}$ can be seen in panel \textbf{b}. }
\end{figure*}

The distributions of masses and core masses of giant stars in our
model sets S1, B1, B19 and B21 are shown in Fig.~\ref{fig:masses-and-core-masses}.
The tail of stars of mass greater than $\Mlimit$ is clear in the
binary-star model sets (B1, B19 and B21), while it is absent in the
single-star model set (S1). The bulk distribution of core masses differs
little between single and binary stars. The small fraction of stars
more massive than $1.3\mathrm{\,M_{\odot}}$ contributes to a high-mass
tail in both mass and core mass.

Model set B21 is exceptional. In this model set we select stars with
$-2\leq\logg\leq3$ to match our $\APO$ data selection. This range
of $\log g$ contains the red clump (core helium burning) stars which
have a core mass of about $0.5\mathrm{\,M_{\odot}}$ so stars with
this core mass are artificially enhanced in number. A good understanding
of selection effects is thus critical to matching any core-mass distribution
measured, e.g.~by asteroseismology, to stellar evolution models.
The distribution of stellar masses does not suffer from such a problem
because the mass of the red clump stars is only slightly less than
that of stars on the red giant branch. This is caused by mass loss
at the tip of the giant branch prior to core helium burning \citep{Miglio2012}.
The peak of the mass distribution of stars in B21 is thus shifted
to slightly lower mass than in S1, B1 or B19.\textbf{ }

\subsection{Comparison with CEMP-\emph{s} stars\label{subsec:Comparison-with-CEMPs}}

The number of massive thick-disc stars in our models, typically $2\pc$
compared to the observed $\APOpc$, may suggest that there are young
interlopers in the observed sample. However, there is circumstantial
evidence from another stellar population that supports significant
binary-star mass transfer. The \emph{s}-process rich, carbon-enhanced
metal-poor (CEMP-\emph{s}) stars of the Galactic halo have metallicities
$\FeH\leq-2$ and are found in intermediate-period binary systems
\citep{2004MmSAI..75..772T,Lucatello2005,2014MNRAS.441.1217S}. Carbon
and $s$-process elements are both made in AGB stars. So CEMP-$s$
stars probably form by mass transfer from an AGB star similarly to
CH and barium stars \citep{1980ApJ...238L..35M,1984ApJ...280L..31M,1999Jorissen}.
CEMP-\emph{s} stars constitute $10$ to $20\pc$ of the extremely-metal
poor (EMP) population \citep{2006ApJ...652L..37L,2014ApJ...788..131L}
while standard binary population synthesis models predict a CEMP/EMP
ratio of about $2\pc$ at $\FeH=-2.3$ \citep{2009A&A...508.1359I}.
The predicted number of CEMP-\emph{s} stars is increased if more are
born with short and intermediate orbital periods which are close enough
for efficient mass transfer. A logarithmically-flat initial-separation
distribution rather than log-normal inital-period distribution does
just this (Appendix~\ref{sec:initial-period-or-separation-distributions}).
Such an initial distribution of binary periods also increases the
number of thick-disc stars with masses exceeding $\Mlimit$, as required
to match the $\APO$ data.

\subsection{Alternatives to our binary-star scenario\label{subsec:Alternatives-to-binary-stars}}

\change{To increase the number of thick-disc stars with masses in
excess of $\Mlimit$ without invoking binary interactions, a young,
metal-poor, $\alpha$-rich population is required. It has been suggested
that Galactic migration is responsible for populating the thick disc
with such stars \citep{Chiappini2015}. We cannot investigate this
claim with our models but we know binary stars exist at all metallicities
so there should be many in the thick disc. So some will interact and
the question is how many. As we discuss above the initial distribution
of orbits is key. To make $10\pc$ of giants more massive than $\Mlimit$
we need a logarithmically-flat initial-separation distribution, so
that there are many more interacting low-mass binaries than predicted
by the solar neighbourhood orbital period distribution. The CEMP-$s$
number problem (Sec.~\ref{subsec:Comparison-with-CEMPs}) also requires
an orbital separation distribution with many interacting binaries,
so qualitatively supports our use of this distribution even if the
evidence is rather circumstantial. It is possible that triple-star
interactions lead to a similar number of interacting binaries even
with an initial orbital period distribution more like that of the
solar neighbourhood. However quantitative predictions are still rather
preliminary \citep{2016ComAC...3....6T}.

The migration scenario cannot solve the CEMP-\emph{s} number problem
either. First these stars are found in the Galactic halo, which is
likely to undergo migration on far longer spatial or temporal scales
than the thick disc. Could enough carbon-rich stars be formed and
migrate into the, especially outer \citep{2012ApJ...744..195C}, halo
in $14\,\Gyr$? Secondly, if CEMP-$s$ stars are young and single,
as in the migration scenario, they are born with $\CFe>+1$ and metallicity
$\FeH\lesssim-2$. Recent star formation under such conditions in
our Galaxy is unknown. Occam's razor therefore favours that interaction
in intermediate-period binaries, which are well known to exist, as
an explanation for CEMP-$s$ stars. 

We have shown that thick-disc stars more massive than $\Mlimit$ can
be formed in interacting binaries or higher order multiple stellar
systems. We have good reason to think such systems exist but there
may well be some contribution to the massive thick-disc population
from migration of younger stars. Differentiation between the two possibilities
may be possible when we better know the binary fraction and orbital
period distribution of the stars more massive than $\Mlimit$. Our
simulations predict that many of these stars are merged, hence single.
They may also be wide binaries which were triples. The migration scenario
would predict a binary fraction and orbital period distribution which
are similar to those of the stars at birth.}

\textbf{}

\subsection{Blue stragglers\label{subsec:Blue-Stragglers}~}

Many of our modelled thick-disc stars with masses greater than $\Mlimit$
accrete mass while they are on the main sequence. They thus pass through
a blue-straggler phase during which they look younger than the main-sequence
lifetime corresponding to their mass. This simple definition of a
blue straggler quietly neglects the many selection effects which plague
quantitative attempts to count such systems (\citealp{2013AJ....145....8G}
and the discussion of \citealp{2014ApJ...780..117S} and references
therein). In most of our model sets, $10$ to $30\pc$ of giants more
massive than $\Mlimit$ are expected to be blue stragglers before
they ascend the giant branch. Most of these form by wind mass transfer
and hence some are expected to be enhanced in carbon and \emph{s-}process
elements. If wind-RLOF is disabled, as in model set B9, the number
of giants that were blue stragglers drops from the \data{B1.7} of
model set B1 to just \data{B9.7}. If the relative number and chemical
properties of blue stragglers in the thick disc can be assessed, perhaps
by barium abundance to detect pollution from a TPAGB donor, the effect
of wind-RLOF may be quantifiable. A first attempt suggests around
$10\pc$ of thick-disc main-sequence stars may be blue stragglers
\citep{2017MNRAS.464.2610F}. Four of the $\APO$ stars more massive
than $1.3\mathrm{\,M_{\odot}}$ are, however, not rich in barium \citep{Yong2016}
suggesting they are post-main sequence mergers.

\subsection{Model uncertainties: binary physics, nucleosynthesis\label{subsec:Model-uncertainties:-binary-nucsyn}}

While our input distributions are rather uncertain, our binary star
model suffers from uncertainty too. As discussed above, wind-RLOF
is required to make a significant number of binary-star giants with
masses in excess of $\Mlimit$. We also test our prescriptions for
common-envelope evolution, companion-reinforced attrition and mass
transfer. Changing the appropriate parameters in our model has limited
effect on our results, certainly less than changing the initial-period
distribution. 

Our nucleosynthesis model, while an improvement on previous versions
of $\binaryc$, is still far from perfect even if it well reproduces
single-star evolution. We assume that stars undergo first dredge up
to a depth the time dependence of which is calculated by the \emph{$\BSE$
}algorithm (\citealp{2000MNRAS.315..543H,2002MNRAS_329_897H}) and
limited to a maximum depth given by our $\STARS$ models. 

For evolved stars on the red giant branch, the entire convective envelope
is mixed to a depth $M_{\mathrm{env}}=M-M_{\mathrm{c}}$. This is
fitted to the mass $M_{0}$ which is defined by \citet{2000MNRAS.315..543H}
as the mass of the star at the base of the giant branch that goes
on to determine its core mass. This is the same approach taken in\emph{
$\BSE$ }to calculate the stellar evolution of red giant stars but
it neglects the fact that the core mass to mass ratio, $M_{\mathrm{c}}/M$,
is different to that of a single star of the same total mass. The
effect of a variable, off-grid, $M_{\mathrm{c}}/M$ on the depth of
first dredge up in such stars remains to be tested with detailed stellar
models.

We further assume that common-envelope evolution mixes the stellar
envelope right down to the core \citep{2011ApJ...730...76I}. This
may be deeper than the convective envelope has mixed at that phase
of evolution, particularly if common-envelope evolution occurs early
on the giant branch. Given the energy injected into the envelope by
orbital decay of the companion star which itself likely descends to
near the core-envelope boundary, this seems reasonable. Unfortunately,
existing stellar evolution models of common-envelope evolution, and
especially stars that merge during this phase, do not go on to predict
the subsequent evolution of the stars except in a few cases \citep[e.g.][]{2013MNRAS.tmp..656Z,2015A&A...579A..49V}.
This issue should be addressed in the future.

\section{Conclusions}

\label{sec:Conclusions}We explore the properties of the binary-star
population of low-mass stars in the thick disc which are more massive
than they should be given their age. Our models naturally contain
a population of red giant stars, from $1$ to $10\pc$ of all giants,
more massive than the maximum $\Mlimit$ that is possible in single-star
evolution. These stars are more likely to be single, merged objects
than binary stars, even when common-envelope ejection is efficient
($\alpha_{\mathrm{CE}}=1$). \change{Some may be wide binaries that
were originally hierarchical triple systems.} Modelling uncertainties
are generally not important, except that wind Roche lobe overflow
or companion-reinforced attrition are required to make binary stars
with $M>\Mlimit$, and that to make as many as seen in $\APO$ we
require a logarithmically-flat initial orbital period distribution
with many initially close binary stars. Single-stars made by merging
processes likely have a peculiar mass to core-mass ratio which could
be measured asteroseismologically by targeting the most massive stars
in $\APO$ or similar samples. It is also crucial to comprehensively
monitor the radial velocities of the thick-disc stars which are anomalously
massive, such as those found in $\APO$, to know their orbital periods
and binary fraction. This work continues in parallel to our theoretical
study.

\section*{Acknowledgements}

\label{sec:Acknowledgements}

\change{We thank the referee for their suggestions which have improved
the clarity of the manuscript.} RGI thanks the STFC for funding his
Rutherford fellowship under grant ST/L003910/1, Churchill College,
Cambridge for his fellowship and access to their library, and Carlo
Abate, Guy Davies, Ilya Mandel, Andrea Miglio and Richard Stancliffe
for useful discussions. RGI and GMH thank the STFC for funding Rutherford
grant ST/M003892/1. HPP thanks the STFC for funding her research studentship.
PJ and TM acknowledge support from the European Union FP7 programme
through ERC grant number 320360. TM acknowledges support provided
by the Spanish Ministry of Economy and Competitiveness (MINECO) under
grant AYA-2014-58082-P and AYA2014-56359-P. CAT thanks Churchill College,
Cambridge for his fellowship. This research has made use of NASA's
Astrophysics Data System Bibliographic Services. We thank Dave Green
for his Cubehelix colour scheme \citep{2011BASI...39..289G} and his
debugging of \emph{MNRAS'} new class file.

\bibliographystyle{mnras}
\bibliography{references}

\vfill{}

\pagebreak{}

\appendix

\section{Stellar winds}

\label{sec:stellar-winds}

In \emph{$\binaryc$} we mostly use the stellar wind mass loss prescription
of \emph{$\SSE$} \citep{2000MNRAS.315..543H}. During giant-branch
and subsequent evolution a mass-loss rate, $\dot{M}_{R}$, is calculated
with the formula of \citet{1978AA-70-227K},
\begin{alignat}{1}
\dot{M}_{\mathrm{R}} & =\eta\,4\times10^{-13}\frac{\left(L/\mathrm{L_{\odot}}\right)\left(R/\mathrm{R_{\odot}}\right)}{\left(M/\mathrm{M_{\odot}}\right)}\mathrm{\,M_{\odot}}\,\mathrm{yr}^{-1},
\end{alignat}
where $\eta=0.5$ and $M$, $L$ and $R$ are the stellar mass, luminosity
and radius.

On the early asymptotic giant branch (EAGB) we use the formula of
\citet{1993ApJ...413..641V},
\begin{alignat}{1}
\dot{M}_{\mathrm{VW}}\left({\cal A}\right) & =10^{\Gamma\left({\cal A}\right)}\mathrm{\,M_{\odot}}\,\mathrm{yr}^{-1},\label{eq:VW-wind}
\end{alignat}
where
\begin{alignat}{2}
\Gamma\left({\cal A}\right) & = & \left[P_{\mathrm{Mira}}/\mathrm{d}-100\,{\cal A}\max\left(\frac{M}{\mathrm{M_{\odot}}}-2.5,0\right)\right]\!\times0.0125\!-\!11.4,
\end{alignat}
and the Mira pulsation period $P_{\mathrm{Mira}}$ is,

The EAGB superwind mass-loss rate is, 
\begin{alignat}{1}
\dot{M}_{\mathrm{superwind},\mathrm{EAGB}} & =1.36\times10^{-9}\left(\frac{L}{\mathrm{L_{\odot}}}\right)\mathrm{\,M_{\odot}}\,\mathrm{yr}^{-1}.\label{eq:VW93-superwind}
\end{alignat}
The wind mass loss rate is then,
\begin{alignat}{1}
\dot{M}_{\mathrm{EAGB}} & =\min\left[\dot{M}_{\mathrm{superwind},\mathrm{EAGB}},\,\dot{M}_{\mathrm{VW}}\left({\cal A}=1\right)\right],\label{eq:VW-wind-1}
\end{alignat}
as used by \citep{2002MNRAS_329_897H}. 

On the thermally-pulsing asymptotic giant branch (TPAGB) we modify
the \citet{1993ApJ...413..641V} formula according to \citet{Karakas2002}
to calculate the wind mass-loss rate, 
\begin{alignat}{1}
\dot{M}_{\mathrm{TPAGB}} & =\min\left[\dot{M}_{\mathrm{superwind},\mathrm{TPAGB}},\,\dot{M}_{\mathrm{VW}}\left({\cal A}=0\right)\right],
\end{alignat}
where
\begin{alignat}{1}
\dot{M}_{\mathrm{superwind},\mathrm{TPAGB}} & =\left\{ \begin{array}{cc}
0, & P_{\mathrm{Mira}}<500\,\mathrm{d},\\
\dfrac{cL}{v_{\mathrm{w}}\mathrm{L_{\odot}}} & P_{\mathrm{Mira}}\geq500\,\mathrm{d},
\end{array}\right.\label{eq:alt-superwind}
\end{alignat}
$c$ is the speed of light\emph{ }in a vacuum and
\begin{alignat}{1}
v_{\mathrm{w}} & =\min\left[15.0,\max\left(-13.5+0.056P_{\mathrm{Mira}}/\mathrm{d},0\right)\right]\,\mathrm{km}\,\mathrm{s}^{-1}\label{eq:vwind}
\end{alignat}
is the wind velocity at infinity \citep{2013A&A...552A..26A}. The
wind mass-loss rate on the asymptotic giant branch $\dot{M}_{\mathrm{AGB}}$
is then $\dot{M}_{\mathrm{EAGB}}$ during the EAGB and $\dot{M}_{\mathrm{TPAGB}}$
during the TPAGB.

The wind of \citet{1990A&A...231..134N} is calculated by 
\begin{alignat}{1}
\dot{M}_{\mathrm{NJ}} & =9.6\times10^{-15}\,\times\nonumber \\
 & \left(\frac{Z}{0.02}\right)^{0.5}\left(\frac{R}{\mathrm{R_{\odot}}}\right)^{0.81}\left(\frac{L}{\mathrm{L_{\odot}}}\right)^{1.24}\left(\frac{M}{\mathrm{M_{\odot}}}\right)^{0.16}\mathrm{\,M_{\odot}}\,\mathrm{yr}^{-1},
\end{alignat}
where $Z$ is the metallicity. A Wolf-Rayet mass-loss rate, suitable
for stars with thin envelopes, is given by
\begin{alignat}{1}
\dot{M}_{\mathrm{WR}} & =10^{-13}\left(\frac{L}{\mathrm{L_{\odot}}}\right)^{1.5}\max\left(0,\,1-\mu\right)\mathrm{\,M_{\odot}}\,\mathrm{yr}^{-1},
\end{alignat}
where
\begin{alignat}{1}
\mu & =\begin{cases}
\left(1-\frac{M_{\mathrm{c}}}{M}\right)\!\min\!\left\{ 5,\max\left[1.2,\!\left(\dfrac{10^{-4}L}{7\mathrm{L_{\odot}}}\right)^{-0.5}\right]\right\} , & \text{H-giants},\\
5\left(1-M_{\mathrm{c}}/\min\left[1.45M-0.31\mathrm{M_{\odot}},\,M\right]\right), & \text{He-giants},
\end{cases}
\end{alignat}
 and $M_{\mathrm{c}}$ is the stellar core mass. The luminous blue
variable (LBV) mass-loss rate is given by,
\begin{alignat}{1}
\dot{M}_{\mathrm{LBV}} & =0.1x^{3}\left(\frac{L}{L_{\mathrm{LBV}}}-1\right)\mathrm{\,M_{\odot}}\,\mathrm{yr}^{-1},
\end{alignat}
if $x\equiv10^{-5}\left(R/\mathrm{R_{\odot}}\right)\left(L/\mathrm{L_{\odot}}\right)^{-0.5}-1>0$,
$L>L_{\mathrm{LBV}}=6\times10^{5}\mathrm{\,L_{\odot}}$ and the star
has left the main sequence. $\dot{M}_{\mathrm{LBV}}=0$ otherwise. 

The total wind mass loss rate is then, for nuclear-burning stars,
\begin{alignat}{1}
\dot{M}_{\mathrm{wind}} & =\max(\dot{M}_{\mathrm{R}},\,\dot{M}_{\mathrm{AGB}},\,\dot{M}_{\mathrm{NJ}},\,\dot{M}_{\mathrm{WR}})+\dot{M}_{\mathrm{LBV}},
\end{alignat}
and zero for stellar remnants (white dwarfs, neutron stars and black
holes).

\section{Initial orbital period and separation distributions}

\label{sec:initial-period-or-separation-distributions}

In this paper we use either a hybrid initial-period distribution which
is a function of initial primary mass, described in section~\ref{subsec:hybrid-period-distribution},
or a logarithmically-flat initial-separation distribution as described
in section~\ref{subsec:opikdist}.

\subsection{Hybrid period distribution}

\label{subsec:hybrid-period-distribution}
\begin{figure*}
\includegraphics{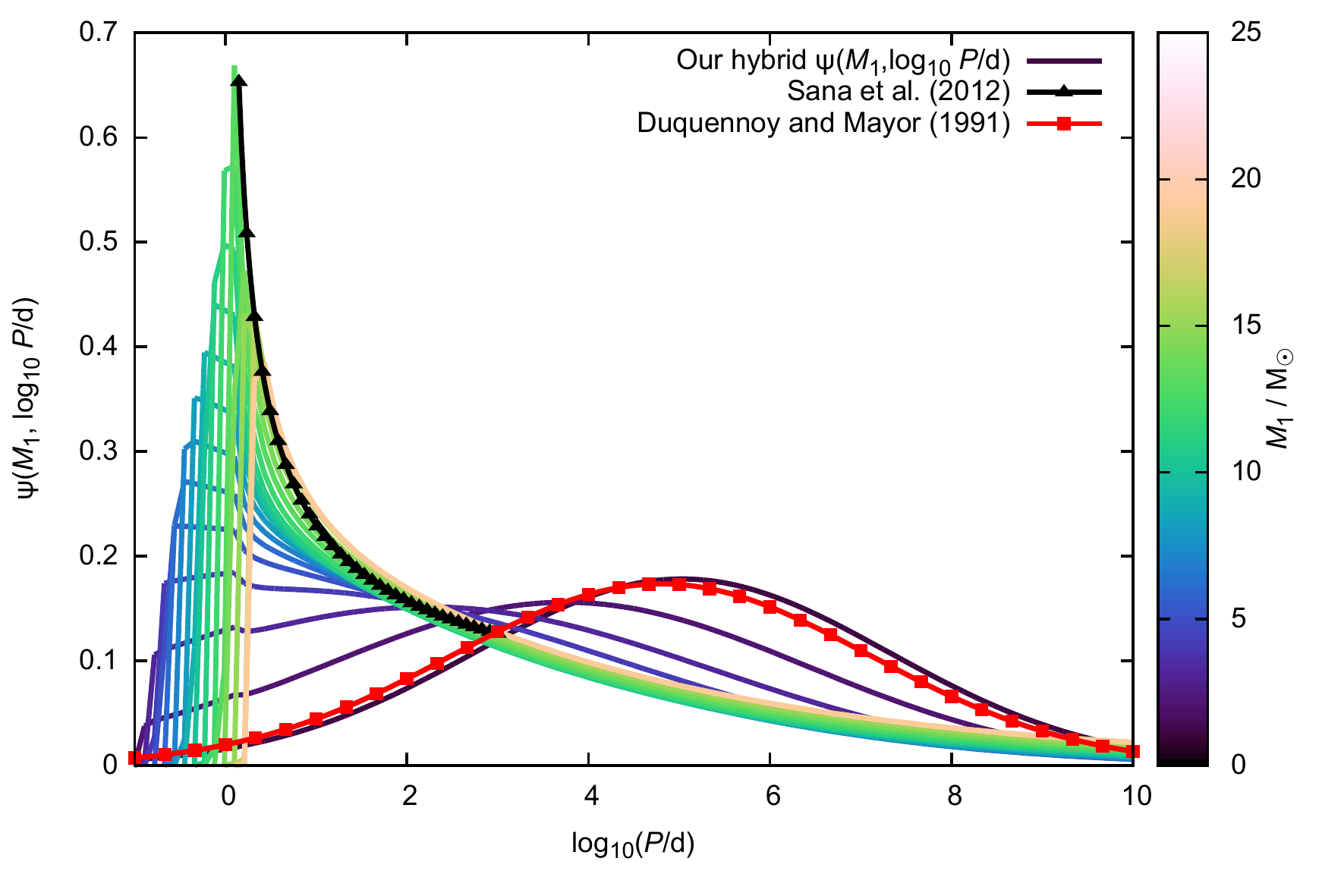}

\caption{\label{fig:Our-hybrid-initial-period-dists}Comparison of our hybrid
initial-period distribution, $\psi\left[M_{1},\log_{10}\left(P/\mathrm{d}\right)\right]$
of Eq.~(\ref{eq:psi}), as a function of primary-star mass, $M_{1}$,
and logarithm of the orbital period, $\log_{10}\left(P/\mathrm{d}\right)$,
to the power-law distribution of \citet[black triangles]{2012Sci...337..444S}
at high mass and the log-normal distribution of \citet[red squares]{1991A&A...248..485D}
at low mass. The curves span{\change{ the primary mass range}} $1$~to~$20\mathrm{\,M_{\odot}}$
in $1\mathrm{\,M_{\odot}}$ increments. \change{When we simulate
a stellar population orbital periods short enough that Roche-lobe
overflow occurs on the main sequence have zero probability even though
they are shown in this figure. The short period limit is a function
of both stellar masses and metallicity.} Note that the observations
of \citet{2012Sci...337..444S} are limited to periods shorter than
$1000\,\mathrm{d}$. At longer periods we extrapolate our Eq.~(\ref{eq:psi}).}
\end{figure*}

\begin{figure*}
\includegraphics{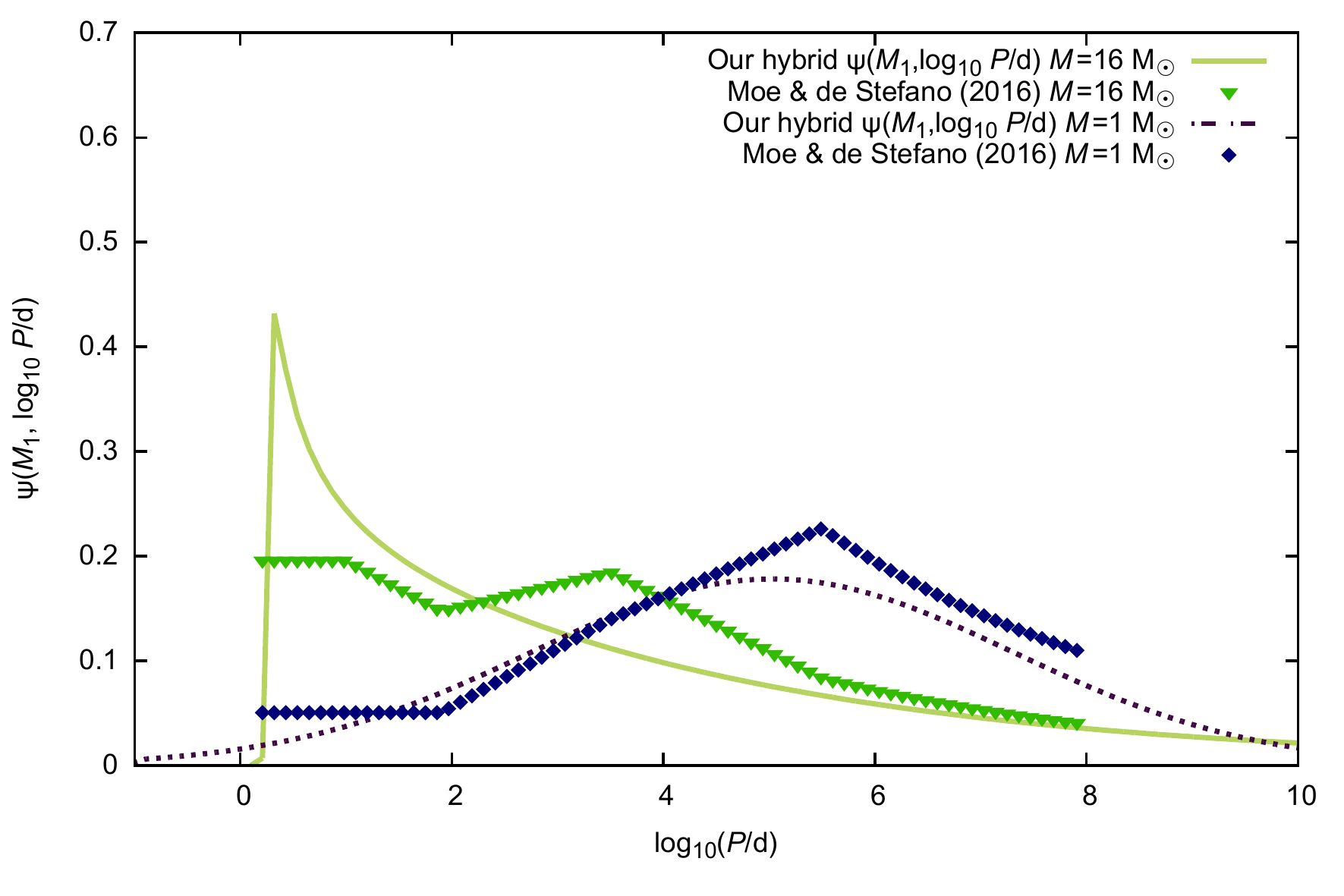}

\caption{{\change{\label{fig:Our-hybrid-initial-period-dists2}As Fig.~\ref{fig:Our-hybrid-initial-period-dists}
at $M=1$ (blue, dot-dashed line) and $16\mathrm{\,M_{\odot}}$ (green
solid line) compared to the distribution of \citet{2017ApJS..230...15M}
at $M=1\mathrm{\,M_{\odot}}$ (blue diamonds) and $M=16\mathrm{\,M_{\odot}}$
(green triangles).}}}
\end{figure*}

The initial orbital period
$P$ distribution is calculated as a function primary mass $M_{1}$
and ${\cal P}=\log_{10}\left(P/\mathrm{d}\right)$ where {\change{
$\max\left({\cal P}_{\min},-1\right)\leq{\cal P}\leq10$. ${\cal P}_{\mathrm{min}}\left(M_{1},\,M_{2},\,Z\right)$
is the shortest orbital period for which Roche-lobe overflow does
not occur at the zero-age main sequence. We obtain this from $\binaryc$.}}
We define the density function of the probability, $p$, of finding
a binary system with the logarithm of its orbital period between ${\cal P}$
and ${\cal P}+d{\cal P}$ given it has primary mass $M_{1}$ by,
\begin{alignat}{1}
\psi & =\psi(M_{1},{\cal P})=\frac{dp}{d{\cal P}}\nonumber \\
 & ={\cal H}{\cal G}\left\{ \exp\left[-\frac{\left({\cal P}-\mu\right)^{2}}{2\sigma^{2}}\right]+\frac{{\cal K}}{\max\left(0.1,{\cal P}\right)}\right\} ,\label{eq:psi}
\end{alignat}
where
\begin{alignat}{1}
{\cal M} & =\max\left[1.15,\min\left(16.3,\,M_{1}/\mathrm{M_{\odot}}\right)\right],
\end{alignat}
\begin{alignat}{1}
f({\cal M},\,b,\,a) & =a+\left(b-a\right)\left(\frac{{\cal M}-1.15}{16.3-1.15}\right),
\end{alignat}
\begin{alignat}{1}
\mu & =\mu\left({\cal M}\right)=f\left({\cal M},-17.8,5.03\right),
\end{alignat}
\begin{alignat}{1}
\sigma & =\sigma\left({\cal M}\right)=f\left({\cal M},9.18,2.28\right),
\end{alignat}
\begin{alignat}{1}
{\cal K} & ={\cal K}\left({\cal M}\right)=f\left({\cal M},0.0693,0\right),
\end{alignat}

\begin{alignat}{1}
{\cal G} & ={\cal G}\left({\cal M}\right)=\left(1+\epsilon^{P-\nu}\right)^{-1},
\end{alignat}
\begin{alignat}{1}
\nu & =\nu\left({\cal M}\right)=f\left({\cal M},\,0.3,\,-1\right),
\end{alignat}
\begin{alignat}{1}
\log_{10}\epsilon & =-30,
\end{alignat}
 and ${\cal H}={\cal H}\left({\cal M}\right)$ is a normalization
factor such that,
\begin{alignat}{1}
{\cal H}\left({\cal M}\right)=\int_{\max\left({\cal P}_{\mathrm{min}},-1\right)}^{10}\psi\left({\cal M},\,{\cal P}\right)\,d{\cal P} & =1\,.
\end{alignat}
Fig.~\ref{fig:Our-hybrid-initial-period-dists} shows how Eq.~(\ref{eq:psi})
compares to the power-law distribution, $\psi({\cal P})\propto{\cal P}^{-0.55}$,
of \citet{2012Sci...337..444S} at high mass ($M_{1}\geq16.3\mathrm{\,M_{\odot}}$)
and the log-normal distribution of \citet{1991A&A...248..485D} at
low mass ($M_{1}\leq1.15\mathrm{\,M_{\odot}}$).{\change{ To produce
Fig.~\ref{fig:Our-hybrid-initial-period-dists} we assume ${\cal P}_{\mathrm{min}}\left(M_{1},\,M_{2},\,Z\right)=-1$
because with $M=1\mathrm{\,M_{\odot}}$, $M_{2}=0.5\mathrm{\,M_{\odot}}$
and $Z=0.008$, ${\cal P}_{\mathrm{min}}=-0.55$, while with $M_{1}=16\mathrm{\,M_{\odot}}$,
$M_{2}=8\mathrm{\,M_{\odot}}$ and $Z=0.008$, ${\cal P}_{\mathrm{min}}=-0.094$.}}
The observations of \citet{2012Sci...337..444S} are limited to $\log_{10}\left(P/\mathrm{d}\right)<3$.
{\change{Fig.~\ref{fig:Our-hybrid-initial-period-dists2} compares
our distribution to that of \citet[their equation 23]{2017ApJS..230...15M}
with $M=1$ and $16\mathrm{\,M_{\odot}}$.}} 

\subsection{Logarithmically-flat initial-separation distribution}

\label{subsec:opikdist}

Our logarithmically-flat initial-separation distribution is based
on that of \citet{1924PTarO..25f...1O} and has been used in many
binary population synthesis studies (e.g.~\citealp{2002MNRAS_329_897H}).
In the range of initial separations $3\leq a/\mathrm{R_{\sun}}\leq10^{4}$
the probability ($p$) density function is a flat function of $\log_{10}\left(a/\mathrm{R_{\odot}}\right)$,
\begin{alignat}{1}
\frac{dp}{d\log_{10}\left(a/\mathrm{R_{\odot}}\right)} & =\mathrm{constant}=4-\log_{10}3,\label{eq:opikdist}
\end{alignat}
and zero otherwise.

\section{Orbital inclination}

\label{sec:orbitalinclination}

Define $i$, the orbital inclination, to be the angle between the
line-of-sight and the normal to the orbital plane of the binary system.
Binaries with $i=0$ and $\pi/2$ are face on and edge on, respectively.
The probability that $i$ lies between $i$ and $i+di$, assuming
all angles are equally likely, is then
\begin{alignat}{1}
P(i)\,di & =\sin i\,di.
\end{alignat}
The projected radial velocity is,
\begin{alignat}{1}
v & =K\sin i,\label{vel}
\end{alignat}
where $K=K_{j}$ is the radial velocity semi-amplitude of star $j=1,\,2$,
\begin{alignat*}{1}
K_{j} & =M_{3-j}\sqrt{\frac{G}{a\left(1-e^{2}\right)\left(M_{1}+M_{2}\right)}},
\end{alignat*}
where $a$ is the semi-major axis of the orbit, $e$ is the orbital
eccentricity and $G$ is Newton's gravitational constant.

If systems with $v>v_{\mathrm{crit}}$ are observed as binary, while
systems with $v<v_{\mathrm{crit}}$ are observed as single, the probability
that a system with radial velocity amplitude $K$ is observed as a
binary is,
\begin{alignat}{1}
P(K) & =\int_{a}^{b}P(i)\,di,
\end{alignat}
where $a=\arcsin\left(v_{\mathrm{crit}}/K\right)$ and $b=\pi/2$.
The probability that a binary system with radial velocity amplitude
$K$ is observed as a binary is thus, 
\begin{alignat}{1}
P(K) & =\cos\left[\arcsin\left(\frac{v_{\mathrm{crit}}}{K}\right)\right].
\end{alignat}
We set $v_{\mathrm{crit}}=1\,\mathrm{km}\,\mathrm{s}^{-1}$ in all
our model sets.
\end{document}

%% file: csnames.tex
\expandafter\newcommand\csname dataS1.1\endcsname{\ensuremath{0}}
\expandafter\newcommand\csname dataS1.2\endcsname{\ensuremath{0}}
\expandafter\newcommand\csname dataS1.3\endcsname{\ensuremath{0}}
\expandafter\newcommand\csname dataS1.4\endcsname{\ensuremath{0}}
\expandafter\newcommand\csname dataS1.5\endcsname{\ensuremath{0}}
\expandafter\newcommand\csname dataS1.6\endcsname{\ensuremath{0}}
\expandafter\newcommand\csname dataS1.7\endcsname{\ensuremath{0}}
\expandafter\newcommand\csname dataS1.8\endcsname{\ensuremath{0}}
\expandafter\newcommand\csname dataS1.algol\endcsname{\ensuremath{0}}
\expandafter\newcommand\csname dataS2.1\endcsname{\ensuremath{0}}
\expandafter\newcommand\csname dataS2.2\endcsname{\ensuremath{0}}
\expandafter\newcommand\csname dataS2.3\endcsname{\ensuremath{0}}
\expandafter\newcommand\csname dataS2.4\endcsname{\ensuremath{0}}
\expandafter\newcommand\csname dataS2.5\endcsname{\ensuremath{0}}
\expandafter\newcommand\csname dataS2.6\endcsname{\ensuremath{0}}
\expandafter\newcommand\csname dataS2.7\endcsname{\ensuremath{0}}
\expandafter\newcommand\csname dataS2.8\endcsname{\ensuremath{0}}
\expandafter\newcommand\csname dataS2.algol\endcsname{\ensuremath{0}}
\expandafter\newcommand\csname dataS3.1\endcsname{\ensuremath{0}}
\expandafter\newcommand\csname dataS3.2\endcsname{\ensuremath{0}}
\expandafter\newcommand\csname dataS3.3\endcsname{\ensuremath{0}}
\expandafter\newcommand\csname dataS3.4\endcsname{\ensuremath{0}}
\expandafter\newcommand\csname dataS3.5\endcsname{\ensuremath{0}}
\expandafter\newcommand\csname dataS3.6\endcsname{\ensuremath{0}}
\expandafter\newcommand\csname dataS3.7\endcsname{\ensuremath{0}}
\expandafter\newcommand\csname dataS3.8\endcsname{\ensuremath{0}}
\expandafter\newcommand\csname dataS3.algol\endcsname{\ensuremath{0}}
\expandafter\newcommand\csname dataS4.1\endcsname{\ensuremath{0}}
\expandafter\newcommand\csname dataS4.2\endcsname{\ensuremath{0}}
\expandafter\newcommand\csname dataS4.3\endcsname{\ensuremath{0}}
\expandafter\newcommand\csname dataS4.4\endcsname{\ensuremath{0}}
\expandafter\newcommand\csname dataS4.5\endcsname{\ensuremath{0}}
\expandafter\newcommand\csname dataS4.6\endcsname{\ensuremath{0}}
\expandafter\newcommand\csname dataS4.7\endcsname{\ensuremath{0}}
\expandafter\newcommand\csname dataS4.8\endcsname{\ensuremath{0}}
\expandafter\newcommand\csname dataS4.algol\endcsname{\ensuremath{0}}
\expandafter\newcommand\csname dataS5.1\endcsname{\ensuremath{0}}
\expandafter\newcommand\csname dataS5.2\endcsname{\ensuremath{0}}
\expandafter\newcommand\csname dataS5.3\endcsname{\ensuremath{0}}
\expandafter\newcommand\csname dataS5.4\endcsname{\ensuremath{0}}
\expandafter\newcommand\csname dataS5.5\endcsname{\ensuremath{0}}
\expandafter\newcommand\csname dataS5.6\endcsname{\ensuremath{0}}
\expandafter\newcommand\csname dataS5.7\endcsname{\ensuremath{0}}
\expandafter\newcommand\csname dataS5.8\endcsname{\ensuremath{0}}
\expandafter\newcommand\csname dataS5.algol\endcsname{\ensuremath{0}}
\expandafter\newcommand\csname dataS6.1\endcsname{\ensuremath{0}}
\expandafter\newcommand\csname dataS6.2\endcsname{\ensuremath{0}}
\expandafter\newcommand\csname dataS6.3\endcsname{\ensuremath{0}}
\expandafter\newcommand\csname dataS6.4\endcsname{\ensuremath{0}}
\expandafter\newcommand\csname dataS6.5\endcsname{\ensuremath{0}}
\expandafter\newcommand\csname dataS6.6\endcsname{\ensuremath{0}}
\expandafter\newcommand\csname dataS6.7\endcsname{\ensuremath{0}}
\expandafter\newcommand\csname dataS6.8\endcsname{\ensuremath{0}}
\expandafter\newcommand\csname dataS6.algol\endcsname{\ensuremath{0}}
\expandafter\newcommand\csname dataB1.1\endcsname{\ensuremath{1.7\pc}}
\expandafter\newcommand\csname dataB1.2\endcsname{\ensuremath{0.88\pc}}
\expandafter\newcommand\csname dataB1.3\endcsname{\ensuremath{99\pc}}
\expandafter\newcommand\csname dataB1.4\endcsname{\ensuremath{0.030\pc}}
\expandafter\newcommand\csname dataB1.5\endcsname{\ensuremath{88\pc}}
\expandafter\newcommand\csname dataB1.6\endcsname{\ensuremath{12\pc}}
\expandafter\newcommand\csname dataB1.7\endcsname{\ensuremath{16\pc}}
\expandafter\newcommand\csname dataB1.8\endcsname{\ensuremath{84\pc}}
\expandafter\newcommand\csname dataB1.algol\endcsname{\ensuremath{0.26\pc}}
\expandafter\newcommand\csname dataB2.1\endcsname{\ensuremath{1.0\pc}}
\expandafter\newcommand\csname dataB2.2\endcsname{\ensuremath{5.1\pc}}
\expandafter\newcommand\csname dataB2.3\endcsname{\ensuremath{95\pc}}
\expandafter\newcommand\csname dataB2.4\endcsname{\ensuremath{0.65\pc}}
\expandafter\newcommand\csname dataB2.5\endcsname{\ensuremath{99\pc}}
\expandafter\newcommand\csname dataB2.6\endcsname{\ensuremath{1.0\pc}}
\expandafter\newcommand\csname dataB2.7\endcsname{\ensuremath{11\pc}}
\expandafter\newcommand\csname dataB2.8\endcsname{\ensuremath{90\pc}}
\expandafter\newcommand\csname dataB2.algol\endcsname{\ensuremath{2.6\pc}}
\expandafter\newcommand\csname dataB3.1\endcsname{\ensuremath{1.2\pc}}
\expandafter\newcommand\csname dataB3.2\endcsname{\ensuremath{1.7\pc}}
\expandafter\newcommand\csname dataB3.3\endcsname{\ensuremath{98\pc}}
\expandafter\newcommand\csname dataB3.4\endcsname{\ensuremath{0.70\pc}}
\expandafter\newcommand\csname dataB3.5\endcsname{\ensuremath{99\pc}}
\expandafter\newcommand\csname dataB3.6\endcsname{\ensuremath{0.68\pc}}
\expandafter\newcommand\csname dataB3.7\endcsname{\ensuremath{9.1\pc}}
\expandafter\newcommand\csname dataB3.8\endcsname{\ensuremath{91\pc}}
\expandafter\newcommand\csname dataB3.algol\endcsname{\ensuremath{2.8\pc}}
\expandafter\newcommand\csname dataB4.1\endcsname{\ensuremath{2.8\pc}}
\expandafter\newcommand\csname dataB4.2\endcsname{\ensuremath{0.79\pc}}
\expandafter\newcommand\csname dataB4.3\endcsname{\ensuremath{99\pc}}
\expandafter\newcommand\csname dataB4.4\endcsname{\ensuremath{0}}
\expandafter\newcommand\csname dataB4.5\endcsname{\ensuremath{68\pc}}
\expandafter\newcommand\csname dataB4.6\endcsname{\ensuremath{32\pc}}
\expandafter\newcommand\csname dataB4.7\endcsname{\ensuremath{34\pc}}
\expandafter\newcommand\csname dataB4.8\endcsname{\ensuremath{66\pc}}
\expandafter\newcommand\csname dataB4.algol\endcsname{\ensuremath{0.16\pc}}
\expandafter\newcommand\csname dataB5.1\endcsname{\ensuremath{1.7\pc}}
\expandafter\newcommand\csname dataB5.2\endcsname{\ensuremath{0.89\pc}}
\expandafter\newcommand\csname dataB5.3\endcsname{\ensuremath{99\pc}}
\expandafter\newcommand\csname dataB5.4\endcsname{\ensuremath{0.030\pc}}
\expandafter\newcommand\csname dataB5.5\endcsname{\ensuremath{88\pc}}
\expandafter\newcommand\csname dataB5.6\endcsname{\ensuremath{12\pc}}
\expandafter\newcommand\csname dataB5.7\endcsname{\ensuremath{16\pc}}
\expandafter\newcommand\csname dataB5.8\endcsname{\ensuremath{84\pc}}
\expandafter\newcommand\csname dataB5.algol\endcsname{\ensuremath{0.26\pc}}
\expandafter\newcommand\csname dataB6.1\endcsname{\ensuremath{1.7\pc}}
\expandafter\newcommand\csname dataB6.2\endcsname{\ensuremath{0.90\pc}}
\expandafter\newcommand\csname dataB6.3\endcsname{\ensuremath{99\pc}}
\expandafter\newcommand\csname dataB6.4\endcsname{\ensuremath{0.030\pc}}
\expandafter\newcommand\csname dataB6.5\endcsname{\ensuremath{88\pc}}
\expandafter\newcommand\csname dataB6.6\endcsname{\ensuremath{12\pc}}
\expandafter\newcommand\csname dataB6.7\endcsname{\ensuremath{15\pc}}
\expandafter\newcommand\csname dataB6.8\endcsname{\ensuremath{85\pc}}
\expandafter\newcommand\csname dataB6.algol\endcsname{\ensuremath{0.26\pc}}
\expandafter\newcommand\csname dataB7.1\endcsname{\ensuremath{1.3\pc}}
\expandafter\newcommand\csname dataB7.2\endcsname{\ensuremath{0.84\pc}}
\expandafter\newcommand\csname dataB7.3\endcsname{\ensuremath{99\pc}}
\expandafter\newcommand\csname dataB7.4\endcsname{\ensuremath{0.030\pc}}
\expandafter\newcommand\csname dataB7.5\endcsname{\ensuremath{84\pc}}
\expandafter\newcommand\csname dataB7.6\endcsname{\ensuremath{16\pc}}
\expandafter\newcommand\csname dataB7.7\endcsname{\ensuremath{21\pc}}
\expandafter\newcommand\csname dataB7.8\endcsname{\ensuremath{79\pc}}
\expandafter\newcommand\csname dataB7.algol\endcsname{\ensuremath{0.27\pc}}
\expandafter\newcommand\csname dataB8.1\endcsname{\ensuremath{0.93\pc}}
\expandafter\newcommand\csname dataB8.2\endcsname{\ensuremath{0.65\pc}}
\expandafter\newcommand\csname dataB8.3\endcsname{\ensuremath{99\pc}}
\expandafter\newcommand\csname dataB8.4\endcsname{\ensuremath{0.020\pc}}
\expandafter\newcommand\csname dataB8.5\endcsname{\ensuremath{79\pc}}
\expandafter\newcommand\csname dataB8.6\endcsname{\ensuremath{21\pc}}
\expandafter\newcommand\csname dataB8.7\endcsname{\ensuremath{27\pc}}
\expandafter\newcommand\csname dataB8.8\endcsname{\ensuremath{73\pc}}
\expandafter\newcommand\csname dataB8.algol\endcsname{\ensuremath{0.25\pc}}
\expandafter\newcommand\csname dataB9.1\endcsname{\ensuremath{1.5\pc}}
\expandafter\newcommand\csname dataB9.2\endcsname{\ensuremath{0.96\pc}}
\expandafter\newcommand\csname dataB9.3\endcsname{\ensuremath{99\pc}}
\expandafter\newcommand\csname dataB9.4\endcsname{\ensuremath{0.030\pc}}
\expandafter\newcommand\csname dataB9.5\endcsname{\ensuremath{99\pc}}
\expandafter\newcommand\csname dataB9.6\endcsname{\ensuremath{0.67\pc}}
\expandafter\newcommand\csname dataB9.7\endcsname{\ensuremath{6.9\pc}}
\expandafter\newcommand\csname dataB9.8\endcsname{\ensuremath{93\pc}}
\expandafter\newcommand\csname dataB9.algol\endcsname{\ensuremath{0.26\pc}}
\expandafter\newcommand\csname dataB10.1\endcsname{\ensuremath{1.5\pc}}
\expandafter\newcommand\csname dataB10.2\endcsname{\ensuremath{0.98\pc}}
\expandafter\newcommand\csname dataB10.3\endcsname{\ensuremath{99\pc}}
\expandafter\newcommand\csname dataB10.4\endcsname{\ensuremath{0.030\pc}}
\expandafter\newcommand\csname dataB10.5\endcsname{\ensuremath{95\pc}}
\expandafter\newcommand\csname dataB10.6\endcsname{\ensuremath{4.7\pc}}
\expandafter\newcommand\csname dataB10.7\endcsname{\ensuremath{10\pc}}
\expandafter\newcommand\csname dataB10.8\endcsname{\ensuremath{90\pc}}
\expandafter\newcommand\csname dataB10.algol\endcsname{\ensuremath{0.26\pc}}
\expandafter\newcommand\csname dataB11.1\endcsname{\ensuremath{1.8\pc}}
\expandafter\newcommand\csname dataB11.2\endcsname{\ensuremath{0.83\pc}}
\expandafter\newcommand\csname dataB11.3\endcsname{\ensuremath{99\pc}}
\expandafter\newcommand\csname dataB11.4\endcsname{\ensuremath{0.030\pc}}
\expandafter\newcommand\csname dataB11.5\endcsname{\ensuremath{89\pc}}
\expandafter\newcommand\csname dataB11.6\endcsname{\ensuremath{11\pc}}
\expandafter\newcommand\csname dataB11.7\endcsname{\ensuremath{15\pc}}
\expandafter\newcommand\csname dataB11.8\endcsname{\ensuremath{85\pc}}
\expandafter\newcommand\csname dataB11.algol\endcsname{\ensuremath{0.50\pc}}
\expandafter\newcommand\csname dataB12.1\endcsname{\ensuremath{1.7\pc}}
\expandafter\newcommand\csname dataB12.2\endcsname{\ensuremath{1.1\pc}}
\expandafter\newcommand\csname dataB12.3\endcsname{\ensuremath{99\pc}}
\expandafter\newcommand\csname dataB12.4\endcsname{\ensuremath{0.040\pc}}
\expandafter\newcommand\csname dataB12.5\endcsname{\ensuremath{88\pc}}
\expandafter\newcommand\csname dataB12.6\endcsname{\ensuremath{12\pc}}
\expandafter\newcommand\csname dataB12.7\endcsname{\ensuremath{16\pc}}
\expandafter\newcommand\csname dataB12.8\endcsname{\ensuremath{84\pc}}
\expandafter\newcommand\csname dataB12.algol\endcsname{\ensuremath{0.15\pc}}
\expandafter\newcommand\csname dataB13.1\endcsname{\ensuremath{1.5\pc}}
\expandafter\newcommand\csname dataB13.2\endcsname{\ensuremath{0.36\pc}}
\expandafter\newcommand\csname dataB13.3\endcsname{\ensuremath{100\pc}}
\expandafter\newcommand\csname dataB13.4\endcsname{\ensuremath{0.010\pc}}
\expandafter\newcommand\csname dataB13.5\endcsname{\ensuremath{81\pc}}
\expandafter\newcommand\csname dataB13.6\endcsname{\ensuremath{19\pc}}
\expandafter\newcommand\csname dataB13.7\endcsname{\ensuremath{23\pc}}
\expandafter\newcommand\csname dataB13.8\endcsname{\ensuremath{77\pc}}
\expandafter\newcommand\csname dataB13.algol\endcsname{\ensuremath{0.19\pc}}
\expandafter\newcommand\csname dataB14.1\endcsname{\ensuremath{1.0\pc}}
\expandafter\newcommand\csname dataB14.2\endcsname{\ensuremath{0.14\pc}}
\expandafter\newcommand\csname dataB14.3\endcsname{\ensuremath{100\pc}}
\expandafter\newcommand\csname dataB14.4\endcsname{\ensuremath{0.010\pc}}
\expandafter\newcommand\csname dataB14.5\endcsname{\ensuremath{65\pc}}
\expandafter\newcommand\csname dataB14.6\endcsname{\ensuremath{35\pc}}
\expandafter\newcommand\csname dataB14.7\endcsname{\ensuremath{38\pc}}
\expandafter\newcommand\csname dataB14.8\endcsname{\ensuremath{62\pc}}
\expandafter\newcommand\csname dataB14.algol\endcsname{\ensuremath{0.13\pc}}
\expandafter\newcommand\csname dataB15.1\endcsname{\ensuremath{1.7\pc}}
\expandafter\newcommand\csname dataB15.2\endcsname{\ensuremath{0.88\pc}}
\expandafter\newcommand\csname dataB15.3\endcsname{\ensuremath{99\pc}}
\expandafter\newcommand\csname dataB15.4\endcsname{\ensuremath{0.030\pc}}
\expandafter\newcommand\csname dataB15.5\endcsname{\ensuremath{88\pc}}
\expandafter\newcommand\csname dataB15.6\endcsname{\ensuremath{12\pc}}
\expandafter\newcommand\csname dataB15.7\endcsname{\ensuremath{16\pc}}
\expandafter\newcommand\csname dataB15.8\endcsname{\ensuremath{84\pc}}
\expandafter\newcommand\csname dataB15.algol\endcsname{\ensuremath{0.26\pc}}
\expandafter\newcommand\csname dataB16.1\endcsname{\ensuremath{1.7\pc}}
\expandafter\newcommand\csname dataB16.2\endcsname{\ensuremath{0.88\pc}}
\expandafter\newcommand\csname dataB16.3\endcsname{\ensuremath{99\pc}}
\expandafter\newcommand\csname dataB16.4\endcsname{\ensuremath{0.030\pc}}
\expandafter\newcommand\csname dataB16.5\endcsname{\ensuremath{88\pc}}
\expandafter\newcommand\csname dataB16.6\endcsname{\ensuremath{12\pc}}
\expandafter\newcommand\csname dataB16.7\endcsname{\ensuremath{16\pc}}
\expandafter\newcommand\csname dataB16.8\endcsname{\ensuremath{84\pc}}
\expandafter\newcommand\csname dataB16.algol\endcsname{\ensuremath{0.26\pc}}
\expandafter\newcommand\csname dataB17.1\endcsname{\ensuremath{1.7\pc}}
\expandafter\newcommand\csname dataB17.2\endcsname{\ensuremath{0.88\pc}}
\expandafter\newcommand\csname dataB17.3\endcsname{\ensuremath{99\pc}}
\expandafter\newcommand\csname dataB17.4\endcsname{\ensuremath{0.030\pc}}
\expandafter\newcommand\csname dataB17.5\endcsname{\ensuremath{88\pc}}
\expandafter\newcommand\csname dataB17.6\endcsname{\ensuremath{12\pc}}
\expandafter\newcommand\csname dataB17.7\endcsname{\ensuremath{16\pc}}
\expandafter\newcommand\csname dataB17.8\endcsname{\ensuremath{84\pc}}
\expandafter\newcommand\csname dataB17.algol\endcsname{\ensuremath{0.26\pc}}
\expandafter\newcommand\csname dataB18.1\endcsname{\ensuremath{1.7\pc}}
\expandafter\newcommand\csname dataB18.2\endcsname{\ensuremath{0.88\pc}}
\expandafter\newcommand\csname dataB18.3\endcsname{\ensuremath{99\pc}}
\expandafter\newcommand\csname dataB18.4\endcsname{\ensuremath{0.030\pc}}
\expandafter\newcommand\csname dataB18.5\endcsname{\ensuremath{88\pc}}
\expandafter\newcommand\csname dataB18.6\endcsname{\ensuremath{12\pc}}
\expandafter\newcommand\csname dataB18.7\endcsname{\ensuremath{16\pc}}
\expandafter\newcommand\csname dataB18.8\endcsname{\ensuremath{84\pc}}
\expandafter\newcommand\csname dataB18.algol\endcsname{\ensuremath{0.26\pc}}
\expandafter\newcommand\csname dataB19.1\endcsname{\ensuremath{11\pc}}
\expandafter\newcommand\csname dataB19.2\endcsname{\ensuremath{0.45\pc}}
\expandafter\newcommand\csname dataB19.3\endcsname{\ensuremath{100\pc}}
\expandafter\newcommand\csname dataB19.4\endcsname{\ensuremath{0.010\pc}}
\expandafter\newcommand\csname dataB19.5\endcsname{\ensuremath{97\pc}}
\expandafter\newcommand\csname dataB19.6\endcsname{\ensuremath{2.9\pc}}
\expandafter\newcommand\csname dataB19.7\endcsname{\ensuremath{12\pc}}
\expandafter\newcommand\csname dataB19.8\endcsname{\ensuremath{88\pc}}
\expandafter\newcommand\csname dataB19.algol\endcsname{\ensuremath{1.9\pc}}
\expandafter\newcommand\csname dataB20.1\endcsname{\ensuremath{0.81\pc}}
\expandafter\newcommand\csname dataB20.2\endcsname{\ensuremath{0.68\pc}}
\expandafter\newcommand\csname dataB20.3\endcsname{\ensuremath{99\pc}}
\expandafter\newcommand\csname dataB20.4\endcsname{\ensuremath{0.070\pc}}
\expandafter\newcommand\csname dataB20.5\endcsname{\ensuremath{99\pc}}
\expandafter\newcommand\csname dataB20.6\endcsname{\ensuremath{1.4\pc}}
\expandafter\newcommand\csname dataB20.7\endcsname{\ensuremath{8.7\pc}}
\expandafter\newcommand\csname dataB20.8\endcsname{\ensuremath{91\pc}}
\expandafter\newcommand\csname dataB20.algol\endcsname{\ensuremath{0.68\pc}}
\expandafter\newcommand\csname dataB21.1\endcsname{\ensuremath{3.0\pc}}
\expandafter\newcommand\csname dataB21.2\endcsname{\ensuremath{0.41\pc}}
\expandafter\newcommand\csname dataB21.3\endcsname{\ensuremath{100\pc}}
\expandafter\newcommand\csname dataB21.4\endcsname{\ensuremath{0}}
\expandafter\newcommand\csname dataB21.5\endcsname{\ensuremath{93\pc}}
\expandafter\newcommand\csname dataB21.6\endcsname{\ensuremath{7.5\pc}}
\expandafter\newcommand\csname dataB21.7\endcsname{\ensuremath{11\pc}}
\expandafter\newcommand\csname dataB21.8\endcsname{\ensuremath{89\pc}}
\expandafter\newcommand\csname dataB21.algol\endcsname{\ensuremath{0.37\pc}}
\expandafter\newcommand\csname dataX.1\endcsname{\ensuremath{0.95\pc}}
\expandafter\newcommand\csname dataX.2\endcsname{\ensuremath{0.88\pc}}
\expandafter\newcommand\csname dataX.3\endcsname{\ensuremath{99\pc}}
\expandafter\newcommand\csname dataX.4\endcsname{\ensuremath{0.030\pc}}
\expandafter\newcommand\csname dataX.5\endcsname{\ensuremath{88\pc}}
\expandafter\newcommand\csname dataX.6\endcsname{\ensuremath{12\pc}}
\expandafter\newcommand\csname dataX.7\endcsname{\ensuremath{16\pc}}
\expandafter\newcommand\csname dataX.8\endcsname{\ensuremath{84\pc}}
\expandafter\newcommand\csname dataX.algol\endcsname{\ensuremath{0.14\pc}}
\expandafter\newcommand\csname dataY.1\endcsname{\ensuremath{1.7\pc}}
\expandafter\newcommand\csname dataY.2\endcsname{\ensuremath{0.41\pc}}
\expandafter\newcommand\csname dataY.3\endcsname{\ensuremath{100\pc}}
\expandafter\newcommand\csname dataY.4\endcsname{\ensuremath{0}}
\expandafter\newcommand\csname dataY.5\endcsname{\ensuremath{93\pc}}
\expandafter\newcommand\csname dataY.6\endcsname{\ensuremath{7.5\pc}}
\expandafter\newcommand\csname dataY.7\endcsname{\ensuremath{11\pc}}
\expandafter\newcommand\csname dataY.8\endcsname{\ensuremath{89\pc}}
\expandafter\newcommand\csname dataY.algol\endcsname{\ensuremath{0.21\pc}}

%% file: CN_extra_data_macros.tex
\expandafter\newcommand\csname dataB20.M1.0\endcsname{\ensuremath{8.91\pc}}
\expandafter\newcommand\csname dataB20.M1.1\endcsname{\ensuremath{2.32\pc}}
\expandafter\newcommand\csname dataB20.M1.2\endcsname{\ensuremath{1.02\pc}}

%% file: results.tex
\begin{tabular}{cc | ccccccc}
\hline
 Model   &   Of all giants   &   \multicolumn{2}{l}{Of giants with $M>1.3\,\mathrm{M}_{\odot}$}   &    &    &    &  \tabularnewline 
  & $M>1.3\,\mathrm{M}_{\odot}$  & $\left[\mathrm{C}/\mathrm{N}\right]\geq 0$  & $\left[\mathrm{C}/\mathrm{N}\right]< 0$  & $\left[\mathrm{C}/\mathrm{N}\right]> 0.5$  & Single  & Binary  & was BSS  & never BSS \tabularnewline 
\hline
S1 & \csname dataS1.1\endcsname  & \csname dataS1.2\endcsname  & \csname dataS1.3\endcsname  & \csname dataS1.4\endcsname  & \csname dataS1.5\endcsname  & \csname dataS1.6\endcsname  & \csname dataS1.7\endcsname  & \csname dataS1.8\endcsname \tabularnewline 
S2 & \csname dataS2.1\endcsname  & \csname dataS2.2\endcsname  & \csname dataS2.3\endcsname  & \csname dataS2.4\endcsname  & \csname dataS2.5\endcsname  & \csname dataS2.6\endcsname  & \csname dataS2.7\endcsname  & \csname dataS2.8\endcsname \tabularnewline 
S3 & \csname dataS3.1\endcsname  & \csname dataS3.2\endcsname  & \csname dataS3.3\endcsname  & \csname dataS3.4\endcsname  & \csname dataS3.5\endcsname  & \csname dataS3.6\endcsname  & \csname dataS3.7\endcsname  & \csname dataS3.8\endcsname \tabularnewline 
S4 & \csname dataS4.1\endcsname  & \csname dataS4.2\endcsname  & \csname dataS4.3\endcsname  & \csname dataS4.4\endcsname  & \csname dataS4.5\endcsname  & \csname dataS4.6\endcsname  & \csname dataS4.7\endcsname  & \csname dataS4.8\endcsname \tabularnewline 
S5 & \csname dataS5.1\endcsname  & \csname dataS5.2\endcsname  & \csname dataS5.3\endcsname  & \csname dataS5.4\endcsname  & \csname dataS5.5\endcsname  & \csname dataS5.6\endcsname  & \csname dataS5.7\endcsname  & \csname dataS5.8\endcsname \tabularnewline 
S6 & \csname dataS6.1\endcsname  & \csname dataS6.2\endcsname  & \csname dataS6.3\endcsname  & \csname dataS6.4\endcsname  & \csname dataS6.5\endcsname  & \csname dataS6.6\endcsname  & \csname dataS6.7\endcsname  & \csname dataS6.8\endcsname \tabularnewline 
B1 & \csname dataB1.1\endcsname  & \csname dataB1.2\endcsname  & \csname dataB1.3\endcsname  & \csname dataB1.4\endcsname  & \csname dataB1.5\endcsname  & \csname dataB1.6\endcsname  & \csname dataB1.7\endcsname  & \csname dataB1.8\endcsname \tabularnewline 
B2 & \csname dataB2.1\endcsname  & \csname dataB2.2\endcsname  & \csname dataB2.3\endcsname  & \csname dataB2.4\endcsname  & \csname dataB2.5\endcsname  & \csname dataB2.6\endcsname  & \csname dataB2.7\endcsname  & \csname dataB2.8\endcsname \tabularnewline 
B3 & \csname dataB3.1\endcsname  & \csname dataB3.2\endcsname  & \csname dataB3.3\endcsname  & \csname dataB3.4\endcsname  & \csname dataB3.5\endcsname  & \csname dataB3.6\endcsname  & \csname dataB3.7\endcsname  & \csname dataB3.8\endcsname \tabularnewline 
B4 & \csname dataB4.1\endcsname  & \csname dataB4.2\endcsname  & \csname dataB4.3\endcsname  & \csname dataB4.4\endcsname  & \csname dataB4.5\endcsname  & \csname dataB4.6\endcsname  & \csname dataB4.7\endcsname  & \csname dataB4.8\endcsname \tabularnewline 
B5 & \csname dataB5.1\endcsname  & \csname dataB5.2\endcsname  & \csname dataB5.3\endcsname  & \csname dataB5.4\endcsname  & \csname dataB5.5\endcsname  & \csname dataB5.6\endcsname  & \csname dataB5.7\endcsname  & \csname dataB5.8\endcsname \tabularnewline 
B6 & \csname dataB6.1\endcsname  & \csname dataB6.2\endcsname  & \csname dataB6.3\endcsname  & \csname dataB6.4\endcsname  & \csname dataB6.5\endcsname  & \csname dataB6.6\endcsname  & \csname dataB6.7\endcsname  & \csname dataB6.8\endcsname \tabularnewline 
B7 & \csname dataB7.1\endcsname  & \csname dataB7.2\endcsname  & \csname dataB7.3\endcsname  & \csname dataB7.4\endcsname  & \csname dataB7.5\endcsname  & \csname dataB7.6\endcsname  & \csname dataB7.7\endcsname  & \csname dataB7.8\endcsname \tabularnewline 
B8 & \csname dataB8.1\endcsname  & \csname dataB8.2\endcsname  & \csname dataB8.3\endcsname  & \csname dataB8.4\endcsname  & \csname dataB8.5\endcsname  & \csname dataB8.6\endcsname  & \csname dataB8.7\endcsname  & \csname dataB8.8\endcsname \tabularnewline 
B9 & \csname dataB9.1\endcsname  & \csname dataB9.2\endcsname  & \csname dataB9.3\endcsname  & \csname dataB9.4\endcsname  & \csname dataB9.5\endcsname  & \csname dataB9.6\endcsname  & \csname dataB9.7\endcsname  & \csname dataB9.8\endcsname \tabularnewline 
B10 & \csname dataB10.1\endcsname  & \csname dataB10.2\endcsname  & \csname dataB10.3\endcsname  & \csname dataB10.4\endcsname  & \csname dataB10.5\endcsname  & \csname dataB10.6\endcsname  & \csname dataB10.7\endcsname  & \csname dataB10.8\endcsname \tabularnewline 
B11 & \csname dataB11.1\endcsname  & \csname dataB11.2\endcsname  & \csname dataB11.3\endcsname  & \csname dataB11.4\endcsname  & \csname dataB11.5\endcsname  & \csname dataB11.6\endcsname  & \csname dataB11.7\endcsname  & \csname dataB11.8\endcsname \tabularnewline 
B12 & \csname dataB12.1\endcsname  & \csname dataB12.2\endcsname  & \csname dataB12.3\endcsname  & \csname dataB12.4\endcsname  & \csname dataB12.5\endcsname  & \csname dataB12.6\endcsname  & \csname dataB12.7\endcsname  & \csname dataB12.8\endcsname \tabularnewline 
B13 & \csname dataB13.1\endcsname  & \csname dataB13.2\endcsname  & \csname dataB13.3\endcsname  & \csname dataB13.4\endcsname  & \csname dataB13.5\endcsname  & \csname dataB13.6\endcsname  & \csname dataB13.7\endcsname  & \csname dataB13.8\endcsname \tabularnewline 
B14 & \csname dataB14.1\endcsname  & \csname dataB14.2\endcsname  & \csname dataB14.3\endcsname  & \csname dataB14.4\endcsname  & \csname dataB14.5\endcsname  & \csname dataB14.6\endcsname  & \csname dataB14.7\endcsname  & \csname dataB14.8\endcsname \tabularnewline 
B15 & \csname dataB15.1\endcsname  & \csname dataB15.2\endcsname  & \csname dataB15.3\endcsname  & \csname dataB15.4\endcsname  & \csname dataB15.5\endcsname  & \csname dataB15.6\endcsname  & \csname dataB15.7\endcsname  & \csname dataB15.8\endcsname \tabularnewline 
B16 & \csname dataB16.1\endcsname  & \csname dataB16.2\endcsname  & \csname dataB16.3\endcsname  & \csname dataB16.4\endcsname  & \csname dataB16.5\endcsname  & \csname dataB16.6\endcsname  & \csname dataB16.7\endcsname  & \csname dataB16.8\endcsname \tabularnewline 
B17 & \csname dataB17.1\endcsname  & \csname dataB17.2\endcsname  & \csname dataB17.3\endcsname  & \csname dataB17.4\endcsname  & \csname dataB17.5\endcsname  & \csname dataB17.6\endcsname  & \csname dataB17.7\endcsname  & \csname dataB17.8\endcsname \tabularnewline 
B18 & \csname dataB18.1\endcsname  & \csname dataB18.2\endcsname  & \csname dataB18.3\endcsname  & \csname dataB18.4\endcsname  & \csname dataB18.5\endcsname  & \csname dataB18.6\endcsname  & \csname dataB18.7\endcsname  & \csname dataB18.8\endcsname \tabularnewline 
B19 & \csname dataB19.1\endcsname  & \csname dataB19.2\endcsname  & \csname dataB19.3\endcsname  & \csname dataB19.4\endcsname  & \csname dataB19.5\endcsname  & \csname dataB19.6\endcsname  & \csname dataB19.7\endcsname  & \csname dataB19.8\endcsname \tabularnewline 
B20 & \csname dataB20.1\endcsname  & \csname dataB20.2\endcsname  & \csname dataB20.3\endcsname  & \csname dataB20.4\endcsname  & \csname dataB20.5\endcsname  & \csname dataB20.6\endcsname  & \csname dataB20.7\endcsname  & \csname dataB20.8\endcsname \tabularnewline 
B21 & \csname dataB21.1\endcsname  & \csname dataB21.2\endcsname  & \csname dataB21.3\endcsname  & \csname dataB21.4\endcsname  & \csname dataB21.5\endcsname  & \csname dataB21.6\endcsname  & \csname dataB21.7\endcsname  & \csname dataB21.8\endcsname \tabularnewline 
X & \csname dataX.1\endcsname  & \csname dataX.2\endcsname  & \csname dataX.3\endcsname  & \csname dataX.4\endcsname  & \csname dataX.5\endcsname  & \csname dataX.6\endcsname  & \csname dataX.7\endcsname  & \csname dataX.8\endcsname \tabularnewline 
Y & \csname dataY.1\endcsname  & \csname dataY.2\endcsname  & \csname dataY.3\endcsname  & \csname dataY.4\endcsname  & \csname dataY.5\endcsname  & \csname dataY.6\endcsname  & \csname dataY.7\endcsname  & \csname dataY.8\endcsname \tabularnewline 
\hline

\end{tabular}